\documentclass[%
 reprint,
superscriptaddress,
 amsmath,amssymb,
 aps,
]{revtex4-2}
\usepackage{mathtools}
\usepackage[colorlinks=true,citecolor=blue,linkcolor=blue,urlcolor=blue]{hyperref}
\usepackage{color}
\DeclarePairedDelimiter\bra{\langle}{\rvert}
\DeclarePairedDelimiter\ket{\lvert}{\rangle}
\DeclarePairedDelimiterX\braket[2]{\langle}{\rangle}
{#1\,\delimsize\vert\,\mathopen{}#2}
 \usepackage{bm}
\usepackage{graphicx}
\usepackage{dcolumn}
\usepackage{bm}
\def\tr{ {\rm{Tr }}\,}

\begin{document}

\preprint{APS/123-QED}

\title{Information acquisition, scrambling, and sensitivity to errors in quantum chaos}






\author{Sreeram PG}
\email{sreerampg@outlook.com}
\affiliation{Department of Physics, Indian Institute of Science Education and Research, Pune 411008,  India}
\author{Abinash Sahu}
\affiliation{Bhabha Atomic Research Centre,  Anushaktinagar, Mumbai 400094, India}
\author{Naga Dileep Varikuti}
\affiliation{Department of Physics, Indian Institute of Technology Madras, Chennai 600036, India}
\author{Bishal Kumar Das}
\affiliation{Department of Physics, Indian Institute of Technology Madras, Chennai 600036, India}
\author{Sourav Manna}
\affiliation{Department of Physics, Indian Institute of Technology Madras, Chennai 600036, India}
\author{Vaibhav Madhok}
\email{madhok@physics.iitm.ac.in}
\affiliation{Department of Physics, Indian Institute of Technology Madras, Chennai 600036, India}

\begin{abstract}
Quantum chaos is the study of footprints of classical chaos in the quantum world. 
The quantum signatures
of chaos can be understood by studying quantum systems whose classical counterpart is chaotic. However, the concepts of integrability, non-integrability and chaos extend to systems without a classical analogue.
Here, we first review the classical route from order into chaos. Since nature is fundamentally quantum, we discuss how chaos manifests in the quantum domain. We briefly describe semi-classical methods, 
 and discuss the consequences of chaos in quantum information processing. 
 We review the quantum version of Lyapunov exponents, as quantified by the out-of-time ordered correlators (OTOC), Kolmogorov-Sinai (KS) entropy and sensitivity to errors. We then review the study of signatures of quantum chaos using quantum tomography. 
Classically, if we know the dynamics exactly,  as we maintain a constant coarse-grained tracking of the trajectory, we gain exponentially fine-grained information about the initial condition. In the quantum setting, as we track the measurement record with fixed signal-to-noise, we gain increasing information about the initial condition. In the process, we have given a new quantification of operator spreading in Krylov subspaces with quantum state reconstruction.
 The study of these signatures is not only of theoretical interest but also of practical importance. 
\end{abstract}

\maketitle


The popular view of chaos, often nothing short of mesmerizing, tells us that even a small flap of butterfly wings can bring unexpected changes like storms a large distance away. One can start by asking the question, what is chaos? In classical physics, a bounded dynamical system exhibiting exponential sensitivity to initial conditions is said to be chaotic \cite{ott2002chaos, strogatz2018nonlinear, devaney2018introduction}.
This means two nearby phase space trajectories separate exponentially 
at a rate given by the characteristic Lyapunov exponent (LE) of the system. As a result, small disturbances in the initial conditions of a perfectly deterministic system may lead to completely different trajectories in the phase space as the system evolves over a sufficient length of time. This is what is popularly known as the ``butterfly effect" --- the flap of wings of a butterfly in Brazil can cause a tornado in China \cite{lorenz1963deterministic}. Therefore, the system loses its long-term predictability even though the equations describing its dynamics are completely deterministic. In such a scenario, the trajectories typically do not settle on fixed points, periodic orbits, or limit cycles \cite{ott2002chaos, strogatz2018nonlinear, devaney2018introduction}. Despite extensive study of chaos in dynamical systems, no universally accepted mathematical definition exists. According to  Devaney's popular text on chaos \cite{devaney2018introduction}, a system can be considered chaotic if it meets any two of the following three criteria: (i) sensitive dependence on initial conditions, (ii) Topological transitivity and (iii) existence of a dense set of periodic orbits. Infact, meeting any two of the above automatically implies the third one \cite{banks1992devaney}. 

 


While classical physics can explain the macroscopic world, quantum mechanics is essential for a completely accurate description of the world at the atomic scale. The double slit experiment, quantum tunneling, photon statistics, and electron diffraction are a few phenomena that have no classical explanations \cite{scully1997quantum, shankar2012principles}. Then, a natural pivotal question to ask is the presence of chaos at the atomic level.
This question has been a much-debated issue due to the fact that the unitary evolution of quantum mechanics preserves the inner product between two initial state vectors. Therefore, the initial disturbance doesn't increase exponentially. Instead, it stays constant throughout the evolution. However, suppose one describes classical mechanics in terms of the evolution of probability densities. In that case, one can show
that even classical dynamics preserve the overlap of initial probability densities throughout evolution \cite{koopman1931hamiltonian, neumann1932operatorenmethode, jordan1961lie}. This seems reasonable as the analogue of a wave function in quantum mechanics is a probability density in classical phase space. Chaos, as characterized by the sensitivity to initial conditions, refers to the exponential departure of nearby \textit{trajectories} and not \textit{probabilities}. Although this explanation motivates one to look beyond the overlap of state vectors in quantum mechanics, this does not answer several pertinent questions: How far does our classical understanding of chaos lead us into the quantum world? Are there ways to characterize chaos in quantum systems the way Lyapunov exponent does for classically chaotic systems? How does chaos emerge from the underlying quantum dynamics in the classical world? If there exists a notion of quantum chaos, what is the route of chaos starting from an integrable system? Classically, this problem has been extensively studied in the last century, starting from Poincaré and culminating in the celebrated Kolmogorov-Arnold-Moser (KAM) theorem \cite{arnold2009proof, kolmogorov1954conservation, moser1962invariant, moser1967convergent, dumas2014kam, Moser+2001, poschel2009lecture}. 

It must be emphasized here that searching for a proper characterization of quantum chaos is not simply fixing a ``definition". There are profound consequences in the macroscopic world that are intimately tied to quantum chaos. Ehrenfest correspondence principle (also known as classical-quantum correspondence) tells us the length of time till which quantum expectation values of observables are going to follow classical equations of motion. The characteristic ``break-time" when these two depart is exponentially small if the underlying dynamics of the system in the classical limit are chaotic as opposed to regular \cite{berman1978condition, toda1987quantal, gu1990evidences}.  Therefore, understanding the dynamics of quantum wave packets as they stretch and fold, causing the interference effects to become significant for the expectation values to depart from classical trajectories, has profound dynamical consequences. This line of thought was picked up by Zurek, who reached provocative conclusions by suggesting that Hyperion, one of the moons of Saturn, should deviate from its trajectory within 20 years \cite{zurek1998decoherence}! His solution to why this has not happened was the role played by environmental decoherence, which suppresses the quantum effects responsible for the breakdown of Ehrenfest correspondence. The role of chaos in the rate of decoherence was studied by Zurek and Juan Pablo Paz, who showed that Lyapunov exponents should determine how rapidly the system decoheres \cite{paz2002environment}. Furthermore, the connections of non-linear dynamics, chaos, and thermalization are at the cornerstone of statistical mechanics. How rapidly and under what conditions an isolated quantum system thermalizes is connected to the underlying chaos in the system dynamics. 


{The study of quantum signatures of classical chaos involves a broad spectrum of overlapping research directions. One studies Hamiltonian systems whose classical analogues are chaotic to unravel the correspondence between quantum and classical theories. One way to see the signatures of chaos is to look at the statistical properties of the Hamiltonian spectra. The distribution of the difference between nearest eigenvalues (level spacing distributions) behaves differently for chaotic and integrable dynamics. If the classical analogue of the quantum system is chaotic, then the level spacing distribution follows that of a suitable random matrix ensemble \citep{bohigas1984characterization}. On the other hand, if classical dynamics is integrable, the level spacings follow a Poissonian distribution \citep{berry1977level}}.
{Random matrix theory has been used to characterize and quantify some properties of chaos in quantum systems. Eugene Wigner pioneered the use of random matrices to model large nuclei \citep{wigner55}. He found that such heavy nuclei are so complicated that their Hamiltonians act like  random matrices. Some spectral properties of the system simply follow from the random matrix. This  carries over to the chaotic Hamiltonians too.}

{The study of level statistics, semiclassical approximations of the spectrum, and connections to the periodic orbit theory have been of focus traditionally \citep{berry1977level, bohigas1971spacing,Haake,delande1986quantum,  atas2013joint, bhosale2018scaling, tekur2018exact, tekur2018higher}. 
{Studying the properties of eigenstates of the Hamiltonian (level statistics) and their localization can tell us a great deal about the system. However, the dynamics reveals interesting physics, like environment assisted decoherence and scrambling of information. Traditionally, there has been a stress on the ``energy domain" since the experimental toolkit to control quantum systems was missing. Hence, one could not perform an experiment in which the system was prepared in a pure state and then monitored with precision. With recent developments in atomic, molecular, and optical physics, and progress towards quantum simulators with superconducting qubits, this is now indeed possible.
	The study of quantum systems is propelled by experiments where focus is on the ability to control quantum systems \citep{Chaudhary, poggi2020quantifying}, thermalization in closed quantum systems \citep{Neill16} and quantum simulations of chaotic and non-integrable Hamiltonians \citep{sieberer2019digital, PhysRevX.7.031011}.} 
	
	{Signatures of  chaos
	based on statistical spectral properties, such as
	nearest-neighbor spacing statistics \citep{Haake}, may not work for systems with small Hilbert spaces, since statistical analysis may result in misleading conclusions for small sample sizes. Alternatively, one can study the dynamics of correlation functions in the system.  Ehrenfest correspondence gives us a time scale upto which quantum expectation values of observables agrees with their classical counterparts. 
 It depends on the the effective Planck constant $h_{\text{eff}}$ of the systems and is inversely proportional to the size of the Hilbert space.
	For chaotic systems, the Ehrenfest correspondence time $E_f\sim \log(1/h_{\text{eff}})/\lambda_C$, where $\lambda_C$ is the classical Lyapunov exponent
	are very short. The quantum correlators have a classical correspondence only in this short period \cite{shepelyansky1983some}.}

Quantum chaos also has a crucial role to play in quantum information processing. This is because, though quantum systems do not show sensitivity to initial states, they can show sensitivity to small changes in the Hamiltonian parameters \cite{peres1984stability}. This sensitivity presents challenges in simulating the quantum chaotic systems on various quantum platforms, especially in the face of hardware errors \cite{chinni2022trotter}. Moreover, in quantum control protocols, one has to reach a target state despite many-body chaos and unavoidable fluctuations in the control dynamics \cite{tomsovic2023controlling}. Therefore, the design of control protocols must consider the consequences of chaos. It is also worth noting that the sensitivity of quantum systems can be a resource in quantum parameter estimation protocols due to its relation with the quantum Fisher information.
Hence, understanding quantum chaos can lead to harnessing the properties of quantum systems to develop superior information processing devices for computation, communication, and metrology. On the other hand, quantum chaos can address several foundational questions regarding the quantum-to-classical transition, defining the quantum-classical border, and the emergence of classical chaos from the underlying quantum mechanics.



This review article aligns with the overarching context we have outlined so far. We do not attempt an exhaustive review of all the signatures of chaos listed in the quantum chaos literature. Rather, we focus on a conceptual understanding of chaos in quantum systems in the first two sections, starting from the mechanism of classical chaos. We give a summary of statistical signatures of chaos based on random matrix theory in section \ref{sec 3}. We then proceed to two important dynamical quantifiers called out-of-time ordered correlators and Loschmidt echo. The last four sections are devoted to the signatures obtained from the tomography of an initially unknown state, which is relatively less popular in the literature. We review the effect of chaos in the information gain and the operator spread while performing tomography. The interplay between chaos and noisy errors in tomography is discussed in the final section before we conclude.  Since even the term ``quantum chaos'' can still confuse the uninitiated, we hope this review gives a flavour of the major research themes in the field. The relevance of understanding the effects of chaos in quantum systems is not merely of theoretical interest but of consequence in the anticipated quantum computing revolution. 

\section{Classical Integrability to chaos --- via KAM theorem}
Integrability is a centuries-old concept and cornerstone of classical physics. A classical Hamiltonian system of $d$-degrees of freedom can be characterized by a $2d$-number of first-order differential equations involving pairs of canonically conjugate variables. The integrability is traditionally referred to as the exact solvability of these equations by means of standard integration techniques, thereby writing the evolution of phase space variables in terms of elementary mathematical functions of time \cite{dumas2014kam}. Whenever a direct integration is not feasible, the conventional approach is to seek solutions by changing variables via canonical transformations. However, whether and when such a transformation exists for the given Hamiltonian is a hard question to answer. In 1853, J. Liouville first provided a rigorous answer to this question, showing that the integrability requires at least $d$-number of independent constants of motion (COMs), which are in complete involution. The COMs are smooth functions over the phase space that do not change with time along the phase space trajectories, and the involution means that their Poisson brackets mutually vanish. It is to be noted that a system of $2d$ independent linear differential equations would require $2d-1$ COMs for complete solvability. Thanks to the symplectic structure of the Hamiltonian dynamics, only $d$-COMs are needed \cite{babelon2003introduction, arnol2013mathematical}. 
Then, the Hamiltonian can be transformed into a new set of action-angle variables, where the COMs play the role of the actions, and the angle coordinates remain cyclic. If $\{I_i\}_{i=1}^{d}$ denote the action variables and $\{\theta_i\}_{i=1}^{d}$ are the corresponding angle variables, Hamilton's equations of motion are given by
$I_i(t)=I_i(0)=\text{const}$, and $\theta_i(t)=(\omega_i t+\theta_0)\mod 1 $, where $\omega_i=\dfrac{\partial H(I_1, I_2, \cdots, I_d)}{\partial \theta_i}$ are interpreted as characteristic frequencies. Fixing each of the COM to a constant value results in a $d$-dimensional surface in the $2d$-dimensional phase space. This surface has the properties of a $d$-torus and is usually termed an invariant torus --- any trajectory set out on this torus traverses on it forever with the characteristic frequencies. These tori are said to be resonant if the trajectories always return to their initial positions after a certain time. Otherwise, if the trajectories never return to their initial conditions, such tori are called non-resonant. The corresponding trajectories are usually referred to as quasi-periodic. When the Hamiltonian is non-degenerate, meaning the characteristic frequencies are non-linear functions of the actions, resonant tori are distributed among non-resonant ones in a manner analogous to the distribution of rational numbers among irrational numbers on the real line $(\mathbb{R})$. Non-degeneracy is crucial in establishing the stability of the integrable systems against small generic perturbations. While integrable systems are generally scarce in nature, they still accurately describe much of the physical world as we know it. Notable examples of these systems include Kepler's two-body problem, the harmonic oscillator, and Euler tops \cite{reichl2021transition}.

\subsection{Kolmogorov-Arnold-Moser stability}

In the early twentieth century, mathematicians and physicists actively investigated the robustness of integrable systems in the presence of weak perturbations. Suppose $H(I)$ denotes the initial integrable Hamiltonian in the actin-angle coordinates. Then, the perturbed Hamiltonian will take the form $H'(I, \theta)=H(I)+\varepsilon f(I, \theta, t)$, where $\varepsilon$ is the perturbation strength and $f(I, \theta, t)$ denotes the perturbation. Then, this pertinent question was asked: \textit{What happens to the conserved quantities when an integrable system is subjected to a smooth generic and weak perturbation that preserves the symplectic structure of Hamilton's equations of motion?}. This is equivalent to asking if the perturbed system admits new COMs derived from the perturbative expansion of the original system's COMs. 
This problem was initially tackled by Poincaré, who observed that the presence of the resonant tori leads to the divergence of the perturbative expansion \cite{poschel2009lecture}. The sketch of this result involves seeking a generator $(\mathcal{\eta})$ of a symplectic transformation, which aims to convert the perturbed Hamiltonian into a new set of action-angle variables $\{(J_i, \phi_i)\}_{i=1}^{d}$ such that the final Hamiltonian remains free of $\phi_i$s up to the first order in the perturbation strength. The result further involves expanding both the generator and the perturbation in the Fourier series, subsequently leading to a relation between the Fourier coefficients of the same: $\eta_{K}(J)=-\dfrac{f_{k}(J)}{(2\pi i)(K.\hat{\omega})}$, where $K\in\mathbb{Z}^{d}-\{\hat{0}\}$ and $\hat{\omega}$ is the vector of characteristic frequencies. The Fourier coefficients diverge whenever \( K \cdot \hat{\omega} = 0 \), which is indeed the case for resonant tori. Additionally, even when the tori are not resonant, there are situations where \( K \cdot \hat{\omega} \) takes very small values, resulting in the non-convergence of the Fourier series expansion of the generator. This result became widely known as the \textit{small-devisor problem}. A naive interpretation of this result without further examination implies that integrable systems are generally unstable under perturbations and tend to become ergodic even when the perturbation is small. Moreover, it indicates that, in general, any generic system is non-integrable \cite{genecand1993transversal, markus1974generic, zehnder1973homoclinic}. However, this interpretation was later proven incorrect, although Poincaré's result itself was not false but rather an incomplete solution. Later advancements due to Kolmogorov, Arnold, and Moser revealed that the integrable systems, under certain conditions, are indeed stable. A definitive set of results, collectively known as the Kolmogorov-Arnold-Moser (KAM) theorem, was formulated to establish this stability, ultimately resolving the problem of the stability of integrable systems \cite{arnold2009proof, kolmogorov1954conservation, moser1962invariant, moser1967convergent, dumas2014kam, Moser+2001, poschel2009lecture}.

The first result of the KAM theorem was due to Kolmogorov, who adopted a different approach to study the small-divisor problem. Recall that the non-resonant tori occupy the phase space with a measure of one in non-degenerate systems, while the resonant tori only constitute a measure zero set. Instead of directly examining the generator function over the entire phase space, Kolmogorov was more interested in studying the stability of individual non-resonant tori and, consequently, the quasiperiodic trajectories over them. This prompted him to study a special set of invariant tori, excluding the resonant and the near-resonant tori. This set is characterized by the frequency vectors as $\mathcal{M}\equiv\{\omega\in \mathbb{R}^{d}| |K\cdot \omega|\geq \alpha/|K|^{\tau}\text{ for all }K\in\mathbb{Z}^{d} \text{ and }K\neq \hat{0}\}$, where $|K|=|K_1|+\cdots +|K_{d}|$, $\alpha>0$, and $\tau>n-1$. It turned out that the invariant tori corresponding to this set are highly non-resonant. Then, the Fourier series expansion of the generating function converges on this set, establishing the stability of the non-resonant tori in the phase space. Under weak perturbations, these tori get slightly deformed. This indicates that the non-degenerate integrable systems generally are stable under perturbations \cite{arnold2009proof, kolmogorov1954conservation, moser1962invariant, moser1967convergent, poschel2009lecture, reichl2021transition}.
Moreover, this set constitutes the full measure over the phase space while the complementary set with resonant and near-resonant tori constitutes measure zero in the phase space. This result was further refined and alternative proofs were provided by Arnold and Moser. The KAM theorem enables the application of perturbation theory to study the dynamics of perturbed integrable systems, which is often applicable to real-world systems.

\subsection{Routes to chaos}
\textbf{KAM route to chaos:}
When the perturbation strength is weak enough, the KAM theorem ensures the regularity of the typical (non-degenerate) near-integrable systems. As the perturbation strength increases, the resonant tori break first. The non-resonant tori follow the resonant ones as the perturbation is further increased. Larger perturbation strengths inevitably lead to the breaking of all invariant tori, rendering the system completely non-integrable. Most of the physical systems existing in nature are generally non-integrable. Examples include the famous three-body problem, double pendulum, and anharmonic oscillators. When no conserved quantities are present, these systems can exhibit ergodic properties. This means that a single phase-space trajectory can fill the entire phase-space region as it evolves indefinitely over time. However, it is important to note that the non-integrability itself is not a sufficient condition for ergodicity, and the latter is slightly a stronger condition for chaos than the former. The following hierarchy characterizes various levels of ergodicity in the dynamical systems \cite{cornfeld2012ergodic, halmos2017lectures, sep-ergodic-hierarchy}:
\begin{widetext}
    \begin{eqnarray*}
\text{Bernoulli}\subset\text{Kolmogorov}\subset\text{Strong mixing}\subset\text{Weak mixing}\subset\text{Ergodic}.
\end{eqnarray*}
\end{widetext}

Recall that the sensitive dependence on the initial conditions is one of the three main characteristics of chaotic systems. Let $\hat{X}(0)$ denote an initial condition of a phase space trajectory corresponding to an arbitrary chaotic system, and $\hat{X}(t)$ be its location after $t$-time steps. Then, another trajectory with a slightly different initial condition $\hat{X}(0)+\delta \hat{X}$ shows an exponential separation from the previous trajectory as it evolves, i.e., $\delta \hat{X}(t)/\delta \hat{X}(0)\sim e^{\lambda_{C} t}$, where $\lambda_{C}$ is the maximum positive Lyapunov exponent. For chaotic systems, the maximum Lyapunov exponent is always positive. In the hierarchy, the Bernoulli and Kolmogorov systems display chaotic behavior.

\textbf{Non-KAM route to chaos:}
It is known that the generic integrable systems are KAM stable. However, the KAM theorem itself relies on a few key assumptions. For instance, the theorem assumes that the unperturbed Hamiltonian is non-degenerate. Here, non-degeneracy means that the Hamiltonian can be expressed as a nonlinear function of only the action variables. Additionally, the frequency ratios must be sufficiently irrational. If these conditions are not met, the phase space tori are likely to break at any finite perturbation, leading to structural instabilities. Near these instabilities, which are also called resonances, the system's integrability gets completely lost. Most integrable systems meet the KAM conditions. However, there exists a class of systems known as non-KAM systems that do not follow the typical KAM route to non-integrability \cite{sankaranarayanan2001quantum, sankaranarayanan2001chaos}. These non-KAM systems exhibit large-scale structural changes at resonances in the presence of perturbations. In the classical phase space, such resonances are often linked to the breaking of invariant tori and the creation of stable and unstable phase space manifolds. This mechanism can lead to diffusive chaos even with very small perturbations. Consequently, non-KAM systems are highly sensitive to small changes in system parameters at resonances. The harmonic oscillator Hamiltonian is degenerate and serves as an example of non-KAM integrable systems. 

\section{From classical to quantum chaos via semi-classics}

How does chaos manifest itself in the quantum world? Characterizing quantum chaos has been a contentious issue given the often-stated fact that classical dynamics can be nonlinear, whereas unitary quantum dynamics is linear.  A deeper analysis, nonetheless, reveals the dynamical signatures of chaos in quantum systems and raises fundamental questions regarding dynamical systems and their properties as described by quantum and {classical mechanics \cite{Haake, gutzwiller2013chaos}}, which we discuss next.

The field of quantum chaos deals with some key questions:
\begin{enumerate}
\item Given a classically chaotic system, what can we say about the corresponding quantum systems, e.g. the energy level statistics, properties of eigenstates, correlation functions, and more recently, quantum correlations and operator spreading. 
\item  Since the underlying reality of this universe is quantum, how does the classical chaos and the ``butterfly effect" arise out of the underlying quantum mechanics?   
\item What are the consequences of chaos at the quantum level for quantum information processing and quantum simulations? 

 \end{enumerate}
 These questions are closely related. However, the first question deals mainly with finding the signatures of chaos by studying the properties of the quantum Hamiltonian, while the second concerns with the dynamical behaviour of quantum states and the emergence of classically chaotic behaviour.    
 
 One of the  first steps in addressing question (1) was taken by Gutzwiller ~\cite{gutzwiller1970energy} by developing the trace formula.  The trace formula relates the density of states of the quantum system in the semiclassical regime to the properties of the classical periodic orbits.  Another central result of quantum chaos is the connection with the theory of random matrices \cite{Haake}.  In the limit of large Hilbert space dimensions (small $\hbar$), for parameters such that the classical description of the dynamics shows global chaos, the eigenstates and eigenvalues of the quantum dynamics have the statistical properties of an ensemble of random matrices.  The appropriate ensemble depends on the properties of the quantum system under time-reversal \cite{Haake}. 
 {In this direction an important quantum signature of chaos was obtained by Bohigas and collaborators \cite{bohigas1984characterization}}, describing the spectral statistics of quantum Hamiltonians whose classical counterparts exhibit complete chaos using random matrix theory. 
 

Semiclassical methods are powerful techniques used in quantum mechanics to analyze systems that are on the boundary between classical and quantum behavior. These methods provide approximations that combine elements of both classical and quantum mechanics, allowing researchers to understand quantum phenomena using classical concepts. Semiclassical approximations are particularly useful in the study of quantum chaos, where the underlying classical system exhibits chaotic dynamics.

In thinking about this, let us recall how classical dynamics appears in quantum formalism.  
\begin{enumerate}
\item The Eikonal ansatz for the wave function, giving the WKB approximation and rigor to the ``old quantum mechanics" of Bohr, Sommerfeld, and Einstein which related the ``stationary states" to ``quantized action".
\item The Feynman path integral, which relates the quantum propagator to the integral over phase factors of ratios of the ``action to $\hbar$". 
\item The Wigner-Weyl representation -- known in modern parlance as the ``coherent state" or ``phase space" representation.
\item The Gutzwiller trace formula, mentioned above, is a central result in semiclassical analysis and is particularly important for quantum chaos. It relates the density of states (quantum energy levels) of a quantum system to the sum over periodic orbits of the corresponding classical system.
This formula demonstrates how quantum mechanical quantities (like the energy spectrum) can be approximated by sums over classical trajectories, especially in systems where classical trajectories are chaotic.
\end{enumerate}
All of these are ways to see some form of the ``classical limit" of quantum mechanics.  The Eikonal ansatz shows the relationship of the phase of the wave function to Hamilton's Principle function (action integral) in the ``$\hbar \rightarrow 0$" limit, and lets us see classical trajectories as the ``ray optics" of wave mechanics (appropriate when the wavelength is short compared to the potential, so the particle nature dominates over the wave nature).  The standard argument about the Feynman path integral is, that in the semiclassical limit, where $S>>\hbar$, only the terms near the classical path survive, since variations of the action are stationary.  This is only possibly true in the large action limit.  Clearly, nonclassical dynamics like tunneling can be and is the dominant dynamics, e.g. a double well tunneling near the ground state.  From the phase-space representation, we see how the generator of dynamics (the Moyal bracket) reduces to the Poisson bracket in the $\hbar \rightarrow 0$ limit.  

{Traditionally, quantum chaos   has mostly focussed on (1) and (2) \cite{gutzwiller2013chaos}.}  Moreover, a majority of that work is about time independent properties, like Green's functions and density of states, for systems described by a classical-like kinetic energy and potential energy. The starting point is the Van-Vleck semiclassical propagator, originally written down in 1928.  This can be derived from exact (non-relativistic) quantum mechanics using WKB or Feynman based of the approximation of ``stationary phase". Gutzwiller used the Van-Vleck formula to arrive at the semiclassical Green's function, whose trace is the density of states.  This is the trace formula which relates the density of states in the semiclassical regime to the properties of the classical periodic orbits.

The semiclassical analysis of the quantum mechanical evolution is based on the mentioned van Vleck-Gutzwiller propagator, which provides the asymptotic form of the time-evolution operator through its matrix elements,
\begin{equation}
\label{eq:JD2}
\langle \psi({\rm final})|\hat{U}(t)|\psi({\rm initial})\rangle\sim \sum_{\gamma}A_{\gamma}{\rm e}^{\frac{i}{\hbar}R_{\gamma}}
\end{equation} 
yielding the probability amplitude of the quantum process that takes an initial state into a final one at given time $t$, as a sum over the possible paths (histories) $\gamma$ that satisfy the classical equations of motion. Quantum mechanical aspects enter here in the fact that the propagator represents a complex amplitude (and therefore describe interference phenomena), but the quantities involved in its calculation (the prefactors $A_{\gamma}$ and the actions $R_{\gamma}$) are purely classical.
   
Traditional quantum mechanics has a strong focus about the time-independent Schr$\ddot{\rm{o}}$dinger's equation -- things like energy spectra, eigenstates, cross-sections. Of course, all information about dynamics is encoded in the properties of the stationary states.   Similarly, studying the properties of the the stationary states (level statistics), localization of eigenstates, etc., can give us a great deal of insight, but the actual dynamics often has surprises, e.g., the idea of decoherence and coherent control. With the advent of controlling quantum systems, evolving and measuring them, semiclassical formalism have a crucial import towards dynamical aspects of many body physics and chaos.
  


Many body quantum systems are usually non-integrable and therefore potentially chaotic. For example, one may try and understand the collective behavior of an ensemble of interacting identical particles that, in the classical limit, can potentially exhibit chaos. There has been a major thrust trying to understand the issues involving thermalization, irreversibility, equilibration and coherent backscattering and effects of quantum interference in such systems. A major motivation currently is to simulate these systems on a quantum machine and on the related issues involving quantifying complexity of these simulations, benchmarking simulations in the presence of noise, error mitigation and exploring connections to chaos \cite{Neill_2016, Eisert2015, Rigol2008, engl2016semiclassical, ThomasPrl2014, Gogolin_2016, Linden2009PRE, Engl2015PRE, Popescu2006}.

\section{Random matrix theory and universal ensembles} \label{sec 3}
In the 1950s, while studying complex heavy nuclei, Eugene Wigner noticed that their spectral fluctuations could be modeled using the eigenvalues of random symmetric matrices with elements drawn independently from the Gaussian distribution $\sim\mathcal{N}(0, 1)$. Essentially, he suggested that the spectra of the complex nuclei share statistical features with the spectra of the random Gaussian symmetric matrices. This idea sparked a widespread development of RMT in subsequent years \cite{mehta2004random}, which finds applications across various branches of physics including atomic and nuclear physics \cite{weidenmuller2009random, mitchell2010random}, statistical mechanics \cite{forrester2010log}, condensed matter physics \cite{beenakker1997random}, and, notably, quantum chaos \cite{haake1991quantum} and quantum information theory \cite{collins2016random}. In the context of quantum chaos, this field was further advanced with the works of Dyson, who formalized the notion of Gaussian ensembles and classified them into three different ensembles, namely, Gaussian unitary ensemble (GUE), Gaussian orthogonal ensemble (GOE), and Gaussian symplectic ensemble (GSE) \cite{porter1965statistical}. The corresponding probability distribution functions of these ensembles are given by
\begin{equation}
  \setlength{\arraycolsep}{0pt}
  P(H) = \left\{ \begin{array}{ l l }
    &{} e^{-n\text{Tr}(H^2)/2}/\mathcal{N}_{\text{GUE}}\quad \text{($H$ is complex Hermitian)}\\
    &{} e^{-n\text{Tr}(H^2)/4}/\mathcal{N}_{\text{GOE}}\quad \text{($H$ is real symmetric)} \\
    &{} e^{-n\text{Tr}(H^2)}/\mathcal{N}_{\text{GSE}} \quad \text{($H$ is Hermitian quaternionic)}
  \end{array} \right.
\end{equation}
The probability measure over GUE is invariant under arbitrary unitary transformations, i.e., $P(H)=P(u^{\dagger}Hu)$ for any $u\in U(d)$, where $U(d)$ is the $d$-dimensional unitary group. Similarly, GOE and GSE are invariant under orthogonal and symplectic transformations, respectively. The matrices drawn from these ensembles display universal spectral characteristics. In particular, the eigenvalues of these matrices repeal each other so that the probability of finding a pair of eigenvalues close to each other is very small. This is evident from the joint distribution function of the eigenvalues, which takes the following form \cite{mehta2004random}:
\begin{multline}
P(\lambda_{1}, \lambda_{2}, \cdots, \lambda_{d})=\\ \dfrac{1}{\mathcal{N}_{\beta}}\exp\left\{-\dfrac{1}{2}\sum_{i=1}^{d}\lambda^2_{i}\right\}\prod_{j<k}\left| \lambda_{j}-\lambda_{k}  \right|^{\beta} ,    
\end{multline}
where
\begin{eqnarray}
\mathcal{N}_{\beta}=(2\pi)^{d/2}\prod_{j=1}^{d} \dfrac{\Gamma(1+j\beta/2)}{\Gamma(1+\beta/2)}.     
\end{eqnarray}
and $\beta=1, 2$ and $4$ correspond to GOE, GUE and GSE. The eigenvalues in the above expression are arranged in descending order, i.e., $\lambda_1\geq \lambda_2\cdots\geq\lambda_d$. The probability of having two identical eigenvalues is clearly zero. While the original purpose of RMT was to study the properties of heavy nuclei, it was later conjectured that quantum chaotic systems display spectral properties resembling those of one of these ensembles, depending on whether the system possesses time-reversal symmetry \cite{bohigas1984characterization}. To be more specific, if $s$ denotes a random variable corresponding to the distance between two nearest neighbor eigenvalues of a matrix chosen randomly from the Gaussian ensembles, its probability density is given by $P(s)\propto s^{\beta}e^{-A_\beta s^2}$, where $A_\beta=\left[\dfrac{\Gamma((\beta+2)/2)}{\Gamma((\beta+1)/2)}\right]^2$. If the given Hamiltonian is not time-reversal symmetric, the spectral statistics follow that of the GUE. In contrast, the GOE ($T^2=1$) and the GSE ($T^2=-1$) admit the time-reversal symmetry, where $T$ is the time-reversal operator.

Following the above classification, Dyson further introduced three analogous unitary ensembles \cite{dyson1962threefold}, namely circular unitary ensemble (CUE), circular orthogonal ensemble (COE), and circular symplectic ensemble (CSE) based on their properties under time-reversal operation. While the Gaussian ensembles are suitable for modeling time-independent systems, the circular ensembles are necessary for modeling time-dependent systems, particularly Floquet systems. The CUE encompasses all unitary matrices from the unitary group $U(d)$, i.e., CUE $\cong U(d)$, and hence a group. On the other hand, the COE consists of the symmetric unitaries, i.e., for any $u\in\text{COE}(d)$, we have $u^{T}=u$. it is to be noted that, unlike the CUE, the COE does not form a group. Moreover, for any $u\in \text{CUE}(d)$, $u^{T}u\in \text{COE}(d)$. The circular ensembles provide an accurate description of the chaotic time-dependent quantum systems. 

RMT description of quantum chaos has been immensely successful. Besides, the systems whose classical limits give rise to regular dynamics have also been studied extensively by the quantum chaos community in RMT context. In their seminal paper, Berry and Tabor \cite{berry1976closed} conjectured that the eigenvalues of the quantum systems with regular classical limits show no correlations among them. They behave as if they are identical and independently distributed (\textit{iid}) random variables. If $s$ is a random variable describing the spacing distribution between two adjacent levels of the Hamiltonian, the probability density of $s$ follows $P(s)=e^{-s}$. In the case of time-dependent systems (Floquet), the eigenphases of the unitary behave as if they were drawn uniformly at random from the unit complex circle.

\section{The Many Facets of Quantum Chaos}
Quantum chaos aims to understand the emergence of chaos and unpredictability in quantum systems. As initially proposed by Michael Berry, one way to tackle this problem is to study quantum systems having chaotic classical limits \cite{berry1989quantum}. Quantum mechanically, the evolution of an isolated system is given by an arbitrary unitary operator. When two different quantum states evolve under the same unitary, the distance between them remains constant throughout the evolution. This renders a naive generalization of the definition of chaos \footnote{Recall that classical chaos is characterized by the rate of separation of two nearby trajectories in the phase space} to the quantum regime meaningless. Hence, searching for suitable signatures of quantum chaos through other means became necessary. Asher Peres, in his seminal paper, introduced Loschmidt echo as a diagnostic tool for chaos in quantum systems \cite{peres1984stability} \footnote{Note that an experimental implementation of the Loschmidt echo was done way before it found applications in quantum chaology \cite{hahn1950spin}.}. The Loschmidt echo measures the overlap between a quantum state that evolved forward in time under a given Hamiltonian ($H$) and then evolved backward in time under a perturbed or modified Hamiltonian ($H'$), expressed as $|\langle\psi |e^{iH't}e^{-iHt} | \psi\rangle|^2$. Subsequently, this quantity has been extensively studied across various physical settings, including classical limits, to understand its behavior in both chaotic and regular systems \cite{jalabert2001environment, jacquod2001golden, cerruti2003uniform, gorin2006dynamics, Goussev2012loschmidt, prosen2002stability}.

Around the time Asher Peres proposed the Loschmidt echo, Bohigas, Giannoni, and Schmit (BGS) conjectured that quantum systems with chaotic classical limits adhere to Wigner-Dyson statistics \cite{bohigas1984characterization}. on the other hand,  the level spacings follow Poisson statistics for the quantum systems with regular classical limits due to the absence of level repulsions \cite{berry1976closed}. 
Shortly afterward, a dynamic measure called spectral form factor (SFF) was studied as a probe for quantum chaos \cite{haake1991quantum}, which complemented the BGS conjecture in examining quantum chaos. The SFF is expressed as $\langle |\text{Tr}(e^{-iH})|^2 \rangle$, where the quantity within the angular brackets is averaged over statically similar system Hamiltonians. Like the level spacing statistics, the SFF shows universal behavior for chaotic quantum systems. In addition, various other related measures, including the average level spacing ratio \cite{atas2013distribution}, have been extensively used to investigate the onset of chaos in both single-body and many-body quantum systems, regardless of the existence of classical limits. While extremely useful in probing quantum chaos, their applicability is often limited due to the presence of symmetries and other aspects. For instance, one must ensure the system is desymmetrized while dealing with spectral statistics in systems with symmetries \cite{tekur2020symmetry}. On the other hand, the spectral form factor is not self-averaging \cite{prange1997spectral}. Hence, physicists have sought to identify various alternative probes of quantum chaos.

With the advent of quantum information theory, the focus has shifted to dynamical correlation functions, especially in the many-body context. These measures are based on the spread of quantum information during the evolution of a local operator (or state) in the Hilbert space.
A local operator in a many-body system grows in its support when it evolves under a chaotic unitary. The operator becomes delocalized and develops overlap with most of the many-body Hilbert space \cite{von2018operator}. The scrambling of quantum information is a probe to quantum 
 chaos, nonintegrability, and thermalization in many-body quantum systems \cite{deutsch1991quantum, srednicki1994chaos, tasaki1998quantum, rigol2008thermalization, rigol2010quantum, torres2013effects}. Operator scrambling is not accessible to local measurements, as the information is delocalized. The study of operator spreading spans over many fields, including 
  black hole physics \cite{hayden2007black, sekino2008fast, hosur2016chaos, shenker2014black, mcginley2022quantifying}, holography \cite{bhattacharyya2022quantum}, integrable systems \cite{xu2020does, rozenbaum2020early, pilatowsky2020positive}, random unitary circuits \cite{nahum2018dynamics, nahum2018operator, khemani2018operator, rakovszky2018diffusive}, quantum field theories \cite{roberts2015diagnosing, stanford2016many, chowdhury2017onset, patel2017quantum}, and chaotic spin-chains \cite{luitz2017information, heyl2018detecting, lin2018out, geller2022quantum}. 
Operator spreading is a definitive signature of chaos in many systems \cite{moudgalya2019operator, omanakuttan2023scrambling}.
\section{Dynamical signatures of chaos} \label{sec 5}
Two important tools to study dynamical signatures of chaos are the Loschmidt echo and the out-of-time-ordered correlators~(OTOCs).  Loschmidt echo captures the sensitivity of the dynamics to perturbations \citep{hahn1950spin,peres1984stability, jalabert2001environment}, while OTOC is a measure of quantum information scrambling \citep{Swingle-2018, shenker2014black}. These two quantities are closely related \citep{yan2020information}, and both employ imperfect time-reversal of the quantum system.  
In quantum mechanics, time-irreversibility arises from the non-controllability of the Hamiltonian \citep{peres1984stability}. This contrasts with classical mechanics, where irreversibility arises because of the  mixing and coarse-graining.  No system is completely isolated from the environment. The environmental perturbations on the Hamiltonian affects a chaotic system much more than a regular one, leading to distinctive dynamics.

\subsection{Loschmidt Echo}
The fact that macroscopic systems follow an arrow of time despite the underlying physical laws being perfectly reversible caused a lot of trouble to scientists in the nineteenth century. Loschmidt thought that the reversal of dynamics should be achievable by reversing the sign of velocities. However, Boltzmann refuted it soon after, noting that it is practically impossible \citep{stephen1966brush}. Reversing the dynamics of a system would require perfect knowledge of its initial conditions, and it is practical only for systems with small degrees of freedom. 

Let us say our system is a large number of gas molecules in a container, expanding into its environment. To reverse this dynamics, one would require the initial knowledge of all the molecules perfectly. Even a slight difference in the initial knowledge of one of the molecules will lead to the disequalization of time very quickly. Not just that, no system is truly isolated in real life. There are always interactions with the environment, which has a large number of degrees of freedom. Imperfect knowledge along with environmental interactions establish an arrow of time for the system. These imperfections can be modeled as a perturbation to the system Hamiltonian. Then the difference in the overlap between the forward evolution and the backward evolution of the system is called Loschmidt echo. It is defined as $F(\tau)=\lvert\bra{\psi_0} e^{iH'\tau/\hbar}e^{-iH\tau/\hbar}\ket{\psi_0}\rvert^2,$ where $\ket{\psi_0}$ is the state undergoing evolution for a time $\tau.$ 

Echo acts as a measure of sensitivity to perturbations. In the absence of perturbations, Loschmidt echo attains the maximum  value of one. The presence of imperfections leads to decay from there. This  immediately suggests that the Loschmidt echo must be closely connected with the decoherence phenomena. The coupling of quantum systems with the environment causes decoherence. It is the process by which the quantum behavior is lost from the system. The connection between echo dynamics and decoherence has been established and the former has been used to quantify the latter \citep{zurek2001sub,cucchietti2003decoherence,cucchietti2004universality}. 

Loschmidt echo is employed in various  fields of physics, including quantum chaos, quantum computation and quantum information, elastic waves, quantum phase transition, statistical mechanics of small systems, etc. It was first implemented experimentally in NMR systems \citep{hahn1950spin}. The time-reversal was achieved using a radio frequency pulse. Perfect reversal is hampered by pulse imperfections and environmental interactions, which act as perturbations.

Echo decay can be used to diagnose chaos in quantum systems. The behavior of the Loschmidt echo depends on the initial state, the underlying dynamics, and the perturbation applied. For single-particle quantum systems whose classical limit is chaotic, the Loschmidt echo dynamics is well-understood \citep{shepelyansky1983some,jalabert2001environment,jacquod2001golden,cerruti2003uniform}. Its typical behavior with respect to time and perturbation strength are as follows. At short times, echo decays parabolically, followed by an asymptotic decay at intermediate times. The type of asymptotic decay is dependent on the perturbation strength. For small perturbations, the decay is Gaussian, whereas, for stronger perturbations, the echo slumps exponentially. The functional form of exponential decay is further dependent on whether the perturbation is global or local. This regime is followed by a saturation region at long times.

\subsection{Out-of-time ordered correlators}

Out-of-time-ordered correlators (OTOCs) are a powerful tool for understanding the scrambling of quantum information, quantum chaos, and thermalization processes in many-body quantum systems. Their role in connecting quantum information theory, statistical mechanics, and quantum gravity makes them a fascinating topic in modern theoretical physics. In classical chaos, sensitivity to initial conditions is characterized by the Lyapunov exponent, which quantifies the exponential growth of small perturbations in phase space. An analogous concept in quantum systems is the growth rate of OTOCs.

\begin{figure*}
    \centering
\includegraphics[scale=0.8]{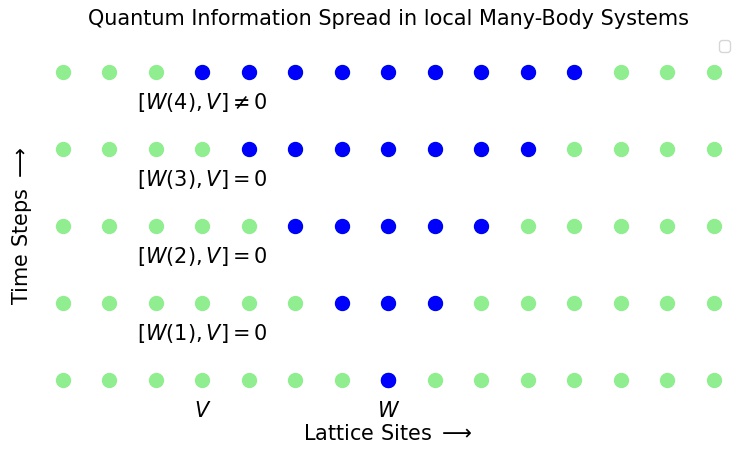}
    \caption{Schematic illustrating how information propagates in a local many-body quantum system with nearest neighbor interactions. }
    \label{fig:scramb1}
\end{figure*}

The OTOCs were initially introduced in the context of superconductivity \cite{larkin} and have recently gained renewed interest in the studies of quantum information scrambling \cite{ope2, ope1, ope4, ope5, lin2018out, shukla2022out}, quantum chaos \cite{chaos1, pawan, seshadri2018tripartite, lakshminarayan2019out, shenker2, moudgalya2019operator, omanakuttan2019out, manybody2, chaos2, prakash2020scrambling, prakash2019out, varikuti2022out, markovic2022detecting, dileep2024}, many-body localization \cite{manybody3, manybody4, manybody1, huang2017out} and holographic systems\cite{shock1, shenker3} due to their connections with various quantities in those respective disciplines. Given two arbitrary Hermitian and/or unitary operators $W$ and $V$, the out-of-time-ordered commutator function in an arbitrary quantum state $|\rho$ can be evaluated as follows:
\begin{eqnarray}\label{commutator}
C_{\rho, \thinspace WV}(t)=\text{Tr}\left(\rho \left[W(t), V\right]^{\dagger}\left[W(t), V\right]\right),
\end{eqnarray}
where $W(t)=\hat{U}^{\dagger}(t)W\hat{U}(t)$ denotes the Heisenberg evolution of the operator $W$ under the evolution generated the system Hamiltonain. For the numerical and the experimental perspectives, it is often convenient to consider $\rho$ to be maximally mixed, i.e., $\rho=\mathbb{I}/d$, where $d$ is the Hilbert space over which $\rho$ is supported. It is also natural to take $W$ and $V$ to be Hermitian if one is interested in seeing classical correspondence of the OTOCs with the maximum Lyapunov exponent. In this case, when expanded, the commutator function in Eq. (\ref{commutator}) contains a two-point and a four-point correlator:
\begin{eqnarray}
 C_{WV}= \dfrac{1}{d}\left[ \text{Tr}\left(W^2(t)V^2\right)-\text{Tr}\left( W(t)VW(t)V \right) \right].  
\end{eqnarray}
Note that we have omitted $\rho$ from the subscript on the left-hand side of the above expression. Due to the unusual time ordering, the four-point correlator is usually referred to as the OTOC. The behavior of $C_{WV}(t)$ depends predominantly on the four-point correlator. Therefore, the terms commutator function and OTOC are frequently used interchangeably to refer to the same quantity, $C_{WV}(t)$.

One of the primary reasons for the OTOCs to gain so much popularity is that it has a direct classical correspondence with the maximum classical Lyapunov exponent. In particular, if the given quantum system has a smooth classical limit, then for an appropriate choice of the initial operators, the OTOCs follow the correspondence principle with the maximum Lyapunov exponent. To understand this, first consider the position and the momentum operators $\hat{X}$ and $\hat{P}$. In the semiclassical limit ($\hbar\rightarrow 0$), the commutators get replaced by the Poisson brackets. One can then see that $C_{XP}(t)|_{\hbar\rightarrow 0} = \{ X(t), P \}^2=(\delta X(t)/\delta X(0))^2\sim e^{2\lambda_C t}$, where $\lambda_C$ is the maximum Lyapunov exponent of the system under consideration. If a system is chaotic, then $\lambda_C$ is positive. Then the correspondence principle implies that for an appropriate choice of the initial operators, the OTOCs of a quantum system, whose classical limit is chaotic, grow exponentially over a short time period. The time period over which the correspondence principle holds is referred to as Ehrenfest’s time $t_{\text{EF}}$. Note that $t_{\text{EF}}$ depends on the dynamics of the system \cite{schubert2012wave, rozenbaum2020early, jalabert2018semiclassical, chen2018operator}. It is now well known that for single particle chaotic systems, the Ehrenfest time $t_{\text{EF}}$ scales logarithmically with the effective Planck constant (or equivalently the Hilbert space dimension) and inversely with the maximum Lyapunov exponent of the corresponding classical limit --- $t_{\text{EF}}\sim \ln(1/\hbar_{\text{eff}})/\lambda_C$. For the time scales beyond $t_{\text{EF}}$ ($t>t_{\text{EF}}$), the corrections to the OTOC arising due to the finiteness of $\hbar$ become non-trivial and cause breakdown of the corresponding principle. In this regime, the OTOC saturates to a finite value depending upon the system Hilbert space dimension. Many studies have shown that the early-time exponential growth rate of OTOCs matches well with the maximum Lyapunov exponent of the classical limits. However, it is important to note that exponential growth does not always indicate true quantum as shown in recent studies \cite{pappalardi2018scrambling, hashimoto2020exponential, pilatowsky2020positive, xu2020does, hummel2019reversible, steinhuber2023dynamical}. This is justified as the positive Lyapunov exponent alone doesn't imply chaos even in the classical limit, as mentioned in the introduction. Only by carefully treating both the short-time and the long-time dynamics of the OTOCs can one verify whether the underlying quantum system is truly chaotic or not.

An interesting fact about OTOCs is that it has intriguing connections to various other probes of quantum chaos and several other quantum information theoretic quantities. For instance, in a bipartite system, when the initial operators are Haar unitary operators and act on disjoint subsystems, then the OTOC, averaging over all the local, boils down to the linear operator entanglement of the time evolution operator \cite{styliaris2021information, zanardi2021information}. To be specific, if one writes the operator Schmidt decomposition for the time evolution operator as $U(t)=\sum_{j=1}^{d}\sqrt{\lambda_{j}(t)}M^{A}_{j}(t)\otimes N^{B}_{j}(t)$, where $\{M^{A}_{j}(t)\}$ and $\{N^{B}_{j}(t)\}$ denote the set of trace orthonormal operators acting on the respective Hilbert spaces and $\{\lambda_{j}\}$s denote the Schmidt coefficients. Then, the average OTOC can be written as
\begin{eqnarray}
 \int_{u, v\in U(d)}d\mu(u) d\mu(v) C_{uv}(t)=1-\dfrac{1}{d^2}\sum_{j=1}^{d}\lambda^2_{j}(t).
\end{eqnarray}
This relation persists even when the Haar random unitaries are replaced with random Hermitian operators chosen at random from the Gaussian ensembles \citep{varikuti2022out}. Apart from this, the OTOCs further share connections with other quantum chaos probes, such as tripartite mutual information \cite{pawan}, quantum coherence \cite{anand2021quantum}, and Loschmidt echo \cite{yan2020information}. Additionally, OTOCs have been explored in the deep quantum regime, where signs of short-time exponential growth have still been observed \cite{sreeram2021out}. For a detailed comparison between OTOCs and observational entropy, a newly introduced concept for studying the thermalization of closed quantum systems, see Ref. \cite{pg2022witnessing} and also \cite{vsafranek2019quantum, vsafranek2019quantum1}.

In many-body quantum systems, the growth of OTOC is connected to how an initially localized operator spreads through the different degrees of freedom in the system as the system evolves. A representative figure, depicted in Fig. \ref{fig:scramb1}, demonstrates how local perturbations—resulting from the application of local unitaries in a typical many-body system with nearest-neighbor interactions—propagate throughout the entire system. Under chaotic quantum dynamics, a simple local operator becomes more complex over time, having its support over many system degrees of freedom. Resultantly, the quantum information that was initially localized becomes increasingly obscured from local measurements \cite{pawan, hayden2007black, moudgalya2019operator, ope1}. Thus, the OTOCs provide a perfect platform to diagnose information scrambling in many-body quantum systems. 

OTOCs can be used to distinguish between different regimes of many-body systems, such as thermalized, integrable, or many-body localized phases. For example, in many-body localized systems (which do not thermalize), OTOCs may show different growth behaviors compared to chaotic systems. They provide a quantitative way to explore how quantum systems approach thermal equilibrium and how deviations from chaotic behavior manifest in different physical systems.

\section{Chaos and quantum tomography}

Another realm where chaos manifests is in the rate of information gain during quantum state tomography, the process of estimating a given, unknown state.  Classically, chaos leads to exponential separation of trajectories starting from two nearby initial conditions, which helps in determining the starting point. Can we find its analogue in quantum state tomography? It turns out we can, as shown in~\citep{PhysRevLett.112.014102}.

In quantum tomography, one makes a series of measurements on a collection of identically prepared systems to gain information about the system. An estimate of the state is obtained by inverting the statistics of measurement records. For the assessment to be a good one, the measurements have to be tomographically complete. That is, the observables measured must span all of the operator space. The straight forward method to carry out tomography is to do projective measurements and then use the Born rule to invert the spectrum. Since projective measurements also destroy the state,  only one observable can be measured after each preparation, and very many replicas of the initial state are required for its reconstruction. 

One can choose to perform weak measurements instead, which prevent the state from collapsing. If the measurements are weak enough to exert minimal disturbance on the states, a good reconstruction fidelity can be achieved with a finite number of copies of the state. There is an added advantage of being able to optimize the measurements for the required fidelity in the estimate. {Silberfab} \emph{et~al.} proposed a weak collective measurement scheme on the identically prepared and collectively evolved ensemble, which will be of interest to us in this article \citep{PhysRevLett.95.030402}. Collective measurements are known to be much more effective and optimal than local measurements \citep{massar2005optimal,vidal1999optimal,gisin1999spin,bagan2006separable,hou2018deterministic}. The reason for this phenomenon is connected to the lower mutual information when measurements are performed locally \citep{bennett1999quantum}. In collective measurements,  the measurement apparatus interacts with the entire ensemble of states as a single quantum system rather than individually. This helps in reducing the effect of measurement induced disturbance on the system.  The measurement backaction gets distributed among the ensemble, leading to negligible effect on any individual system. 

In 2006, {Greg A Smith} \textit{et al.} performed  the protocol described above of weak measurements on a cloud of Cs atoms, coupled to an off-resonant optical probe \citep{PhysRevLett.97.180403}. Here the Stokes vector of the optical probe gets correlated to the atomic spin. Measuring the polarization of the probe weakly measures a collective spin observable. This is a standard Von-Neumann type measurement, where one couples the system with a meter and strongly measures the latter. Information completeness is achieved by keeping the Cs cloud in a time varying magnetic field. The varying magnetic field couples with the spin system, changing the observable measured each time.

Quantum tomography is useful in estimating the correctness of various control protocols. Estimating the quantum state gives information on its dynamics and spread across the Hilbert space during evolution and helps characterize quantum chaos. State tomography is also an essential tool in estimating quantum channels in process tomography.  

Suppose we are given an ensemble of $N_s$ identical systems $\rho^{\otimes N_s}_0$.    A probe is coupled to the ensemble of states that will generate the measurement record by performing weak continuous measurement of the collective observable $\mathcal{O}$. Thus, the ensemble is coherently evolved by a unitary $U(t)$ and collectively probed to give the estimate of the state in a single shot. The operator, evolving in Heisenberg fashion, is measured at time $t$ as
\begin{equation}
 \mathcal{O}(t)=U^{\dag}(t)\mathcal{O}U(t)
\end{equation}
We can exploit this choice of dynamics for time evolution and explore many properties of the quantum systems, as we will see in the subsequent chapters.

The positive operator valued measurement (POVM) elements for measurement outcomes $X(t)$ at time $t$ are \cite{silberfarb2005quantum,madhok2014information}
\begin{equation}
 E_{X(t)}=\frac {1}{\sqrt{2\pi\sigma^2}}\mathrm{exp}\  \left\{-\frac{1}{2\sigma^2}[X(t)-\mathcal{O}(t)]^2 \right\}.
\end{equation}
The standard deviation $\sigma$ in $E_{X(t)}$ is because of the probe noise (shot noise). 
Approximating the entire dynamics as a unitary evolution is an excellent approximation \cite{smith2013quantum, deutsch2010quantum}.
The system and probe coupling is sufficiently weak that the entangling effect of measurement backaction is negligible, and the ensemble is well approximated by a product state at all times.  
The shot noise sets the fundamental resolution of the probe. When the randomness of the measurement outcomes is dominated by the quantum noise in the probe rather than the measurement uncertainty, i.e., the projection noise, quantum backaction is negligible, and the state remains approximately separable. Thus, the measurement records can be approximated to be 
\begin{equation}
 M(t)=X(t)/N_s=\mathrm{Tr}[\mathcal{O}(t)\rho_0]+W(t)
 \label{BG_tom_noise}
\end{equation}
where $W(t)$ is a Gaussian white noise with spread $\sigma/N_s$.

The density matrix of any arbitrary state having Hilbert space dimension $d$ can be expressed in the orthonormal basis of $d^2-1$ traceless and Hermitian operators $\{E_\alpha\},$ and the state lies on the generalized Bloch sphere parametrized by the Bloch vector $\bf r$. Thus, the density matrix can be represented as 
\begin{equation}
\rho_0=\mathbb{I}/d+\Sigma^{d^2-1}_{\alpha=1}\ r_\alpha E_\alpha,  
\label{dens_matrix}
\end{equation}
where $$\Sigma^{d^2-1}_{\alpha=1}\ r_\alpha^2 = 1 - 1/d.$$

We consider the measurement records at discrete times 
\begin{equation}
M_n=M(t_n)=N_s\sum_{\alpha}r_\alpha \mathrm{Tr}[\mathcal{O}_{n}E_\alpha]+W_n,
\end{equation}
 where $\mathcal{O}_n=U^{\dagger n}\mathcal{O}U^{n}$. The measurement history can be re-cast in the vector form as
  \begin{equation}
  {\bf M}=\tilde{\mathcal{O}}{\bf r}+{\bf W},
  \label{ms} 
  \end{equation}
  i.e.
 \begin{equation}
 \begin{pmatrix}
M_1 \\
M_2 \\
..  \\
..  \\
M_n
\end{pmatrix}
=\begin{pmatrix}
\mathcal{\tilde{O}}_{11} & \mathcal{\tilde{O}}_{12} & .. & .. & \mathcal{\tilde{O}}_{1d^2-1}\\
\mathcal{\tilde{O}}_{21} & \mathcal{\tilde{O}}_{22} & .. & .. & \mathcal{\tilde{O}}_{2d^2-1}\\
.. & .. & .. & .. & ..\\
.. & .. & .. & .. & ..\\
\mathcal{\tilde{O}}_{n1} & \mathcal{\tilde{O}}_{n2} & .. & .. & \mathcal{\tilde{O}}_{nd^2-1}
\end{pmatrix}
\begin{pmatrix}
r_1 \\
r_2 \\
..  \\
..  \\
r_{d^2-1}
\end{pmatrix}
+
\begin{pmatrix}
W_1 \\
W_2 \\
..  \\
..  \\
W_n
\end{pmatrix}.
\end{equation}
When the backaction is insignificant, the probability distribution corresponding to measurement record $\bf M$ for a given Bloch vector $\bf r$ is 
\begin{equation} 
\begin{split}
p({\bf M|r}) & \varpropto \mathrm{exp}\  \left\{-\frac {N^2_s}{2\sigma^2}\sum_{i}[M_i-\sum_{\alpha}\tilde{\mathcal{O}}_{i\alpha}r_\alpha]^2 \right\}\\
& \varpropto \mathrm{exp}\  \left\{-\frac {N^2_s}{2\sigma^2}\sum_{\alpha,\beta}({\bf r-r_{ML}})_\alpha\ C^{-1}_{\alpha\beta}\ ({\bf r-r_{ML}})_\beta \right\},
\label{pdf_uniform_prior}
\end{split}
\end{equation}
where ${\bf C}=\left(\tilde{\mathcal{O}}^T\tilde{\mathcal{O}}\right)^{-1}$ is the covariance matrix with $\tilde{\mathcal{O}}_{n\alpha}=\mathrm{Tr}[\mathcal{O}_{n}E_\alpha]$ \cite{silberfarb2005quantum,smith2006efficient}. Given the measurement record and the knowledge of the dynamics, one can invert this measurement record to get an estimate of the parameters characterizing the unknown quantum state in Eq. (\ref{dens_matrix}).
The least-square fit of the Gaussian distribution in the parameter space is the maximum-likelihood (ML) estimation of the Bloch vector, ${\bf r}_{ML}=\bf C\tilde{\mathcal{O}}^{T}M.$ The measurement record is informationally complete if the covariance matrix is of full rank. If the covariance matrix is not of full rank, the inverse of the covariance matrix is replaced by Moore-Penrose pseudo inverse \cite{ben2003generalized}, inverting over the subspace where the covariance matrix has support. The eigenvalues of the $\bf C^{-1}$ determine the relative signal-to-noise ratio with which different observables have been measured. The estimated Bloch vector ${\bf r}_{ML}$ may not represent a physical density matrix with non-negative eigenvalues because of the noise present (having a finite signal-to-noise ratio). Therefore, we impose the constraint of positive semidefiniteness \cite{baldwin2016strictly}  on the reconstructed density matrix and obtain the physical state closest to the maximum likelihood estimate.

To do this ,one can employ a convex optimization \cite{vandenberghe1996semidefinite} procedure where the final estimate of the Bloch vector $\bf \bar{r}$ is obtained by minimizing the argument 
\begin{equation}
 \|{\bf r}_{ML}-{\bf \bar{r}}\|^2=({\bf r}_{ML}-{\bf \bar{r}})^T{\bf C^{-1}}({\bf r}_{ML}-\bf \bar{r})
\end{equation}
subject to the constraint $$\mathbb{I}/d+\Sigma^{d^2-1}_{\alpha=1}\ \bar{r}_\alpha E_\alpha\geq0.$$ {The positivity constraint  plays a crucial role in compressed sensing tomography.  Any optimization heuristic with positivity constraint is effectively a compressed sensing protocol, provided that the measurements are within the special class associated with compressed sensing \cite{kalev2015quantum}. }

The straightforward evaluation of the reconstruction procedure is given by the fidelity of the reconstructed state $\bar{\rho}$ relative to the actual state $\ket{\psi_0},$ $\mathcal{F}=\bra{\psi_0}\bar{\rho}\ket{\psi_0}$ as a function of time. The reconstruction fidelity $\mathcal{F}$ depends on the informational completeness of the measurement record \cite{merkel2010random, madhok2014information, sreeram2021quantum}, the choice of observables and quantum states \cite{sahu2022effect} as well as the presence of noise in the measurement outcomes \cite{sahu2022quantum}.

 In the next section, we document the information-theoretic quantifiers in detail. These metrics  quantify the information gain in quantum tomography and the amount of operator spreading. 
\section{Signatures of chaos from tomography: The Information-theoretic quantifiers}
\label{MI_Shannon}


\subsection{Shannon entropy}
In information theory, the uncertainty of a random variable is quantified by entropy known as Shannon entropy in this context. This also quantifies the information one has about the random variable, drawn from a given probability distribution, when it is observed.
The Shannon entropy $\mathcal{S}$ is defined as 
\begin{equation}
 \mathcal{S}= - \sum_{i} p(i) \log p(i),
 \label{Eq:HKCT}
 \end{equation}
where some random variable sets the probability vector $\{p(1), p(2), ..., p(i)...\}$. While solving some tasks for information processing, Shannon entropy, which measures the amount of information, relates to the physical resources required, for example, the amount of memory needed to store that information.  
Shannon's noiseless channel coding theorem gives an operational meaning to the Shannon entropy by establishing statistical limits to data compression where the source of data is an \textit{i.i.d} random variable \cite{shannon1948mathematical}. 

Shannon entropy of the dynamics can be obtained from the inverse covariance matrix, $\bf{C}^{-1}.$ As noted before, the eigenvalues of $\bf{C^{-1}}$ are signal-to noise ratios in orthogonal directions. A flat distribution of the eigenvalues mean a  rather uniform spreading of the initial operator. On the other hand, a skewed distribution of eigenvalues point to an uneven spreading during the evolution.
In the superoperator picture,
\begin{equation}
 {\bf C^{-1}}=\sum_{n=1}^N{|{\mathcal{O}_n})({\mathcal{O}_n}|},
\end{equation}
where $|{\mathcal{O}_n})$ denote the evolved operator $\mathcal{O}$ at time $t=n.$  The covariance matrix helps us to define certain information-theoretic quantifiers for information gain in continuous measurement tomography.

Mutual information quantifies the knowledge about the quantum state obtained from the measurement record $M$ \cite{coverm2006elements}
\begin{equation}
    \mathcal{I}[\textbf{r};\textbf{M}]=\mathcal{S}(\textbf{M})-\mathcal{S}(\textbf{M}|\textbf{r}),
\end{equation}
The Shannon entropy of the measurements $\mathcal{S}(\textbf{M}$ is due to shot noise associated with the probe, which is independent of the quantum state in question.  Neglecting this, the mutual information is determined by the conditional distribution:
\begin{equation}
    \mathcal{I}[\textbf{r};\textbf{M}]=-\mathcal{S}(\textbf{M}|\textbf{r}) = -\frac{1}{2}\ln(\det(\textbf{C})) = \ln\left(\frac{1}{V}\right).
\end{equation}
Here, V is the volume of the error ellipsoid whose semimajor axes are defined by the covariance matrix. Mutual information is maximized when 
 $1/V=\sqrt{\det(\bf{C^{-1}})}$ has the largest value.  At the same time, the trace of $\bf{C^{-1}}$ obeys the constraint equation  \cite{madhok2014information}:
\begin{equation} \mathrm{Tr}({\bf{C^{-1}}})=\sum_{i,\alpha}\tilde{\mathcal{O}}_{i\alpha}^2=n\|\mathcal{O}\|^2,
    \label{tr_const}
\end{equation}
where $n$ is the time of evolution, and $\|\mathcal{O}\|^2=\sum_{\alpha}\mathrm{Tr}(\mathcal{O}E_{\alpha})^2$ is the Euclidean square norm of $\mathcal{O}$, the initial operator. From Eq. (\ref{tr_const}), it is clear that $\mathrm{Tr}(\textbf{C}^{-1})$ is independent of the nature of the dynamics, and it grows linearly in time. Invoking the relation between the geometric and the arithmetic mean, we get
\begin{equation}
    \det({\bf{C^{-1}}})\le \left(\frac{1}{d^2-1}\mathrm{Tr}({\bf{C^{-1}}}) \right)^{d^2-1}= \left(\frac{n}{d^2-1}\|\mathcal{O}\|^2 \right)^{d^2-1},
\end{equation}
where  $d^2-1$ is the rank of the regularized $\bf{C^{-1}}$. Since $\bf C^{-1}$ does not saturate the full rank $d^2-1$ in a single unitary evolution, one has to apply a Tikhonov regularization by adding a small constant times the identity matrix with $\bf C^{-1}$ to avoid an ill defined entropy \cite{ng2004feature}. For the above inequality to be saturated, all the eigenvalues of $\bf{C^{-1}}$ have to be equal to each other. That is when the operator is spread uniformly through the operator space \cite{madhok2014information}. 
 Converting the eigenvalue spectrum to a probability distribution by normalizing, we can calculate the Shannon entropy as follows:
 
\begin{equation}
 \mathcal{S}_c=-\sum_i{\lambda_i}\ln\lambda_i,
 \label{shn_ent}
\end{equation}
where $\{\lambda_i\}$ is the normalized eigenvalues of $\bf C^{-1}$ \cite{madhok2014information,sreeram2021quantum}. The Shannon entropy raises and saturates when the operator reaches the maximum support in the Hilbert space, aided by the unitary dynamics.  The saturation value of $\mathcal{S}_c$ for a chaotic dynamics is higher than that of a regular dynamics.  

\begin{figure*}
    \centering
    \includegraphics[width= \linewidth]{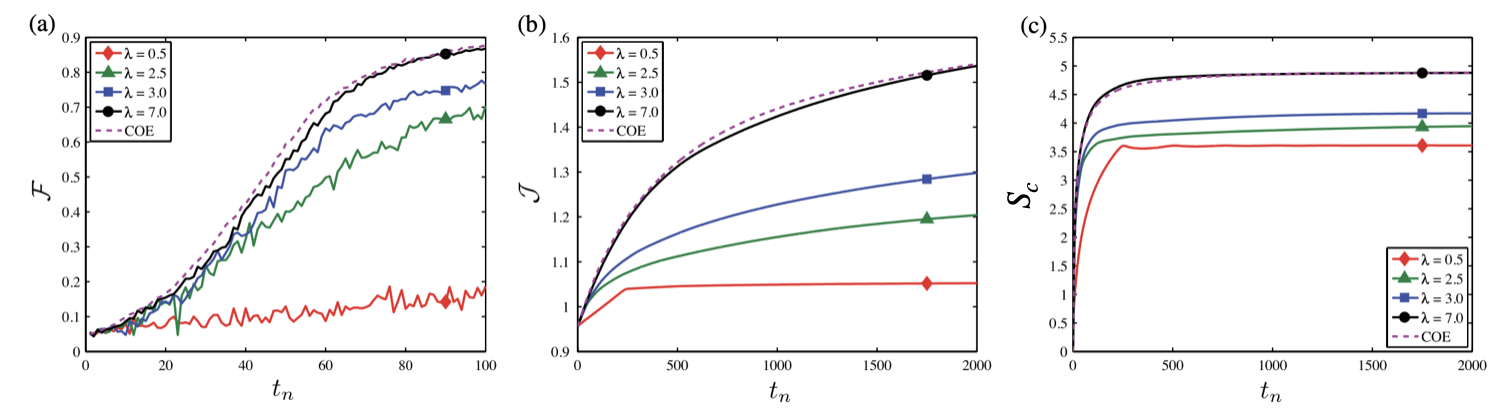}
    \caption{ A random pure quantum state is evolved with the kicked top unitary $U = \exp (-i \frac{\lambda}{2j} J_z^2 ) \exp(-i \alpha  J_x)$ and reconstruction is done using the stroboscopic measurement records at each time step. The angular momentum $j=10$, and the averages are made over 100 random initial states. The information gain is quantified using (a) Average fidelity of reconstruction (b) Fisher information and (c) Shannon entropy of the normalized eigenvalues of $\bf{C}^{-1}.$ All the three metrics are positively correlated with the kicking strength $\lambda.$
The COE results, averaged over 100 random pure states, are
plotted as the dashed line in (a)–(c). Figure is taken with permission from \cite{madhok2014information}.}
    \label{fig:enter-label}
\end{figure*}
\subsection{Fisher information}
 Let $X$ be a random variable, dependent on an underlying parameter $\theta.$ Then, the Fisher information captures the amount of knowledge about $\theta$ contained in a realization of the random variable $X.$ If the probability density of the realizations denoted by $f(X;\theta)$ is sharply peaked, then a small change in the parameter leads to a large change in the probability density. This means that a given outcome of the random variable contains a lot of information about the underlying parameter. On the other hand, if the density function is flat, then a particular realization of the random variable does not give much information about $\theta.$ Then, it takes a large sample of $X$ outcome data to reach a conclusion about the actual value of the parameter. This urges a quantification of the amount of information gain in terms of the variance with respect to the parameter. The formal definition of the Fisher information is given as follows \citep{fisher1922mathematical, cover2012elements},
 \begin{equation}
     \mathcal{J}(\theta)= \mathrm{E}\left[\left( \left.\dfrac{\partial}{\partial \theta} \mathrm{log}f(X;\theta)\right)^2\right|\theta\right].
 \end{equation}
 When there are multiple parameters $\theta=\lbrace \theta_1,\theta_2,...\theta_n \rbrace$ determining $X,$ the Fisher information becomes a matrix. An element of the matrix is given as
 \begin{equation}
      \mathcal{J}(\theta)_{i,j}= \mathrm{E}\left[\left( \left.\dfrac{\partial}{\partial \theta_i} \mathrm{log}f(X;\theta)\right) \left( \dfrac{\partial}{\partial \theta_j} \mathrm{log}f(X;\theta)\right) \right|\theta \right].
 \end{equation}
 An important relation concerning Fisher information was stated in 1946 by {Herald Cramer} and {C.R Rao}, called the Cramer-Rao bound \citep{cramir1946mathematical,rao1992information}. Consider $\hat{\theta}$, an unbiased estimator~(a rule/function) for calculating the estimate of $\theta$, whose expectation is equal to the true value of the parameter. Then the Cramer-Rao bound states that the inverse of the Fisher information is a lower bound for the error in the estimator,
 \begin{equation}
     \mathrm{Var}(\theta) \geq \frac{1}{\mathcal{J}(\theta)}.
 \end{equation}Thus Fisher information sets a fundamental limitation to the precision of the parameter estimation.

The variance between  the true and estimated states during tomography, quantified by the Hilbert Schmidt distance, is equal to the sum of the uncertainties in the orthogonal directions \cite{hradil02}
\begin{equation}
 \mathcal{D}_{HS} =\langle \mathrm{Tr}[(\rho_0 - \bar{\rho})^2]\rangle = \sum_\alpha\langle (\Delta r_\alpha)^2 \rangle,
\end{equation}
  where$\Delta r_\alpha=r_\alpha- \bar{r}_{\alpha}$. According to Cramer-Rao inequality, $\langle (\Delta r_\alpha)^2 \rangle \ge \left[ \bf{F}^{-1} \right]_{\alpha \alpha}$, where $\bf{F}$ is the Fisher information matrix. For the state estimation procedure with the  multivariate- Gaussian conditional probability in Eq. (\ref{pdf_uniform_prior}), it turns out that $\bf{F} = \bf{C}^{-1}$ \cite{madhok2014information}. One can compute the total information-gain in the process by defining a collective Fisher information $\mathcal{J}$ as follows:

\begin{equation}
\mathcal{J} = \frac{1}{\mathrm{Tr}[\bf{C}]}  
\end{equation}
  
 The inverse covariance matrix is not full rank in this protocol. One can regularize $\bf{C}^{-1}$ by adding to it a small fraction of the identity matrix. This is to avoid an ill-defined Fisher information.  The collective Fisher information is correlated to the average fidelity of reconstruction $\langle \mathcal{F} \rangle.$  For pure states,  the average Hilbert-Schmidt distance $\mathcal{D}_{HS} =1/\mathcal{J}= 1-\langle \mathrm{Tr}\bar{\rho}^2\rangle -2 \langle \mathcal{F} \rangle$ \cite{hradil02}. The collective Fisher information keeps increasing with more measurements for all dynamics. The rise is faster for a chaotic dynamics, compared to a regular one. 

We described three quantities in tomography so far: fidelity, Fisher information and Shannon entropy. Figure (\ref{fig:enter-label}) illustrates how these measures diagnose chaos in the case of random state reconstruction with a kicked top evolution. Kicked top evolution is described by $U = \exp (-i \frac{\lambda}{2j} J_z^2 ) \exp(-i \alpha  J_x),$ where $\lambda$ determines the amount of chaos in the system. As $\lambda$ increases, the system transitions from regular to chaotic dynamics. Information gain about the random states is more when the unitary drive is chaotic as seen in Fig. (\ref{fig:enter-label}). Thus, depending on the value of these measures, one can determine whether the nature of the dynamics.

\section{Quantifying operator spreading and chaos in Krylov subspaces with quantum tomography}

One of the characteristics of chaos in quantum systems is operator spreading \cite{moudgalya2019operator, omanakuttan2023scrambling}. OTOC is a popular quantifier of operator growth. However, it is not an experimentally friendly quantity to measure, as OTOC involves time reversal. Backward evolution is imperfect in practice, and some of the recent OTOC measurement protocols have tried to get around this problem \cite{vermersch2019probing, blocher2022measuring, sundar2022proposal}.

Some of the other candidates that capture operator spreading include Krylov complexity \cite{parker2019universal, yates2021strong, rabinovici2021operator,  noh2021operator, dymarsky2021krylov, caputa2022geometry, rabinovici2022krylov, avdoshkin2022krylov, rabinovici2022k, bhattacharya2022operator, bhattacharya2023krylov, suchsland2023krylov}, operator entanglement \cite{nie2019signature, wang2019barrier, alba2019operator, styliaris2021information} and memory matrix formalism \cite{mcculloch2022operator}. Krylov complexity, which has recently seen a flurry of interest, associates a complexity with the operator in terms of the number and probability amplitudes of basis states needed to describe it. Suppose the initial operator undergoing evolution is $|O_0),$ a $d^2$ dimensional vector in the superoperator representation. At a later time $t,$ the evolved operator can be represented by $|O_t).$ Then there exists a set of Krylov orthogonal operators $\{|K_1), |K_2)...|K_n)\},\: n<d^2$ which form the basis vectors that completely describe $|O_t)$ at any time $t.$ The Krylov complexity is given by 
\begin{equation}
    \mathcal{K}_c(t)= \sum_{i=1}^n c_n|(K_i|O_t)|^2.
\end{equation}
While such a measure has an obvious dependence on the chosen basis, the complexity when the operator is expressed in Krylov basis is the global minimum over all the orthogonal bases of the Hilbert space \cite{balasubramanian2022quantum}. The Krylov basis is found by an iterative orthonormalization process, called the Laczos algorithm \cite{parker2019universal, caputa2022geometry}. 

\begin{figure*}[htbp]
  	\centering
    \includegraphics[scale=0.62]{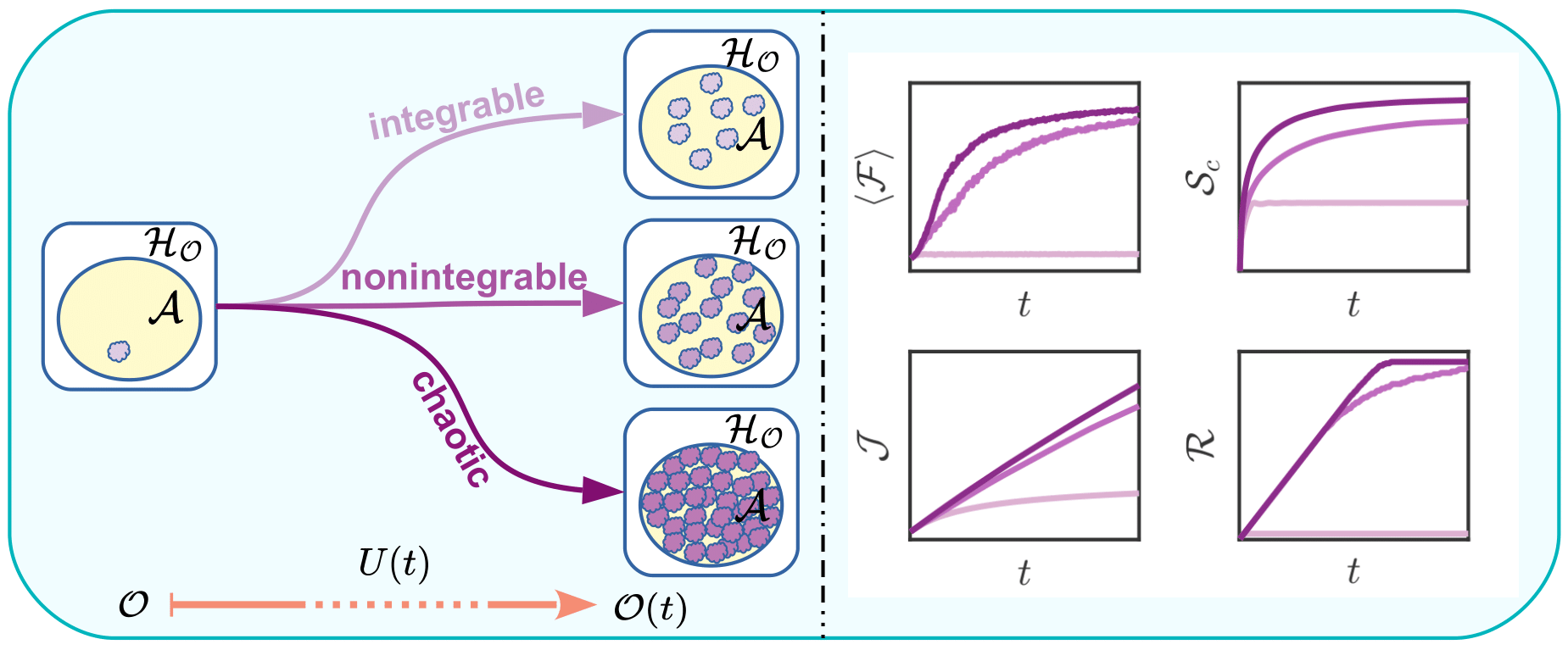}
  	
  	\caption {A local operator $\mathcal{O}$ in the operator space $\mathcal{H}_\mathcal{O}$ is unitarily evolved by a unitary $U(t)$.  Let $\mathcal{A}$ denote the largest subspace $\mathcal{O}$ can evolve into.  Then $\mathrm{dim}(\mathcal{A})=d^2-d+1$. Only a fully chaotic dynamics helps to span all of $\mathcal{A}$. If the unitary is not chaotic, $\mathcal{O}(t)$ develops nonzero support over a smaller subspace. the  Operator spreading is quantified through the rate of reconstruction with certain information-theoretic quantifiers like average reconstruction fidelity $\langle\mathcal{F}\rangle$, Shannon entropy $\mathcal{S}_c$, Fisher information $\mathcal{J}$, and rank of covariance matrix $\mathcal{R}$. Figure taken from \cite{sahu2023quantifying}}
  	\label{fig:OPSpread}
  \end{figure*}

The drawback of Krylov complexity is that it depends on the choice of the initial operator, which leads to ambiguous conclusions about chaos. For instance, the saturation value of Krylov complexity for the integrable and chaotic regimes in the  spin chain model considered in \cite{espanol2023assessing} is inverted for a different initial observable. Furthermore, some of the integrable systems showed an initial exponential growth of $\mathcal{K}_c,$ which was only expected from chaotic dynamics \cite{dymarsky2021krylov, avdoshkin2022krylov}. All of this has dampened the enthusiasm about Krylov complexity in the quantum chaos community.

An alternate way to capture the operator spread is through the continuous weak measurement tomography process we previously considered. Chaotic and regular dynamics would lead to different levels of operator spread, which would reflect in the rate of information gain. The amount of operator spreading in the operator space is more as the dynamics becomes strongly non-integrable which is verified in different dime-dependent and time-independent spin models for local, global and local random initial observables in a recent article \cite{sahu2023quantifying}. Figure \ref{fig:OPSpread} illustrates the visualization of operator spread through various information theoretic metrics.  An initially local observable in a $d^2$ dimensional operator space evolved with a chaotic unitary spans a $d^2-d+1$ dimensional subspace. If the dynamics is fully chaotic, the Krylov subspace corresponding to this initial operator would also contain $d^2-d+1$ orthogonal vectors \cite{sahu2023quantifying}. Thus, a stroboscopically generated set of evolved operators starting from the initial one, spans the Krylov subspace. The advantage of the tomographic process is that it is experimentally accessible and leads to an unambiguous quantification of chaos.

\begin{figure*}
    \centering
\includegraphics[scale=0.6]{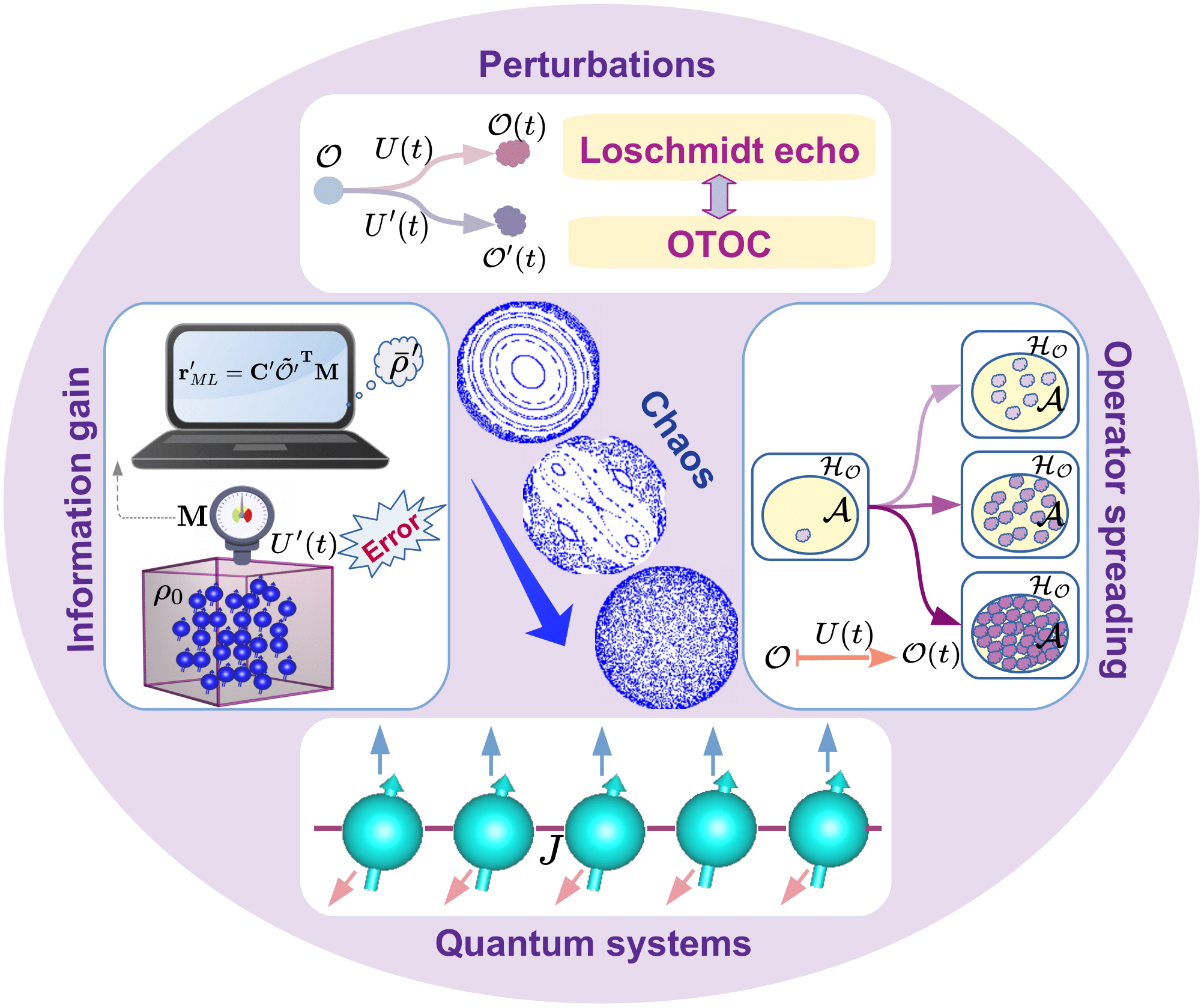}
    \caption{A graphical summary illustrating various signatures of chaos and their relationship when studied under the paradigm of quantum tomography. The continuous measurement quantum tomography helps to connect two quantifiers of chaos - OTOC and Loschmidt echo. The information gain in tomography also quantifies the operator spreading in quantum dynamical systems.}
    \label{fig:scramb}
\end{figure*}

Chaos, which leads to unpredictability aiding tomography could be surprising at first. However, they are two sides of the same coin. The Kolmogorov Sinai (KS) entropy quantifies the exponential divergence of nearby trajectories in a classically chaotic system \cite{pesin1977characteristic}. KS entropy is equal to the sum of positive Lyapunov exponents. However, the KS entropy is also the rate at which one gains information about the initial state by making measurements along the evolution. It is this effect of chaos that quantum tomography also benefits from.


\section{Noisy tomography and chaos}

Quantum systems are sensitive to errors that naturally occur while implementing many-body Hamiltonians. It is obvious to ask a question in this regard: Can we trust our analog quantum simulators \cite{ hauke2012can, chinni2022trotter}?
Fault tolerant digital quantum simulators are at least theoretically possible with the support of fully developed quantum error correction formalism. Howevere, an analog quantum simulator does not come with this guarantee while studying phase transitions in condensed matter systems or structural instabilities in chaotic systems \cite{chinni2022trotter, chaudhury2009quantum}. 
The imperfect dynamical evolution and the trotter errors can take the system dynamics away from the intended dynamics. The accuracy of any real physical experiment will drop as a function of time in the presence of some imperfections.


The chaotic dynamics helps in faster information gain which is quantified by the positive KS entropy. The chaos in the dynamics also makes the dynamics sensitive to errors which is captured by Loschmidt echo. Thus, in the presence of both chaos and errors in the dynamics, one can expect an interplay between the two. It is also interesting to study the interplay between the KS entropy and Loschmidt echo where the former causes a rapid information gain and the latter is associated with a rapid accumulation of errors. 
  

Continuous weak measurement tomography is used as a paradigm to explore and explain the sensitivity to errors in the quantum simulations of chaotic Hamiltonians. In \cite{sahu2022quantum}, the authors reconstructed random quantum states with random initial operators as priors for the evolution. They observed that the reconstruction fidelity increases initially despite errors in the dynamics regardless of the level of chaos in the dynamics.  However,  the fidelity drops later on and the drop rate in fidelity is inversely correlated with the extent of chaos in the dynamics. This behaviour gives us an operational interpretation of Loschmidt echo for operators $\mathcal{F_O}$ by connecting it to the performance of quantum tomography, where 
\begin{equation}
	\mathcal{F_O}(t_n)=\frac{\tr(\mathcal{O}^{\dagger}_n\mathcal{O}'_n)}{\tr(\mathcal{O}^2)}.
\end{equation}
Here, $\mathcal{O}_n'$ is in the trajectory of a noisy evolution under a perturbed unitary $U'(t).$

Unlike the scrambling of information, the scrambling of error is captured by a quantity similar to an out-of-time-ordered correlator (OTOC) between two operators, where one is evolved under an \emph{effective error unitary}.
\begin{equation}
	\mathcal{I}_{\mathcal{O}}(t_n)=\frac{1}{2j^4} \tr(|[\mathcal{O},\mathcal{U}^{\dagger}_n\mathcal{O}\mathcal{U}_n]|^2), 
	\label{error_otoc}
\end{equation}
where $\mathcal{U}_{n}=U'^{n}_{\tau}U^{\dagger n}_{\tau}$ is the effective {error unitary.} 
This result illustrates a fundamental link between two different quantifiers of chaos namely, Loschmidt echo, and OTOC while having some operational consequences in quantum information processing. Figure \ref{fig:scramb} gives a graphical summary of the various signatures of chaos from tomography. This study motivates further investigations in the performance of analog simulation while the system dynamics suffers from some unavoidable perturbations.


 \section{Conclusion}

The hallmark of a physical theory is its ability to predict. Classical physics, governed by Newton's laws of motion, is deterministic in nature. This means that given all the information in the present, the future course is unique. But is the future predictable? At the heart of this question lies the concept of chaos - the phenomena which makes even a deterministic system unpredictable. 
Chaos is ubiquitous in nature; it is more of a rule than an exception.  Weather, simple ecological systems, EEG signals from the brain, the simple double pendulum, and the famous three-body problem inspired by Newton's Solar system studies are a small subset of examples of chaotic behaviour. Chaos is also popularly known as the ``butterfly effect" due to the extremely sensitive nature of the dynamics to small perturbations.
In classical physics, chaos refers to the sensitivity of a system to initial conditions. Even small differences in starting conditions can lead to vastly different outcomes over time, making long-term prediction impossible. This is often characterized by exponential divergence of nearby trajectories in phase space (Lyapunov exponents).

 Quantum chaos explores the behavior of quantum systems whose classical counterparts exhibit chaotic dynamics. It bridges the gap between classical chaos theory and quantum mechanics, providing insights into how chaotic behavior emerges in quantum systems and how it differs from classical chaos.
How does chaos manifest itself on the atomic scale where quantum physics instead of Newton's laws is the correct theory to describe the physics? How does one characterize chaos in the quantum world? 

 In quantum systems, one way to identify chaos is by examining the statistics of energy levels. For chaotic systems, the distribution of energy level spacings often follows the Wigner-Dyson distribution, which indicates level repulsion, as opposed to the Poisson distribution typically seen in integrable (non-chaotic) systems.
 While the concept of Lyapunov exponents is well-defined in classical chaos, defining an analogous quantity in quantum mechanics is more challenging. Recent research has focused on defining quantum Lyapunov exponents that quantify the exponential growth of certain out-of-time-order correlators (OTOCs). Another manifestation of chaos is ``mixing".
A drop of dye in a glass of water completely spreads in the entire cup.

 We discussed the different pathways to chaos in classical physics, followed by a discussion on quantum signatures of chaos. Eigenvalue statistics combined with Random matrix theory was one of the first diagnostics of quantum chaos.  We then focused on dynamical signatures uncovered by popular tools such as OTOC and Loschmidt echo. The quantum information scrambling can be  considered the quantum analogue of spreading a drop of ink in water. The ``quantum drop of dye" or the quantum operators smear out and spread in the entire Hilbert space, which is analogous to the water in a cup. While studying  tomography, we ask the completely counter-intuitive and opposite question - is this unpredictability a source of useful information? At first, unpredictability and information seem to be at odds. However, a deeper analysis reveals an exact correspondence. If everything can be accurately forecasted or already known, we obtain no new information. We only acquire new information after the outcome of an event when its occurrence is not known. In other words, we learn from occurrence of an event, the posterior probabilities of the \textit{missing} information by constantly updating the priors. 
 In quantum mechanics, the art of obtaining this \textit{missing} information from unpredictable outcomes is ``quantum tomography". We relate the amount of information acquired in quantum tomography 
to the degree of chaos or unpredictability in the quantum system.
We give an unambiguous criterion for characterizing and quantifying the smearing of ``quantum ink," and our protocol can be experimentally realized in experiments involving trapped cold atoms and light-matter interaction between the atomic ensemble and a laser pulse. Reconstructing an unknown state with a chaotic unitary also picks up the vestiges of chaos, which becomes apparent in the fidelity as well. To unravel these effects, one can also use information theoretic measures such as Fisher information and Shannon entropy. 

The origin of chaos from the quantum world is a fundamental question in physics.  
 Quantum chaos helps understand the behavior of electrons in disordered systems, such as Anderson localization and metal-insulator transitions. Quantum chaos plays a role in understanding thermalization, information scrambling, and the behavior of entanglement in complex quantum systems, which is crucial for quantum computing and quantum simulation.
Furthermore, quantum chaos could determine the usability of quantum devices such as quantum computers. Therefore chaos in quantum systems warrants further understanding from a practical perspective as well.

\section{Acknowledgments}
This work was supported in part by grant  DST/ICPS/QusT/Theme-3/2019/Q69, New faculty Seed Grant from IIT Madras, and National Science Foundation grant PHY-2112890. The authors were supported, in part, by a grant from Mphasis to the Centre for Quantum Information, Communication, and Computing (CQuICC) at IIT Madras. On behalf of all authors, the corresponding author states that there is no conflict of interest.

\bibliography{references}

\begin{thebibliography}{216}%
\makeatletter
\providecommand \@ifxundefined [1]{%
 \@ifx{#1\undefined}
}%
\providecommand \@ifnum [1]{%
 \ifnum #1\expandafter \@firstoftwo
 \else \expandafter \@secondoftwo
 \fi
}%
\providecommand \@ifx [1]{%
 \ifx #1\expandafter \@firstoftwo
 \else \expandafter \@secondoftwo
 \fi
}%
\providecommand \natexlab [1]{#1}%
\providecommand \enquote  [1]{``#1''}%
\providecommand \bibnamefont  [1]{#1}%
\providecommand \bibfnamefont [1]{#1}%
\providecommand \citenamefont [1]{#1}%
\providecommand \href@noop [0]{\@secondoftwo}%
\providecommand \href [0]{\begingroup \@sanitize@url \@href}%
\providecommand \@href[1]{\@@startlink{#1}\@@href}%
\providecommand \@@href[1]{\endgroup#1\@@endlink}%
\providecommand \@sanitize@url [0]{\catcode `\\12\catcode `\$12\catcode `\&12\catcode `\#12\catcode `\^12\catcode `\_12\catcode `\%12\relax}%
\providecommand \@@startlink[1]{}%
\providecommand \@@endlink[0]{}%
\providecommand \url  [0]{\begingroup\@sanitize@url \@url }%
\providecommand \@url [1]{\endgroup\@href {#1}{\urlprefix }}%
\providecommand \urlprefix  [0]{URL }%
\providecommand \Eprint [0]{\href }%
\providecommand \doibase [0]{https://doi.org/}%
\providecommand \selectlanguage [0]{\@gobble}%
\providecommand \bibinfo  [0]{\@secondoftwo}%
\providecommand \bibfield  [0]{\@secondoftwo}%
\providecommand \translation [1]{[#1]}%
\providecommand \BibitemOpen [0]{}%
\providecommand \bibitemStop [0]{}%
\providecommand \bibitemNoStop [0]{.\EOS\space}%
\providecommand \EOS [0]{\spacefactor3000\relax}%
\providecommand \BibitemShut  [1]{\csname bibitem#1\endcsname}%
\let\auto@bib@innerbib\@empty
\bibitem [{\citenamefont {Ott}(2002)}]{ott2002chaos}%
  \BibitemOpen
  \bibfield  {author} {\bibinfo {author} {\bibfnamefont {E.}~\bibnamefont {Ott}},\ }\href@noop {} {\emph {\bibinfo {title} {Chaos in dynamical systems}}}\ (\bibinfo  {publisher} {Cambridge university press},\ \bibinfo {year} {2002})\BibitemShut {NoStop}%
\bibitem [{\citenamefont {Strogatz}(2018)}]{strogatz2018nonlinear}%
  \BibitemOpen
  \bibfield  {author} {\bibinfo {author} {\bibfnamefont {S.~H.}\ \bibnamefont {Strogatz}},\ }\href@noop {} {\emph {\bibinfo {title} {Nonlinear dynamics and chaos with student solutions manual: With applications to physics, biology, chemistry, and engineering}}}\ (\bibinfo  {publisher} {CRC press},\ \bibinfo {year} {2018})\BibitemShut {NoStop}%
\bibitem [{\citenamefont {Devaney}(2018)}]{devaney2018introduction}%
  \BibitemOpen
  \bibfield  {author} {\bibinfo {author} {\bibfnamefont {R.}~\bibnamefont {Devaney}},\ }\href@noop {} {\emph {\bibinfo {title} {An introduction to chaotic dynamical systems}}}\ (\bibinfo  {publisher} {CRC press},\ \bibinfo {year} {2018})\BibitemShut {NoStop}%
\bibitem [{\citenamefont {Lorenz}(1963)}]{lorenz1963deterministic}%
  \BibitemOpen
  \bibfield  {author} {\bibinfo {author} {\bibfnamefont {E.~N.}\ \bibnamefont {Lorenz}},\ }\bibfield  {title} {\bibinfo {title} {Deterministic nonperiodic flow},\ }\href@noop {} {\bibfield  {journal} {\bibinfo  {journal} {Journal of atmospheric sciences}\ }\textbf {\bibinfo {volume} {20}},\ \bibinfo {pages} {130} (\bibinfo {year} {1963})}\BibitemShut {NoStop}%
\bibitem [{\citenamefont {Banks}\ \emph {et~al.}(1992)\citenamefont {Banks}, \citenamefont {Brooks}, \citenamefont {Cairns}, \citenamefont {Davis},\ and\ \citenamefont {Stacey}}]{banks1992devaney}%
  \BibitemOpen
  \bibfield  {author} {\bibinfo {author} {\bibfnamefont {J.}~\bibnamefont {Banks}}, \bibinfo {author} {\bibfnamefont {J.}~\bibnamefont {Brooks}}, \bibinfo {author} {\bibfnamefont {G.}~\bibnamefont {Cairns}}, \bibinfo {author} {\bibfnamefont {G.}~\bibnamefont {Davis}},\ and\ \bibinfo {author} {\bibfnamefont {P.}~\bibnamefont {Stacey}},\ }\bibfield  {title} {\bibinfo {title} {On devaney's definition of chaos},\ }\href@noop {} {\bibfield  {journal} {\bibinfo  {journal} {The American mathematical monthly}\ }\textbf {\bibinfo {volume} {99}},\ \bibinfo {pages} {332} (\bibinfo {year} {1992})}\BibitemShut {NoStop}%
\bibitem [{\citenamefont {Scully}\ and\ \citenamefont {Zubairy}(1997)}]{scully1997quantum}%
  \BibitemOpen
  \bibfield  {author} {\bibinfo {author} {\bibfnamefont {M.~O.}\ \bibnamefont {Scully}}\ and\ \bibinfo {author} {\bibfnamefont {M.~S.}\ \bibnamefont {Zubairy}},\ }\href@noop {} {\emph {\bibinfo {title} {Quantum optics}}}\ (\bibinfo  {publisher} {Cambridge university press},\ \bibinfo {year} {1997})\BibitemShut {NoStop}%
\bibitem [{\citenamefont {Shankar}(2012)}]{shankar2012principles}%
  \BibitemOpen
  \bibfield  {author} {\bibinfo {author} {\bibfnamefont {R.}~\bibnamefont {Shankar}},\ }\href@noop {} {\emph {\bibinfo {title} {Principles of quantum mechanics}}}\ (\bibinfo  {publisher} {Springer Science \& Business Media},\ \bibinfo {year} {2012})\BibitemShut {NoStop}%
\bibitem [{\citenamefont {Koopman}(1931)}]{koopman1931hamiltonian}%
  \BibitemOpen
  \bibfield  {author} {\bibinfo {author} {\bibfnamefont {B.~O.}\ \bibnamefont {Koopman}},\ }\bibfield  {title} {\bibinfo {title} {Hamiltonian systems and transformation in hilbert space},\ }\href@noop {} {\bibfield  {journal} {\bibinfo  {journal} {Proceedings of the National Academy of Sciences}\ }\textbf {\bibinfo {volume} {17}},\ \bibinfo {pages} {315} (\bibinfo {year} {1931})}\BibitemShut {NoStop}%
\bibitem [{\citenamefont {Neumann}(1932)}]{neumann1932operatorenmethode}%
  \BibitemOpen
  \bibfield  {author} {\bibinfo {author} {\bibfnamefont {J.~v.}\ \bibnamefont {Neumann}},\ }\bibfield  {title} {\bibinfo {title} {Zur operatorenmethode in der klassischen mechanik},\ }\href@noop {} {\bibfield  {journal} {\bibinfo  {journal} {Annals of Mathematics}\ ,\ \bibinfo {pages} {587}} (\bibinfo {year} {1932})}\BibitemShut {NoStop}%
\bibitem [{\citenamefont {Jordan}\ and\ \citenamefont {Sudarshan}(1961)}]{jordan1961lie}%
  \BibitemOpen
  \bibfield  {author} {\bibinfo {author} {\bibfnamefont {T.~F.}\ \bibnamefont {Jordan}}\ and\ \bibinfo {author} {\bibfnamefont {E.}~\bibnamefont {Sudarshan}},\ }\bibfield  {title} {\bibinfo {title} {Lie group dynamical formalism and the relation between quantum mechanics and classical mechanics},\ }\href@noop {} {\bibfield  {journal} {\bibinfo  {journal} {Reviews of Modern Physics}\ }\textbf {\bibinfo {volume} {33}},\ \bibinfo {pages} {515} (\bibinfo {year} {1961})}\BibitemShut {NoStop}%
\bibitem [{\citenamefont {Arnold}(2009)}]{arnold2009proof}%
  \BibitemOpen
  \bibfield  {author} {\bibinfo {author} {\bibfnamefont {V.~I.}\ \bibnamefont {Arnold}},\ }\bibfield  {title} {\bibinfo {title} {Proof of a theorem of an kolmogorov on the invariance of quasi-periodic motions under small perturbations of the hamiltonian},\ }\href@noop {} {\bibfield  {journal} {\bibinfo  {journal} {Collected Works: Representations of Functions, Celestial Mechanics and KAM Theory, 1957--1965}\ ,\ \bibinfo {pages} {267}} (\bibinfo {year} {2009})}\BibitemShut {NoStop}%
\bibitem [{\citenamefont {Kolmogorov}(1954)}]{kolmogorov1954conservation}%
  \BibitemOpen
  \bibfield  {author} {\bibinfo {author} {\bibfnamefont {A.~N.}\ \bibnamefont {Kolmogorov}},\ }\bibfield  {title} {\bibinfo {title} {On conservation of conditionally periodic motions for a small change in hamilton's function},\ }in\ \href@noop {} {\emph {\bibinfo {booktitle} {Dokl. Akad. Nauk SSSR}}},\ Vol.~\bibinfo {volume} {98}\ (\bibinfo {year} {1954})\ pp.\ \bibinfo {pages} {527--530}\BibitemShut {NoStop}%
\bibitem [{\citenamefont {M{\"o}ser}(1962)}]{moser1962invariant}%
  \BibitemOpen
  \bibfield  {author} {\bibinfo {author} {\bibfnamefont {J.}~\bibnamefont {M{\"o}ser}},\ }\bibfield  {title} {\bibinfo {title} {On invariant curves of area-preserving mappings of an annulus},\ }\href@noop {} {\bibfield  {journal} {\bibinfo  {journal} {Nachr. Akad. Wiss. G{\"o}ttingen, II}\ ,\ \bibinfo {pages} {1}} (\bibinfo {year} {1962})}\BibitemShut {NoStop}%
\bibitem [{\citenamefont {Moser}(1967)}]{moser1967convergent}%
  \BibitemOpen
  \bibfield  {author} {\bibinfo {author} {\bibfnamefont {J.}~\bibnamefont {Moser}},\ }\bibfield  {title} {\bibinfo {title} {Convergent series expansions for quasi-periodic motions},\ }\href@noop {} {\bibfield  {journal} {\bibinfo  {journal} {Mathematische Annalen}\ }\textbf {\bibinfo {volume} {169}},\ \bibinfo {pages} {136} (\bibinfo {year} {1967})}\BibitemShut {NoStop}%
\bibitem [{\citenamefont {Dumas}(2014)}]{dumas2014kam}%
  \BibitemOpen
  \bibfield  {author} {\bibinfo {author} {\bibfnamefont {H.~S.}\ \bibnamefont {Dumas}},\ }\href@noop {} {\emph {\bibinfo {title} {Kam Story, The: A Friendly Introduction To The Content, History, And Significance Of Classical Kolmogorov-arnold-moser Theory}}}\ (\bibinfo  {publisher} {World Scientific Publishing Company},\ \bibinfo {year} {2014})\BibitemShut {NoStop}%
\bibitem [{\citenamefont {Moser}(2001)}]{Moser+2001}%
  \BibitemOpen
  \bibfield  {author} {\bibinfo {author} {\bibfnamefont {J.}~\bibnamefont {Moser}},\ }\href {https://doi.org/doi:10.1515/9781400882694} {\emph {\bibinfo {title} {Stable and Random Motions in Dynamical Systems}}}\ (\bibinfo  {publisher} {Princeton University Press},\ \bibinfo {address} {Princeton},\ \bibinfo {year} {2001})\BibitemShut {NoStop}%
\bibitem [{\citenamefont {P{\"o}schel}(2009)}]{poschel2009lecture}%
  \BibitemOpen
  \bibfield  {author} {\bibinfo {author} {\bibfnamefont {J.}~\bibnamefont {P{\"o}schel}},\ }\bibfield  {title} {\bibinfo {title} {A lecture on the classical kam theorem},\ }\href@noop {} {\bibfield  {journal} {\bibinfo  {journal} {arXiv preprint arXiv:0908.2234}\ } (\bibinfo {year} {2009})}\BibitemShut {NoStop}%
\bibitem [{\citenamefont {Berman}\ and\ \citenamefont {Zaslavsky}(1978)}]{berman1978condition}%
  \BibitemOpen
  \bibfield  {author} {\bibinfo {author} {\bibfnamefont {G.~P.}\ \bibnamefont {Berman}}\ and\ \bibinfo {author} {\bibfnamefont {G.~M.}\ \bibnamefont {Zaslavsky}},\ }\bibfield  {title} {\bibinfo {title} {Condition of stochasticity in quantum nonlinear systems},\ }\href@noop {} {\bibfield  {journal} {\bibinfo  {journal} {Physica A: Statistical Mechanics and its Applications}\ }\textbf {\bibinfo {volume} {91}},\ \bibinfo {pages} {450} (\bibinfo {year} {1978})}\BibitemShut {NoStop}%
\bibitem [{\citenamefont {Toda}\ and\ \citenamefont {Ikeda}(1987)}]{toda1987quantal}%
  \BibitemOpen
  \bibfield  {author} {\bibinfo {author} {\bibfnamefont {M.}~\bibnamefont {Toda}}\ and\ \bibinfo {author} {\bibfnamefont {K.}~\bibnamefont {Ikeda}},\ }\bibfield  {title} {\bibinfo {title} {Quantal lyapunov exponent},\ }\href@noop {} {\bibfield  {journal} {\bibinfo  {journal} {Physics Letters A}\ }\textbf {\bibinfo {volume} {124}},\ \bibinfo {pages} {165} (\bibinfo {year} {1987})}\BibitemShut {NoStop}%
\bibitem [{\citenamefont {Gu}(1990)}]{gu1990evidences}%
  \BibitemOpen
  \bibfield  {author} {\bibinfo {author} {\bibfnamefont {Y.}~\bibnamefont {Gu}},\ }\bibfield  {title} {\bibinfo {title} {Evidences of classical and quantum chaos in the time evolution of nonequilibrium ensembles},\ }\href@noop {} {\bibfield  {journal} {\bibinfo  {journal} {Physics Letters A}\ }\textbf {\bibinfo {volume} {149}},\ \bibinfo {pages} {95} (\bibinfo {year} {1990})}\BibitemShut {NoStop}%
\bibitem [{\citenamefont {Zurek}(1998)}]{zurek1998decoherence}%
  \BibitemOpen
  \bibfield  {author} {\bibinfo {author} {\bibfnamefont {W.~H.}\ \bibnamefont {Zurek}},\ }\bibfield  {title} {\bibinfo {title} {Decoherence, chaos, quantum-classical correspondence, and the algorithmic arrow of time},\ }\href@noop {} {\bibfield  {journal} {\bibinfo  {journal} {Physica Scripta}\ }\textbf {\bibinfo {volume} {1998}},\ \bibinfo {pages} {186} (\bibinfo {year} {1998})}\BibitemShut {NoStop}%
\bibitem [{\citenamefont {Paz}\ and\ \citenamefont {Zurek}(2002)}]{paz2002environment}%
  \BibitemOpen
  \bibfield  {author} {\bibinfo {author} {\bibfnamefont {J.~P.}\ \bibnamefont {Paz}}\ and\ \bibinfo {author} {\bibfnamefont {W.~H.}\ \bibnamefont {Zurek}},\ }\bibfield  {title} {\bibinfo {title} {Environment-induced decoherence and the transition from quantum to classical},\ }in\ \href@noop {} {\emph {\bibinfo {booktitle} {Fundamentals of quantum information: quantum computation, communication, decoherence and all that}}}\ (\bibinfo  {publisher} {Springer},\ \bibinfo {year} {2002})\ pp.\ \bibinfo {pages} {77--148}\BibitemShut {NoStop}%
\bibitem [{\citenamefont {Bohigas}\ \emph {et~al.}(1984)\citenamefont {Bohigas}, \citenamefont {Giannoni},\ and\ \citenamefont {Schmit}}]{bohigas1984characterization}%
  \BibitemOpen
  \bibfield  {author} {\bibinfo {author} {\bibfnamefont {O.}~\bibnamefont {Bohigas}}, \bibinfo {author} {\bibfnamefont {M.-J.}\ \bibnamefont {Giannoni}},\ and\ \bibinfo {author} {\bibfnamefont {C.}~\bibnamefont {Schmit}},\ }\bibfield  {title} {\bibinfo {title} {Characterization of chaotic quantum spectra and universality of level fluctuation laws},\ }\href {https://journals.aps.org/prl/abstract/10.1103/PhysRevLett.52.1} {\bibfield  {journal} {\bibinfo  {journal} {Physical review letters}\ }\textbf {\bibinfo {volume} {52}},\ \bibinfo {pages} {1} (\bibinfo {year} {1984})}\BibitemShut {NoStop}%
\bibitem [{\citenamefont {Berry}\ and\ \citenamefont {Tabor}(1977)}]{berry1977level}%
  \BibitemOpen
  \bibfield  {author} {\bibinfo {author} {\bibfnamefont {M.~V.}\ \bibnamefont {Berry}}\ and\ \bibinfo {author} {\bibfnamefont {M.}~\bibnamefont {Tabor}},\ }\bibfield  {title} {\bibinfo {title} {Level clustering in the regular spectrum},\ }\href@noop {} {\bibfield  {journal} {\bibinfo  {journal} {Proceedings of the Royal Society of London. A. Mathematical and Physical Sciences}\ }\textbf {\bibinfo {volume} {356}},\ \bibinfo {pages} {375} (\bibinfo {year} {1977})}\BibitemShut {NoStop}%
\bibitem [{\citenamefont {Wigner}(1955)}]{wigner55}%
  \BibitemOpen
  \bibfield  {author} {\bibinfo {author} {\bibfnamefont {E.~P.}\ \bibnamefont {Wigner}},\ }\bibfield  {title} {\bibinfo {title} {Characteristic vectors of bordered matrices with infinite dimensions},\ }\href {http://www.jstor.org/stable/1970079} {\bibfield  {journal} {\bibinfo  {journal} {Annals of Mathematics}\ }\textbf {\bibinfo {volume} {62}},\ \bibinfo {pages} {548} (\bibinfo {year} {1955})}\BibitemShut {NoStop}%
\bibitem [{\citenamefont {Bohigas}\ and\ \citenamefont {Flores}(1971)}]{bohigas1971spacing}%
  \BibitemOpen
  \bibfield  {author} {\bibinfo {author} {\bibfnamefont {O.}~\bibnamefont {Bohigas}}\ and\ \bibinfo {author} {\bibfnamefont {J.}~\bibnamefont {Flores}},\ }\bibfield  {title} {\bibinfo {title} {Spacing and individual eigenvalue distributions of two-body random hamiltonians},\ }\href@noop {} {\bibfield  {journal} {\bibinfo  {journal} {Physics Letters B}\ }\textbf {\bibinfo {volume} {35}},\ \bibinfo {pages} {383} (\bibinfo {year} {1971})}\BibitemShut {NoStop}%
\bibitem [{\citenamefont {Haake}(1991{\natexlab{a}})}]{Haake}%
  \BibitemOpen
  \bibfield  {author} {\bibinfo {author} {\bibfnamefont {F.}~\bibnamefont {Haake}},\ }\href@noop {} {\emph {\bibinfo {title} {Quantum Signatures of Chaos}}}\ (\bibinfo  {publisher} {Spring-Verlag, Berlin},\ \bibinfo {year} {1991})\BibitemShut {NoStop}%
\bibitem [{\citenamefont {Delande}\ and\ \citenamefont {Gay}(1986)}]{delande1986quantum}%
  \BibitemOpen
  \bibfield  {author} {\bibinfo {author} {\bibfnamefont {D.}~\bibnamefont {Delande}}\ and\ \bibinfo {author} {\bibfnamefont {J.}~\bibnamefont {Gay}},\ }\bibfield  {title} {\bibinfo {title} {Quantum chaos and statistical properties of energy levels: Numerical study of the hydrogen atom in a magnetic field},\ }\href@noop {} {\bibfield  {journal} {\bibinfo  {journal} {Physical review letters}\ }\textbf {\bibinfo {volume} {57}},\ \bibinfo {pages} {2006} (\bibinfo {year} {1986})}\BibitemShut {NoStop}%
\bibitem [{\citenamefont {Atas}\ \emph {et~al.}(2013{\natexlab{a}})\citenamefont {Atas}, \citenamefont {Bogomolny}, \citenamefont {Giraud}, \citenamefont {Vivo},\ and\ \citenamefont {Vivo}}]{atas2013joint}%
  \BibitemOpen
  \bibfield  {author} {\bibinfo {author} {\bibfnamefont {Y.}~\bibnamefont {Atas}}, \bibinfo {author} {\bibfnamefont {E.}~\bibnamefont {Bogomolny}}, \bibinfo {author} {\bibfnamefont {O.}~\bibnamefont {Giraud}}, \bibinfo {author} {\bibfnamefont {P.}~\bibnamefont {Vivo}},\ and\ \bibinfo {author} {\bibfnamefont {E.}~\bibnamefont {Vivo}},\ }\bibfield  {title} {\bibinfo {title} {Joint probability densities of level spacing ratios in random matrices},\ }\href@noop {} {\bibfield  {journal} {\bibinfo  {journal} {Journal of Physics A: Mathematical and Theoretical}\ }\textbf {\bibinfo {volume} {46}},\ \bibinfo {pages} {355204} (\bibinfo {year} {2013}{\natexlab{a}})}\BibitemShut {NoStop}%
\bibitem [{\citenamefont {Bhosale}\ \emph {et~al.}(2018)\citenamefont {Bhosale}, \citenamefont {Tekur},\ and\ \citenamefont {Santhanam}}]{bhosale2018scaling}%
  \BibitemOpen
  \bibfield  {author} {\bibinfo {author} {\bibfnamefont {U.~T.}\ \bibnamefont {Bhosale}}, \bibinfo {author} {\bibfnamefont {S.~H.}\ \bibnamefont {Tekur}},\ and\ \bibinfo {author} {\bibfnamefont {M.}~\bibnamefont {Santhanam}},\ }\bibfield  {title} {\bibinfo {title} {Scaling in the eigenvalue fluctuations of correlation matrices},\ }\href@noop {} {\bibfield  {journal} {\bibinfo  {journal} {Physical Review E}\ }\textbf {\bibinfo {volume} {98}},\ \bibinfo {pages} {052133} (\bibinfo {year} {2018})}\BibitemShut {NoStop}%
\bibitem [{\citenamefont {Tekur}\ \emph {et~al.}(2018{\natexlab{a}})\citenamefont {Tekur}, \citenamefont {Kumar},\ and\ \citenamefont {Santhanam}}]{tekur2018exact}%
  \BibitemOpen
  \bibfield  {author} {\bibinfo {author} {\bibfnamefont {S.~H.}\ \bibnamefont {Tekur}}, \bibinfo {author} {\bibfnamefont {S.}~\bibnamefont {Kumar}},\ and\ \bibinfo {author} {\bibfnamefont {M.}~\bibnamefont {Santhanam}},\ }\bibfield  {title} {\bibinfo {title} {Exact distribution of spacing ratios for random and localized states in quantum chaotic systems},\ }\href@noop {} {\bibfield  {journal} {\bibinfo  {journal} {Physical Review E}\ }\textbf {\bibinfo {volume} {97}},\ \bibinfo {pages} {062212} (\bibinfo {year} {2018}{\natexlab{a}})}\BibitemShut {NoStop}%
\bibitem [{\citenamefont {Tekur}\ \emph {et~al.}(2018{\natexlab{b}})\citenamefont {Tekur}, \citenamefont {Bhosale},\ and\ \citenamefont {Santhanam}}]{tekur2018higher}%
  \BibitemOpen
  \bibfield  {author} {\bibinfo {author} {\bibfnamefont {S.~H.}\ \bibnamefont {Tekur}}, \bibinfo {author} {\bibfnamefont {U.~T.}\ \bibnamefont {Bhosale}},\ and\ \bibinfo {author} {\bibfnamefont {M.}~\bibnamefont {Santhanam}},\ }\bibfield  {title} {\bibinfo {title} {Higher-order spacing ratios in random matrix theory and complex quantum systems},\ }\href@noop {} {\bibfield  {journal} {\bibinfo  {journal} {Physical Review B}\ }\textbf {\bibinfo {volume} {98}},\ \bibinfo {pages} {104305} (\bibinfo {year} {2018}{\natexlab{b}})}\BibitemShut {NoStop}%
\bibitem [{\citenamefont {Chaudhury}\ \emph {et~al.}(2009{\natexlab{a}})\citenamefont {Chaudhury}, \citenamefont {Smith}, \citenamefont {Anderson}, \citenamefont {Ghose},\ and\ \citenamefont {Jessen}}]{Chaudhary}%
  \BibitemOpen
  \bibfield  {author} {\bibinfo {author} {\bibfnamefont {S.}~\bibnamefont {Chaudhury}}, \bibinfo {author} {\bibfnamefont {A.}~\bibnamefont {Smith}}, \bibinfo {author} {\bibfnamefont {B.~E.}\ \bibnamefont {Anderson}}, \bibinfo {author} {\bibfnamefont {S.}~\bibnamefont {Ghose}},\ and\ \bibinfo {author} {\bibfnamefont {P.~S.}\ \bibnamefont {Jessen}},\ }\bibfield  {title} {\bibinfo {title} {Quantum signatures of chaos in a kicked top},\ }\href@noop {} {\bibfield  {journal} {\bibinfo  {journal} {Nature}\ }\textbf {\bibinfo {volume} {461}},\ \bibinfo {pages} {768} (\bibinfo {year} {2009}{\natexlab{a}})}\BibitemShut {NoStop}%
\bibitem [{\citenamefont {Poggi}\ \emph {et~al.}(2020)\citenamefont {Poggi}, \citenamefont {Lysne}, \citenamefont {Kuper}, \citenamefont {Deutsch},\ and\ \citenamefont {Jessen}}]{poggi2020quantifying}%
  \BibitemOpen
  \bibfield  {author} {\bibinfo {author} {\bibfnamefont {P.~M.}\ \bibnamefont {Poggi}}, \bibinfo {author} {\bibfnamefont {N.~K.}\ \bibnamefont {Lysne}}, \bibinfo {author} {\bibfnamefont {K.~W.}\ \bibnamefont {Kuper}}, \bibinfo {author} {\bibfnamefont {I.~H.}\ \bibnamefont {Deutsch}},\ and\ \bibinfo {author} {\bibfnamefont {P.~S.}\ \bibnamefont {Jessen}},\ }\bibfield  {title} {\bibinfo {title} {Quantifying the sensitivity to errors in analog quantum simulation},\ }\href@noop {} {\bibfield  {journal} {\bibinfo  {journal} {PRX Quantum}\ }\textbf {\bibinfo {volume} {1}},\ \bibinfo {pages} {020308} (\bibinfo {year} {2020})}\BibitemShut {NoStop}%
\bibitem [{\citenamefont {Neill}\ \emph {et~al.}(2016{\natexlab{a}})\citenamefont {Neill}, \citenamefont {Roushan}, \citenamefont {Fang}, \citenamefont {Chen}, \citenamefont {Kolodrubetz}, \citenamefont {Chen}, \citenamefont {Megrant}, \citenamefont {Barends}, \citenamefont {Campbell}, \citenamefont {Chiaro}, \citenamefont {Dunsworth}, \citenamefont {Jeffrey}, \citenamefont {Kelly}, \citenamefont {Mutus}, \citenamefont {O’Malley}, \citenamefont {Quintana}, \citenamefont {Sank}, \citenamefont {Vainsencher}, \citenamefont {Wenner}, \citenamefont {White}, \citenamefont {Polkovnikov},\ and\ \citenamefont {Martinis}}]{Neill16}%
  \BibitemOpen
  \bibfield  {author} {\bibinfo {author} {\bibfnamefont {C.}~\bibnamefont {Neill}}, \bibinfo {author} {\bibfnamefont {P.}~\bibnamefont {Roushan}}, \bibinfo {author} {\bibfnamefont {M.}~\bibnamefont {Fang}}, \bibinfo {author} {\bibfnamefont {Y.}~\bibnamefont {Chen}}, \bibinfo {author} {\bibfnamefont {M.}~\bibnamefont {Kolodrubetz}}, \bibinfo {author} {\bibfnamefont {Z.}~\bibnamefont {Chen}}, \bibinfo {author} {\bibfnamefont {A.}~\bibnamefont {Megrant}}, \bibinfo {author} {\bibfnamefont {R.}~\bibnamefont {Barends}}, \bibinfo {author} {\bibfnamefont {B.}~\bibnamefont {Campbell}}, \bibinfo {author} {\bibfnamefont {B.}~\bibnamefont {Chiaro}}, \bibinfo {author} {\bibfnamefont {A.}~\bibnamefont {Dunsworth}}, \bibinfo {author} {\bibfnamefont {E.}~\bibnamefont {Jeffrey}}, \bibinfo {author} {\bibfnamefont {J.}~\bibnamefont {Kelly}}, \bibinfo {author} {\bibfnamefont {J.}~\bibnamefont {Mutus}}, \bibinfo {author} {\bibfnamefont {P.~J.~J.}\ \bibnamefont {O’Malley}}, \bibinfo {author} {\bibfnamefont {C.}~\bibnamefont
  {Quintana}}, \bibinfo {author} {\bibfnamefont {D.}~\bibnamefont {Sank}}, \bibinfo {author} {\bibfnamefont {A.}~\bibnamefont {Vainsencher}}, \bibinfo {author} {\bibfnamefont {J.}~\bibnamefont {Wenner}}, \bibinfo {author} {\bibfnamefont {T.~C.}\ \bibnamefont {White}}, \bibinfo {author} {\bibfnamefont {A.}~\bibnamefont {Polkovnikov}},\ and\ \bibinfo {author} {\bibfnamefont {J.~M.}\ \bibnamefont {Martinis}},\ }\bibfield  {title} {\bibinfo {title} {Ergodic dynamics and thermalization in an isolated quantum system},\ }\href@noop {} {\bibfield  {journal} {\bibinfo  {journal} {Nature Physics}\ }\textbf {\bibinfo {volume} {12}},\ \bibinfo {pages} {1037} (\bibinfo {year} {2016}{\natexlab{a}})}\BibitemShut {NoStop}%
\bibitem [{\citenamefont {Sieberer}\ \emph {et~al.}(2019)\citenamefont {Sieberer}, \citenamefont {Olsacher}, \citenamefont {Elben}, \citenamefont {Heyl}, \citenamefont {Hauke}, \citenamefont {Haake},\ and\ \citenamefont {Zoller}}]{sieberer2019digital}%
  \BibitemOpen
  \bibfield  {author} {\bibinfo {author} {\bibfnamefont {L.~M.}\ \bibnamefont {Sieberer}}, \bibinfo {author} {\bibfnamefont {T.}~\bibnamefont {Olsacher}}, \bibinfo {author} {\bibfnamefont {A.}~\bibnamefont {Elben}}, \bibinfo {author} {\bibfnamefont {M.}~\bibnamefont {Heyl}}, \bibinfo {author} {\bibfnamefont {P.}~\bibnamefont {Hauke}}, \bibinfo {author} {\bibfnamefont {F.}~\bibnamefont {Haake}},\ and\ \bibinfo {author} {\bibfnamefont {P.}~\bibnamefont {Zoller}},\ }\bibfield  {title} {\bibinfo {title} {Digital quantum simulation, trotter errors, and quantum chaos of the kicked top},\ }\href@noop {} {\bibfield  {journal} {\bibinfo  {journal} {npj Quantum Information}\ }\textbf {\bibinfo {volume} {5}},\ \bibinfo {pages} {1} (\bibinfo {year} {2019})}\BibitemShut {NoStop}%
\bibitem [{\citenamefont {Li}\ \emph {et~al.}(2017)\citenamefont {Li}, \citenamefont {Fan}, \citenamefont {Wang}, \citenamefont {Ye}, \citenamefont {Zeng}, \citenamefont {Zhai}, \citenamefont {Peng},\ and\ \citenamefont {Du}}]{PhysRevX.7.031011}%
  \BibitemOpen
  \bibfield  {author} {\bibinfo {author} {\bibfnamefont {J.}~\bibnamefont {Li}}, \bibinfo {author} {\bibfnamefont {R.}~\bibnamefont {Fan}}, \bibinfo {author} {\bibfnamefont {H.}~\bibnamefont {Wang}}, \bibinfo {author} {\bibfnamefont {B.}~\bibnamefont {Ye}}, \bibinfo {author} {\bibfnamefont {B.}~\bibnamefont {Zeng}}, \bibinfo {author} {\bibfnamefont {H.}~\bibnamefont {Zhai}}, \bibinfo {author} {\bibfnamefont {X.}~\bibnamefont {Peng}},\ and\ \bibinfo {author} {\bibfnamefont {J.}~\bibnamefont {Du}},\ }\bibfield  {title} {\bibinfo {title} {Measuring out-of-time-order correlators on a nuclear magnetic resonance quantum simulator},\ }\href {https://doi.org/10.1103/PhysRevX.7.031011} {\bibfield  {journal} {\bibinfo  {journal} {Phys. Rev. X}\ }\textbf {\bibinfo {volume} {7}},\ \bibinfo {pages} {031011} (\bibinfo {year} {2017})}\BibitemShut {NoStop}%
\bibitem [{\citenamefont {Shepelyansky}(1983)}]{shepelyansky1983some}%
  \BibitemOpen
  \bibfield  {author} {\bibinfo {author} {\bibfnamefont {D.~L.}\ \bibnamefont {Shepelyansky}},\ }\bibfield  {title} {\bibinfo {title} {Some statistical properties of simple classically stochastic quantum systems},\ }\href@noop {} {\bibfield  {journal} {\bibinfo  {journal} {Physica D: Nonlinear Phenomena}\ }\textbf {\bibinfo {volume} {8}},\ \bibinfo {pages} {208} (\bibinfo {year} {1983})}\BibitemShut {NoStop}%
\bibitem [{\citenamefont {Peres}(1984)}]{peres1984stability}%
  \BibitemOpen
  \bibfield  {author} {\bibinfo {author} {\bibfnamefont {A.}~\bibnamefont {Peres}},\ }\bibfield  {title} {\bibinfo {title} {Stability of quantum motion in chaotic and regular systems},\ }\href@noop {} {\bibfield  {journal} {\bibinfo  {journal} {Physical Review A}\ }\textbf {\bibinfo {volume} {30}},\ \bibinfo {pages} {1610} (\bibinfo {year} {1984})}\BibitemShut {NoStop}%
\bibitem [{\citenamefont {Chinni}\ \emph {et~al.}(2022)\citenamefont {Chinni}, \citenamefont {Mu{\~n}oz-Arias}, \citenamefont {Deutsch},\ and\ \citenamefont {Poggi}}]{chinni2022trotter}%
  \BibitemOpen
  \bibfield  {author} {\bibinfo {author} {\bibfnamefont {K.}~\bibnamefont {Chinni}}, \bibinfo {author} {\bibfnamefont {M.~H.}\ \bibnamefont {Mu{\~n}oz-Arias}}, \bibinfo {author} {\bibfnamefont {I.~H.}\ \bibnamefont {Deutsch}},\ and\ \bibinfo {author} {\bibfnamefont {P.~M.}\ \bibnamefont {Poggi}},\ }\bibfield  {title} {\bibinfo {title} {Trotter errors from dynamical structural instabilities of floquet maps in quantum simulation},\ }\href@noop {} {\bibfield  {journal} {\bibinfo  {journal} {PRX Quantum}\ }\textbf {\bibinfo {volume} {3}},\ \bibinfo {pages} {010351} (\bibinfo {year} {2022})}\BibitemShut {NoStop}%
\bibitem [{\citenamefont {Tomsovic}\ \emph {et~al.}(2023)\citenamefont {Tomsovic}, \citenamefont {Urbina},\ and\ \citenamefont {Richter}}]{tomsovic2023controlling}%
  \BibitemOpen
  \bibfield  {author} {\bibinfo {author} {\bibfnamefont {S.}~\bibnamefont {Tomsovic}}, \bibinfo {author} {\bibfnamefont {J.~D.}\ \bibnamefont {Urbina}},\ and\ \bibinfo {author} {\bibfnamefont {K.}~\bibnamefont {Richter}},\ }\bibfield  {title} {\bibinfo {title} {Controlling quantum chaos: Optimal coherent targeting},\ }\href@noop {} {\bibfield  {journal} {\bibinfo  {journal} {Physical Review Letters}\ }\textbf {\bibinfo {volume} {130}},\ \bibinfo {pages} {020201} (\bibinfo {year} {2023})}\BibitemShut {NoStop}%
\bibitem [{\citenamefont {Babelon}\ \emph {et~al.}(2003)\citenamefont {Babelon}, \citenamefont {Bernard},\ and\ \citenamefont {Talon}}]{babelon2003introduction}%
  \BibitemOpen
  \bibfield  {author} {\bibinfo {author} {\bibfnamefont {O.}~\bibnamefont {Babelon}}, \bibinfo {author} {\bibfnamefont {D.}~\bibnamefont {Bernard}},\ and\ \bibinfo {author} {\bibfnamefont {M.}~\bibnamefont {Talon}},\ }\href@noop {} {\emph {\bibinfo {title} {Introduction to classical integrable systems}}}\ (\bibinfo  {publisher} {Cambridge University Press},\ \bibinfo {year} {2003})\BibitemShut {NoStop}%
\bibitem [{\citenamefont {Arnol'd}(2013)}]{arnol2013mathematical}%
  \BibitemOpen
  \bibfield  {author} {\bibinfo {author} {\bibfnamefont {V.~I.}\ \bibnamefont {Arnol'd}},\ }\href@noop {} {\emph {\bibinfo {title} {Mathematical methods of classical mechanics}}},\ Vol.~\bibinfo {volume} {60}\ (\bibinfo  {publisher} {Springer Science \& Business Media},\ \bibinfo {year} {2013})\BibitemShut {NoStop}%
\bibitem [{\citenamefont {Reichl}(2021)}]{reichl2021transition}%
  \BibitemOpen
  \bibfield  {author} {\bibinfo {author} {\bibfnamefont {L.~E.}\ \bibnamefont {Reichl}},\ }\href@noop {} {\emph {\bibinfo {title} {Transition to Chaos}}}\ (\bibinfo  {publisher} {Springer},\ \bibinfo {year} {2021})\BibitemShut {NoStop}%
\bibitem [{\citenamefont {Genecand}(1993)}]{genecand1993transversal}%
  \BibitemOpen
  \bibfield  {author} {\bibinfo {author} {\bibfnamefont {C.}~\bibnamefont {Genecand}},\ }\bibfield  {title} {\bibinfo {title} {Transversal homoclinic orbits near elliptic fixed points of area-preserving diffeomorphisms of the plane},\ }in\ \href@noop {} {\emph {\bibinfo {booktitle} {Dynamics Reported: Expositions in Dynamical Systems}}}\ (\bibinfo  {publisher} {Springer},\ \bibinfo {year} {1993})\ pp.\ \bibinfo {pages} {1--30}\BibitemShut {NoStop}%
\bibitem [{\citenamefont {Markus}\ and\ \citenamefont {Meyer}(1974)}]{markus1974generic}%
  \BibitemOpen
  \bibfield  {author} {\bibinfo {author} {\bibfnamefont {L.}~\bibnamefont {Markus}}\ and\ \bibinfo {author} {\bibfnamefont {K.~R.}\ \bibnamefont {Meyer}},\ }\href@noop {} {\emph {\bibinfo {title} {Generic Hamiltonian dynamical systems are neither integrable nor ergodic}}},\ Vol.\ \bibinfo {volume} {144}\ (\bibinfo  {publisher} {American Mathematical Society Providence},\ \bibinfo {year} {1974})\BibitemShut {NoStop}%
\bibitem [{\citenamefont {Zehnder}(1973)}]{zehnder1973homoclinic}%
  \BibitemOpen
  \bibfield  {author} {\bibinfo {author} {\bibfnamefont {E.}~\bibnamefont {Zehnder}},\ }\bibfield  {title} {\bibinfo {title} {Homoclinic points near elliptic fixed points},\ }\href@noop {} {\bibfield  {journal} {\bibinfo  {journal} {Communications on pure and applied mathematics}\ }\textbf {\bibinfo {volume} {26}},\ \bibinfo {pages} {131} (\bibinfo {year} {1973})}\BibitemShut {NoStop}%
\bibitem [{\citenamefont {Cornfeld}\ \emph {et~al.}(2012)\citenamefont {Cornfeld}, \citenamefont {Fomin},\ and\ \citenamefont {Sinai}}]{cornfeld2012ergodic}%
  \BibitemOpen
  \bibfield  {author} {\bibinfo {author} {\bibfnamefont {I.~P.}\ \bibnamefont {Cornfeld}}, \bibinfo {author} {\bibfnamefont {S.~V.}\ \bibnamefont {Fomin}},\ and\ \bibinfo {author} {\bibfnamefont {Y.~G.}\ \bibnamefont {Sinai}},\ }\href@noop {} {\emph {\bibinfo {title} {Ergodic theory}}},\ Vol.\ \bibinfo {volume} {245}\ (\bibinfo  {publisher} {Springer Science \& Business Media},\ \bibinfo {year} {2012})\BibitemShut {NoStop}%
\bibitem [{\citenamefont {Halmos}(2017)}]{halmos2017lectures}%
  \BibitemOpen
  \bibfield  {author} {\bibinfo {author} {\bibfnamefont {P.~R.}\ \bibnamefont {Halmos}},\ }\href@noop {} {\emph {\bibinfo {title} {Lectures on ergodic theory}}}\ (\bibinfo  {publisher} {Courier Dover Publications},\ \bibinfo {year} {2017})\BibitemShut {NoStop}%
\bibitem [{\citenamefont {Frigg}\ \emph {et~al.}(2020)\citenamefont {Frigg}, \citenamefont {Berkovitz},\ and\ \citenamefont {Kronz}}]{sep-ergodic-hierarchy}%
  \BibitemOpen
  \bibfield  {author} {\bibinfo {author} {\bibfnamefont {R.}~\bibnamefont {Frigg}}, \bibinfo {author} {\bibfnamefont {J.}~\bibnamefont {Berkovitz}},\ and\ \bibinfo {author} {\bibfnamefont {F.}~\bibnamefont {Kronz}},\ }\bibfield  {title} {\bibinfo {title} {{The Ergodic Hierarchy}},\ }in\ \href@noop {} {\emph {\bibinfo {booktitle} {The {Stanford} Encyclopedia of Philosophy}}},\ \bibinfo {editor} {edited by\ \bibinfo {editor} {\bibfnamefont {E.~N.}\ \bibnamefont {Zalta}}}\ (\bibinfo  {publisher} {Metaphysics Research Lab, Stanford University},\ \bibinfo {year} {2020})\ \bibinfo {edition} {{F}all 2020}\ ed.\BibitemShut {Stop}%
\bibitem [{\citenamefont {Sankaranarayanan}\ \emph {et~al.}(2001{\natexlab{a}})\citenamefont {Sankaranarayanan}, \citenamefont {Lakshminarayan},\ and\ \citenamefont {Sheorey}}]{sankaranarayanan2001quantum}%
  \BibitemOpen
  \bibfield  {author} {\bibinfo {author} {\bibfnamefont {R.}~\bibnamefont {Sankaranarayanan}}, \bibinfo {author} {\bibfnamefont {A.}~\bibnamefont {Lakshminarayan}},\ and\ \bibinfo {author} {\bibfnamefont {V.}~\bibnamefont {Sheorey}},\ }\bibfield  {title} {\bibinfo {title} {Quantum chaos of a particle in a square well: Competing length scales and dynamical localization},\ }\href@noop {} {\bibfield  {journal} {\bibinfo  {journal} {Physical Review E}\ }\textbf {\bibinfo {volume} {64}},\ \bibinfo {pages} {046210} (\bibinfo {year} {2001}{\natexlab{a}})}\BibitemShut {NoStop}%
\bibitem [{\citenamefont {Sankaranarayanan}\ \emph {et~al.}(2001{\natexlab{b}})\citenamefont {Sankaranarayanan}, \citenamefont {Lakshminarayan},\ and\ \citenamefont {Sheorey}}]{sankaranarayanan2001chaos}%
  \BibitemOpen
  \bibfield  {author} {\bibinfo {author} {\bibfnamefont {R.}~\bibnamefont {Sankaranarayanan}}, \bibinfo {author} {\bibfnamefont {A.}~\bibnamefont {Lakshminarayan}},\ and\ \bibinfo {author} {\bibfnamefont {V.}~\bibnamefont {Sheorey}},\ }\bibfield  {title} {\bibinfo {title} {Chaos in a well: effects of competing length scales},\ }\href@noop {} {\bibfield  {journal} {\bibinfo  {journal} {Physics Letters A}\ }\textbf {\bibinfo {volume} {279}},\ \bibinfo {pages} {313} (\bibinfo {year} {2001}{\natexlab{b}})}\BibitemShut {NoStop}%
\bibitem [{\citenamefont {Gutzwiller}(2013)}]{gutzwiller2013chaos}%
  \BibitemOpen
  \bibfield  {author} {\bibinfo {author} {\bibfnamefont {M.~C.}\ \bibnamefont {Gutzwiller}},\ }\href@noop {} {\emph {\bibinfo {title} {Chaos in classical and quantum mechanics}}},\ Vol.~\bibinfo {volume} {1}\ (\bibinfo  {publisher} {Springer Science \& Business Media},\ \bibinfo {year} {2013})\BibitemShut {NoStop}%
\bibitem [{\citenamefont {Gutzwiller}(1970)}]{gutzwiller1970energy}%
  \BibitemOpen
  \bibfield  {author} {\bibinfo {author} {\bibfnamefont {M.~C.}\ \bibnamefont {Gutzwiller}},\ }\bibfield  {title} {\bibinfo {title} {Energy spectrum according to classical mechanics},\ }\href@noop {} {\bibfield  {journal} {\bibinfo  {journal} {Journal of Mathematical Physics}\ }\textbf {\bibinfo {volume} {11}},\ \bibinfo {pages} {1791} (\bibinfo {year} {1970})}\BibitemShut {NoStop}%
\bibitem [{\citenamefont {Neill}\ \emph {et~al.}(2016{\natexlab{b}})\citenamefont {Neill}, \citenamefont {Roushan}, \citenamefont {Fang}, \citenamefont {Chen}, \citenamefont {Kolodrubetz}, \citenamefont {Chen}, \citenamefont {Megrant}, \citenamefont {Barends}, \citenamefont {Campbell}, \citenamefont {Chiaro}, \citenamefont {Dunsworth}, \citenamefont {Jeffrey}, \citenamefont {Kelly}, \citenamefont {Mutus}, \citenamefont {O’Malley}, \citenamefont {Quintana}, \citenamefont {Sank}, \citenamefont {Vainsencher}, \citenamefont {Wenner}, \citenamefont {White}, \citenamefont {Polkovnikov},\ and\ \citenamefont {Martinis}}]{Neill_2016}%
  \BibitemOpen
  \bibfield  {author} {\bibinfo {author} {\bibfnamefont {C.}~\bibnamefont {Neill}}, \bibinfo {author} {\bibfnamefont {P.}~\bibnamefont {Roushan}}, \bibinfo {author} {\bibfnamefont {M.}~\bibnamefont {Fang}}, \bibinfo {author} {\bibfnamefont {Y.}~\bibnamefont {Chen}}, \bibinfo {author} {\bibfnamefont {M.}~\bibnamefont {Kolodrubetz}}, \bibinfo {author} {\bibfnamefont {Z.}~\bibnamefont {Chen}}, \bibinfo {author} {\bibfnamefont {A.}~\bibnamefont {Megrant}}, \bibinfo {author} {\bibfnamefont {R.}~\bibnamefont {Barends}}, \bibinfo {author} {\bibfnamefont {B.}~\bibnamefont {Campbell}}, \bibinfo {author} {\bibfnamefont {B.}~\bibnamefont {Chiaro}}, \bibinfo {author} {\bibfnamefont {A.}~\bibnamefont {Dunsworth}}, \bibinfo {author} {\bibfnamefont {E.}~\bibnamefont {Jeffrey}}, \bibinfo {author} {\bibfnamefont {J.}~\bibnamefont {Kelly}}, \bibinfo {author} {\bibfnamefont {J.}~\bibnamefont {Mutus}}, \bibinfo {author} {\bibfnamefont {P.~J.~J.}\ \bibnamefont {O’Malley}}, \bibinfo {author} {\bibfnamefont {C.}~\bibnamefont
  {Quintana}}, \bibinfo {author} {\bibfnamefont {D.}~\bibnamefont {Sank}}, \bibinfo {author} {\bibfnamefont {A.}~\bibnamefont {Vainsencher}}, \bibinfo {author} {\bibfnamefont {J.}~\bibnamefont {Wenner}}, \bibinfo {author} {\bibfnamefont {T.~C.}\ \bibnamefont {White}}, \bibinfo {author} {\bibfnamefont {A.}~\bibnamefont {Polkovnikov}},\ and\ \bibinfo {author} {\bibfnamefont {J.~M.}\ \bibnamefont {Martinis}},\ }\bibfield  {title} {\bibinfo {title} {Ergodic dynamics and thermalization in an isolated quantum system},\ }\href {https://doi.org/10.1038/nphys3830} {\bibfield  {journal} {\bibinfo  {journal} {Nature Physics}\ }\textbf {\bibinfo {volume} {12}},\ \bibinfo {pages} {1037–1041} (\bibinfo {year} {2016}{\natexlab{b}})}\BibitemShut {NoStop}%
\bibitem [{\citenamefont {Eisert}\ \emph {et~al.}(2015)\citenamefont {Eisert}, \citenamefont {Friesdorf},\ and\ \citenamefont {Gogolin}}]{Eisert2015}%
  \BibitemOpen
  \bibfield  {author} {\bibinfo {author} {\bibfnamefont {J.}~\bibnamefont {Eisert}}, \bibinfo {author} {\bibfnamefont {M.}~\bibnamefont {Friesdorf}},\ and\ \bibinfo {author} {\bibfnamefont {C.}~\bibnamefont {Gogolin}},\ }\bibfield  {title} {\bibinfo {title} {Quantum many-body systems out of equilibrium},\ }\href {https://doi.org/10.1038/nphys3215} {\bibfield  {journal} {\bibinfo  {journal} {Nature Physics}\ }\textbf {\bibinfo {volume} {11}},\ \bibinfo {pages} {124} (\bibinfo {year} {2015})}\BibitemShut {NoStop}%
\bibitem [{\citenamefont {Rigol}\ \emph {et~al.}(2008{\natexlab{a}})\citenamefont {Rigol}, \citenamefont {Dunjko},\ and\ \citenamefont {Olshanii}}]{Rigol2008}%
  \BibitemOpen
  \bibfield  {author} {\bibinfo {author} {\bibfnamefont {M.}~\bibnamefont {Rigol}}, \bibinfo {author} {\bibfnamefont {V.}~\bibnamefont {Dunjko}},\ and\ \bibinfo {author} {\bibfnamefont {M.}~\bibnamefont {Olshanii}},\ }\bibfield  {title} {\bibinfo {title} {Thermalization and its mechanism for generic isolated quantum systems},\ }\href {https://doi.org/10.1038/nature06838} {\bibfield  {journal} {\bibinfo  {journal} {Nature}\ }\textbf {\bibinfo {volume} {452}},\ \bibinfo {pages} {854} (\bibinfo {year} {2008}{\natexlab{a}})}\BibitemShut {NoStop}%
\bibitem [{\citenamefont {Engl}\ \emph {et~al.}(2016)\citenamefont {Engl}, \citenamefont {Urbina},\ and\ \citenamefont {Richter}}]{engl2016semiclassical}%
  \BibitemOpen
  \bibfield  {author} {\bibinfo {author} {\bibfnamefont {T.}~\bibnamefont {Engl}}, \bibinfo {author} {\bibfnamefont {J.~D.}\ \bibnamefont {Urbina}},\ and\ \bibinfo {author} {\bibfnamefont {K.}~\bibnamefont {Richter}},\ }\bibfield  {title} {\bibinfo {title} {The semiclassical propagator in fock space: dynamical echo and many-body interference},\ }\href@noop {} {\bibfield  {journal} {\bibinfo  {journal} {Philosophical Transactions of the Royal Society A: Mathematical, Physical and Engineering Sciences}\ }\textbf {\bibinfo {volume} {374}},\ \bibinfo {pages} {20150159} (\bibinfo {year} {2016})}\BibitemShut {NoStop}%
\bibitem [{\citenamefont {Engl}\ \emph {et~al.}(2014)\citenamefont {Engl}, \citenamefont {Dujardin}, \citenamefont {Arg\"uelles}, \citenamefont {Schlagheck}, \citenamefont {Richter},\ and\ \citenamefont {Urbina}}]{ThomasPrl2014}%
  \BibitemOpen
  \bibfield  {author} {\bibinfo {author} {\bibfnamefont {T.}~\bibnamefont {Engl}}, \bibinfo {author} {\bibfnamefont {J.}~\bibnamefont {Dujardin}}, \bibinfo {author} {\bibfnamefont {A.}~\bibnamefont {Arg\"uelles}}, \bibinfo {author} {\bibfnamefont {P.}~\bibnamefont {Schlagheck}}, \bibinfo {author} {\bibfnamefont {K.}~\bibnamefont {Richter}},\ and\ \bibinfo {author} {\bibfnamefont {J.~D.}\ \bibnamefont {Urbina}},\ }\bibfield  {title} {\bibinfo {title} {Coherent backscattering in fock space: A signature of quantum many-body interference in interacting bosonic systems},\ }\href {https://doi.org/10.1103/PhysRevLett.112.140403} {\bibfield  {journal} {\bibinfo  {journal} {Phys. Rev. Lett.}\ }\textbf {\bibinfo {volume} {112}},\ \bibinfo {pages} {140403} (\bibinfo {year} {2014})}\BibitemShut {NoStop}%
\bibitem [{\citenamefont {Gogolin}\ and\ \citenamefont {Eisert}(2016)}]{Gogolin_2016}%
  \BibitemOpen
  \bibfield  {author} {\bibinfo {author} {\bibfnamefont {C.}~\bibnamefont {Gogolin}}\ and\ \bibinfo {author} {\bibfnamefont {J.}~\bibnamefont {Eisert}},\ }\bibfield  {title} {\bibinfo {title} {Equilibration, thermalisation, and the emergence of statistical mechanics in closed quantum systems},\ }\href {https://doi.org/10.1088/0034-4885/79/5/056001} {\bibfield  {journal} {\bibinfo  {journal} {Reports on Progress in Physics}\ }\textbf {\bibinfo {volume} {79}},\ \bibinfo {pages} {056001} (\bibinfo {year} {2016})}\BibitemShut {NoStop}%
\bibitem [{\citenamefont {Linden}\ \emph {et~al.}(2009)\citenamefont {Linden}, \citenamefont {Popescu}, \citenamefont {Short},\ and\ \citenamefont {Winter}}]{Linden2009PRE}%
  \BibitemOpen
  \bibfield  {author} {\bibinfo {author} {\bibfnamefont {N.}~\bibnamefont {Linden}}, \bibinfo {author} {\bibfnamefont {S.}~\bibnamefont {Popescu}}, \bibinfo {author} {\bibfnamefont {A.~J.}\ \bibnamefont {Short}},\ and\ \bibinfo {author} {\bibfnamefont {A.}~\bibnamefont {Winter}},\ }\bibfield  {title} {\bibinfo {title} {Quantum mechanical evolution towards thermal equilibrium},\ }\href {https://doi.org/10.1103/PhysRevE.79.061103} {\bibfield  {journal} {\bibinfo  {journal} {Phys. Rev. E}\ }\textbf {\bibinfo {volume} {79}},\ \bibinfo {pages} {061103} (\bibinfo {year} {2009})}\BibitemShut {NoStop}%
\bibitem [{\citenamefont {Engl}\ \emph {et~al.}(2015)\citenamefont {Engl}, \citenamefont {Urbina},\ and\ \citenamefont {Richter}}]{Engl2015PRE}%
  \BibitemOpen
  \bibfield  {author} {\bibinfo {author} {\bibfnamefont {T.}~\bibnamefont {Engl}}, \bibinfo {author} {\bibfnamefont {J.~D.}\ \bibnamefont {Urbina}},\ and\ \bibinfo {author} {\bibfnamefont {K.}~\bibnamefont {Richter}},\ }\bibfield  {title} {\bibinfo {title} {Periodic mean-field solutions and the spectra of discrete bosonic fields: Trace formula for bose-hubbard models},\ }\href {https://doi.org/10.1103/PhysRevE.92.062907} {\bibfield  {journal} {\bibinfo  {journal} {Phys. Rev. E}\ }\textbf {\bibinfo {volume} {92}},\ \bibinfo {pages} {062907} (\bibinfo {year} {2015})}\BibitemShut {NoStop}%
\bibitem [{\citenamefont {Popescu}\ \emph {et~al.}(2006)\citenamefont {Popescu}, \citenamefont {Short},\ and\ \citenamefont {Winter}}]{Popescu2006}%
  \BibitemOpen
  \bibfield  {author} {\bibinfo {author} {\bibfnamefont {S.}~\bibnamefont {Popescu}}, \bibinfo {author} {\bibfnamefont {A.~J.}\ \bibnamefont {Short}},\ and\ \bibinfo {author} {\bibfnamefont {A.}~\bibnamefont {Winter}},\ }\bibfield  {title} {\bibinfo {title} {Entanglement and the foundations of statistical mechanics},\ }\href {https://doi.org/10.1038/nphys444} {\bibfield  {journal} {\bibinfo  {journal} {Nature Physics}\ }\textbf {\bibinfo {volume} {2}},\ \bibinfo {pages} {754} (\bibinfo {year} {2006})}\BibitemShut {NoStop}%
\bibitem [{\citenamefont {Mehta}(2004)}]{mehta2004random}%
  \BibitemOpen
  \bibfield  {author} {\bibinfo {author} {\bibfnamefont {M.~L.}\ \bibnamefont {Mehta}},\ }\href@noop {} {\emph {\bibinfo {title} {Random matrices}}}\ (\bibinfo  {publisher} {Elsevier},\ \bibinfo {year} {2004})\BibitemShut {NoStop}%
\bibitem [{\citenamefont {Weidenm{\"u}ller}\ and\ \citenamefont {Mitchell}(2009)}]{weidenmuller2009random}%
  \BibitemOpen
  \bibfield  {author} {\bibinfo {author} {\bibfnamefont {H.}~\bibnamefont {Weidenm{\"u}ller}}\ and\ \bibinfo {author} {\bibfnamefont {G.}~\bibnamefont {Mitchell}},\ }\bibfield  {title} {\bibinfo {title} {Random matrices and chaos in nuclear physics: Nuclear structure},\ }\href@noop {} {\bibfield  {journal} {\bibinfo  {journal} {Reviews of Modern Physics}\ }\textbf {\bibinfo {volume} {81}},\ \bibinfo {pages} {539} (\bibinfo {year} {2009})}\BibitemShut {NoStop}%
\bibitem [{\citenamefont {Mitchell}\ \emph {et~al.}(2010)\citenamefont {Mitchell}, \citenamefont {Richter},\ and\ \citenamefont {Weidenm{\"u}ller}}]{mitchell2010random}%
  \BibitemOpen
  \bibfield  {author} {\bibinfo {author} {\bibfnamefont {G.~E.}\ \bibnamefont {Mitchell}}, \bibinfo {author} {\bibfnamefont {A.}~\bibnamefont {Richter}},\ and\ \bibinfo {author} {\bibfnamefont {H.~A.}\ \bibnamefont {Weidenm{\"u}ller}},\ }\bibfield  {title} {\bibinfo {title} {Random matrices and chaos in nuclear physics: Nuclear reactions},\ }\href@noop {} {\bibfield  {journal} {\bibinfo  {journal} {Reviews of Modern Physics}\ }\textbf {\bibinfo {volume} {82}},\ \bibinfo {pages} {2845} (\bibinfo {year} {2010})}\BibitemShut {NoStop}%
\bibitem [{\citenamefont {Forrester}(2010)}]{forrester2010log}%
  \BibitemOpen
  \bibfield  {author} {\bibinfo {author} {\bibfnamefont {P.~J.}\ \bibnamefont {Forrester}},\ }\href@noop {} {\emph {\bibinfo {title} {Log-gases and random matrices (LMS-34)}}}\ (\bibinfo  {publisher} {Princeton university press},\ \bibinfo {year} {2010})\BibitemShut {NoStop}%
\bibitem [{\citenamefont {Beenakker}(1997)}]{beenakker1997random}%
  \BibitemOpen
  \bibfield  {author} {\bibinfo {author} {\bibfnamefont {C.~W.}\ \bibnamefont {Beenakker}},\ }\bibfield  {title} {\bibinfo {title} {Random-matrix theory of quantum transport},\ }\href@noop {} {\bibfield  {journal} {\bibinfo  {journal} {Reviews of modern physics}\ }\textbf {\bibinfo {volume} {69}},\ \bibinfo {pages} {731} (\bibinfo {year} {1997})}\BibitemShut {NoStop}%
\bibitem [{\citenamefont {Haake}(1991{\natexlab{b}})}]{haake1991quantum}%
  \BibitemOpen
  \bibfield  {author} {\bibinfo {author} {\bibfnamefont {F.}~\bibnamefont {Haake}},\ }\bibfield  {title} {\bibinfo {title} {Quantum signatures of chaos},\ }in\ \href@noop {} {\emph {\bibinfo {booktitle} {Quantum Coherence in Mesoscopic Systems}}}\ (\bibinfo  {publisher} {Springer},\ \bibinfo {year} {1991})\ pp.\ \bibinfo {pages} {583--595}\BibitemShut {NoStop}%
\bibitem [{\citenamefont {Collins}\ and\ \citenamefont {Nechita}(2016)}]{collins2016random}%
  \BibitemOpen
  \bibfield  {author} {\bibinfo {author} {\bibfnamefont {B.}~\bibnamefont {Collins}}\ and\ \bibinfo {author} {\bibfnamefont {I.}~\bibnamefont {Nechita}},\ }\bibfield  {title} {\bibinfo {title} {Random matrix techniques in quantum information theory},\ }\href@noop {} {\bibfield  {journal} {\bibinfo  {journal} {Journal of Mathematical Physics}\ }\textbf {\bibinfo {volume} {57}} (\bibinfo {year} {2016})}\BibitemShut {NoStop}%
\bibitem [{\citenamefont {Porter}(1965)}]{porter1965statistical}%
  \BibitemOpen
  \bibfield  {author} {\bibinfo {author} {\bibfnamefont {C.~E.}\ \bibnamefont {Porter}},\ }\bibfield  {title} {\bibinfo {title} {Statistical theories of spectra: fluctuations, a collection of reprints and original papers, with an introductory review},\ }\href@noop {} {\bibfield  {journal} {\bibinfo  {journal} {(No Title)}\ } (\bibinfo {year} {1965})}\BibitemShut {NoStop}%
\bibitem [{\citenamefont {Dyson}(1962)}]{dyson1962threefold}%
  \BibitemOpen
  \bibfield  {author} {\bibinfo {author} {\bibfnamefont {F.~J.}\ \bibnamefont {Dyson}},\ }\bibfield  {title} {\bibinfo {title} {The threefold way. algebraic structure of symmetry groups and ensembles in quantum mechanics},\ }\href@noop {} {\bibfield  {journal} {\bibinfo  {journal} {Journal of Mathematical Physics}\ }\textbf {\bibinfo {volume} {3}},\ \bibinfo {pages} {1199} (\bibinfo {year} {1962})}\BibitemShut {NoStop}%
\bibitem [{\citenamefont {Berry}\ and\ \citenamefont {Tabor}(1976)}]{berry1976closed}%
  \BibitemOpen
  \bibfield  {author} {\bibinfo {author} {\bibfnamefont {M.~V.}\ \bibnamefont {Berry}}\ and\ \bibinfo {author} {\bibfnamefont {M.}~\bibnamefont {Tabor}},\ }\bibfield  {title} {\bibinfo {title} {Closed orbits and the regular bound spectrum},\ }\href@noop {} {\bibfield  {journal} {\bibinfo  {journal} {Proceedings of the Royal Society of London. A. Mathematical and Physical Sciences}\ }\textbf {\bibinfo {volume} {349}},\ \bibinfo {pages} {101} (\bibinfo {year} {1976})}\BibitemShut {NoStop}%
\bibitem [{\citenamefont {Berry}(1989)}]{berry1989quantum}%
  \BibitemOpen
  \bibfield  {author} {\bibinfo {author} {\bibfnamefont {M.}~\bibnamefont {Berry}},\ }\bibfield  {title} {\bibinfo {title} {Quantum chaology, not quantum chaos},\ }\href@noop {} {\bibfield  {journal} {\bibinfo  {journal} {Physica Scripta}\ }\textbf {\bibinfo {volume} {40}},\ \bibinfo {pages} {335} (\bibinfo {year} {1989})}\BibitemShut {NoStop}%
\bibitem [{Note1()}]{Note1}%
  \BibitemOpen
  \bibinfo {note} {Recall that classical chaos is characterized by the rate of separation of two nearby trajectories in the phase space}\BibitemShut {NoStop}%
\bibitem [{Note2()}]{Note2}%
  \BibitemOpen
  \bibinfo {note} {Note that an experimental implementation of the Loschmidt echo was done way before it found applications in quantum chaology \cite {hahn1950spin}.}\BibitemShut {Stop}%
\bibitem [{\citenamefont {Jalabert}\ and\ \citenamefont {Pastawski}(2001)}]{jalabert2001environment}%
  \BibitemOpen
  \bibfield  {author} {\bibinfo {author} {\bibfnamefont {R.~A.}\ \bibnamefont {Jalabert}}\ and\ \bibinfo {author} {\bibfnamefont {H.~M.}\ \bibnamefont {Pastawski}},\ }\bibfield  {title} {\bibinfo {title} {Environment-independent decoherence rate in classically chaotic systems},\ }\href@noop {} {\bibfield  {journal} {\bibinfo  {journal} {Physical review letters}\ }\textbf {\bibinfo {volume} {86}},\ \bibinfo {pages} {2490} (\bibinfo {year} {2001})}\BibitemShut {NoStop}%
\bibitem [{\citenamefont {Jacquod}\ \emph {et~al.}(2001)\citenamefont {Jacquod}, \citenamefont {Silvestrov},\ and\ \citenamefont {Beenakker}}]{jacquod2001golden}%
  \BibitemOpen
  \bibfield  {author} {\bibinfo {author} {\bibfnamefont {P.}~\bibnamefont {Jacquod}}, \bibinfo {author} {\bibfnamefont {P.~G.}\ \bibnamefont {Silvestrov}},\ and\ \bibinfo {author} {\bibfnamefont {C.~W.}\ \bibnamefont {Beenakker}},\ }\bibfield  {title} {\bibinfo {title} {Golden rule decay versus lyapunov decay of the quantum loschmidt echo},\ }\href@noop {} {\bibfield  {journal} {\bibinfo  {journal} {Physical Review E}\ }\textbf {\bibinfo {volume} {64}},\ \bibinfo {pages} {055203} (\bibinfo {year} {2001})}\BibitemShut {NoStop}%
\bibitem [{\citenamefont {Cerruti}\ and\ \citenamefont {Tomsovic}(2003)}]{cerruti2003uniform}%
  \BibitemOpen
  \bibfield  {author} {\bibinfo {author} {\bibfnamefont {N.~R.}\ \bibnamefont {Cerruti}}\ and\ \bibinfo {author} {\bibfnamefont {S.}~\bibnamefont {Tomsovic}},\ }\bibfield  {title} {\bibinfo {title} {A uniform approximation for the fidelity in chaotic systems},\ }\href@noop {} {\bibfield  {journal} {\bibinfo  {journal} {Journal of Physics A: Mathematical and General}\ }\textbf {\bibinfo {volume} {36}},\ \bibinfo {pages} {3451} (\bibinfo {year} {2003})}\BibitemShut {NoStop}%
\bibitem [{\citenamefont {Gorin}\ \emph {et~al.}(2006)\citenamefont {Gorin}, \citenamefont {Prosen}, \citenamefont {Seligman},\ and\ \citenamefont {{\v{Z}}nidari{\v{c}}}}]{gorin2006dynamics}%
  \BibitemOpen
  \bibfield  {author} {\bibinfo {author} {\bibfnamefont {T.}~\bibnamefont {Gorin}}, \bibinfo {author} {\bibfnamefont {T.}~\bibnamefont {Prosen}}, \bibinfo {author} {\bibfnamefont {T.~H.}\ \bibnamefont {Seligman}},\ and\ \bibinfo {author} {\bibfnamefont {M.}~\bibnamefont {{\v{Z}}nidari{\v{c}}}},\ }\bibfield  {title} {\bibinfo {title} {Dynamics of loschmidt echoes and fidelity decay},\ }\href@noop {} {\bibfield  {journal} {\bibinfo  {journal} {Physics Reports}\ }\textbf {\bibinfo {volume} {435}},\ \bibinfo {pages} {33} (\bibinfo {year} {2006})}\BibitemShut {NoStop}%
\bibitem [{\citenamefont {Goussev}\ \emph {et~al.}(2012)\citenamefont {Goussev}, \citenamefont {Jalabert}, \citenamefont {Pastawski},\ and\ \citenamefont {Wisniacki}}]{Goussev2012loschmidt}%
  \BibitemOpen
  \bibfield  {author} {\bibinfo {author} {\bibfnamefont {A.}~\bibnamefont {Goussev}}, \bibinfo {author} {\bibfnamefont {R.~A.}\ \bibnamefont {Jalabert}}, \bibinfo {author} {\bibfnamefont {H.~M.}\ \bibnamefont {Pastawski}},\ and\ \bibinfo {author} {\bibfnamefont {D.}~\bibnamefont {Wisniacki}},\ }\bibfield  {title} {\bibinfo {title} {Loschmidt echo},\ }\href@noop {} {\bibfield  {journal} {\bibinfo  {journal} {arXiv preprint arXiv:1206.6348}\ } (\bibinfo {year} {2012})}\BibitemShut {NoStop}%
\bibitem [{\citenamefont {Prosen}\ and\ \citenamefont {Znidaric}(2002)}]{prosen2002stability}%
  \BibitemOpen
  \bibfield  {author} {\bibinfo {author} {\bibfnamefont {T.}~\bibnamefont {Prosen}}\ and\ \bibinfo {author} {\bibfnamefont {M.}~\bibnamefont {Znidaric}},\ }\bibfield  {title} {\bibinfo {title} {Stability of quantum motion and correlation decay},\ }\href@noop {} {\bibfield  {journal} {\bibinfo  {journal} {Journal of Physics A: Mathematical and General}\ }\textbf {\bibinfo {volume} {35}},\ \bibinfo {pages} {1455} (\bibinfo {year} {2002})}\BibitemShut {NoStop}%
\bibitem [{\citenamefont {Atas}\ \emph {et~al.}(2013{\natexlab{b}})\citenamefont {Atas}, \citenamefont {Bogomolny}, \citenamefont {Giraud},\ and\ \citenamefont {Roux}}]{atas2013distribution}%
  \BibitemOpen
  \bibfield  {author} {\bibinfo {author} {\bibfnamefont {Y.~Y.}\ \bibnamefont {Atas}}, \bibinfo {author} {\bibfnamefont {E.}~\bibnamefont {Bogomolny}}, \bibinfo {author} {\bibfnamefont {O.}~\bibnamefont {Giraud}},\ and\ \bibinfo {author} {\bibfnamefont {G.}~\bibnamefont {Roux}},\ }\bibfield  {title} {\bibinfo {title} {Distribution of the ratio of consecutive level spacings in random matrix ensembles},\ }\href@noop {} {\bibfield  {journal} {\bibinfo  {journal} {Physical review letters}\ }\textbf {\bibinfo {volume} {110}},\ \bibinfo {pages} {084101} (\bibinfo {year} {2013}{\natexlab{b}})}\BibitemShut {NoStop}%
\bibitem [{\citenamefont {Tekur}\ and\ \citenamefont {Santhanam}(2020)}]{tekur2020symmetry}%
  \BibitemOpen
  \bibfield  {author} {\bibinfo {author} {\bibfnamefont {S.~H.}\ \bibnamefont {Tekur}}\ and\ \bibinfo {author} {\bibfnamefont {M.}~\bibnamefont {Santhanam}},\ }\bibfield  {title} {\bibinfo {title} {Symmetry deduction from spectral fluctuations in complex quantum systems},\ }\href@noop {} {\bibfield  {journal} {\bibinfo  {journal} {Physical Review Research}\ }\textbf {\bibinfo {volume} {2}},\ \bibinfo {pages} {032063} (\bibinfo {year} {2020})}\BibitemShut {NoStop}%
\bibitem [{\citenamefont {Prange}(1997)}]{prange1997spectral}%
  \BibitemOpen
  \bibfield  {author} {\bibinfo {author} {\bibfnamefont {R.}~\bibnamefont {Prange}},\ }\bibfield  {title} {\bibinfo {title} {The spectral form factor is not self-averaging},\ }\href@noop {} {\bibfield  {journal} {\bibinfo  {journal} {Physical review letters}\ }\textbf {\bibinfo {volume} {78}},\ \bibinfo {pages} {2280} (\bibinfo {year} {1997})}\BibitemShut {NoStop}%
\bibitem [{\citenamefont {Von~Keyserlingk}\ \emph {et~al.}(2018{\natexlab{a}})\citenamefont {Von~Keyserlingk}, \citenamefont {Rakovszky}, \citenamefont {Pollmann},\ and\ \citenamefont {Sondhi}}]{von2018operator}%
  \BibitemOpen
  \bibfield  {author} {\bibinfo {author} {\bibfnamefont {C.}~\bibnamefont {Von~Keyserlingk}}, \bibinfo {author} {\bibfnamefont {T.}~\bibnamefont {Rakovszky}}, \bibinfo {author} {\bibfnamefont {F.}~\bibnamefont {Pollmann}},\ and\ \bibinfo {author} {\bibfnamefont {S.~L.}\ \bibnamefont {Sondhi}},\ }\bibfield  {title} {\bibinfo {title} {Operator hydrodynamics, otocs, and entanglement growth in systems without conservation laws},\ }\href@noop {} {\bibfield  {journal} {\bibinfo  {journal} {Physical Review X}\ }\textbf {\bibinfo {volume} {8}},\ \bibinfo {pages} {021013} (\bibinfo {year} {2018}{\natexlab{a}})}\BibitemShut {NoStop}%
\bibitem [{\citenamefont {Deutsch}(1991)}]{deutsch1991quantum}%
  \BibitemOpen
  \bibfield  {author} {\bibinfo {author} {\bibfnamefont {J.~M.}\ \bibnamefont {Deutsch}},\ }\bibfield  {title} {\bibinfo {title} {Quantum statistical mechanics in a closed system},\ }\href {https://journals.aps.org/pra/abstract/10.1103/PhysRevA.43.2046} {\bibfield  {journal} {\bibinfo  {journal} {Physical review a}\ }\textbf {\bibinfo {volume} {43}},\ \bibinfo {pages} {2046} (\bibinfo {year} {1991})}\BibitemShut {NoStop}%
\bibitem [{\citenamefont {Srednicki}(1994)}]{srednicki1994chaos}%
  \BibitemOpen
  \bibfield  {author} {\bibinfo {author} {\bibfnamefont {M.}~\bibnamefont {Srednicki}},\ }\bibfield  {title} {\bibinfo {title} {Chaos and quantum thermalization},\ }\href {https://journals.aps.org/pre/abstract/10.1103/PhysRevE.50.888} {\bibfield  {journal} {\bibinfo  {journal} {Physical review e}\ }\textbf {\bibinfo {volume} {50}},\ \bibinfo {pages} {888} (\bibinfo {year} {1994})}\BibitemShut {NoStop}%
\bibitem [{\citenamefont {Tasaki}(1998)}]{tasaki1998quantum}%
  \BibitemOpen
  \bibfield  {author} {\bibinfo {author} {\bibfnamefont {H.}~\bibnamefont {Tasaki}},\ }\bibfield  {title} {\bibinfo {title} {From quantum dynamics to the canonical distribution: general picture and a rigorous example},\ }\href@noop {} {\bibfield  {journal} {\bibinfo  {journal} {Physical review letters}\ }\textbf {\bibinfo {volume} {80}},\ \bibinfo {pages} {1373} (\bibinfo {year} {1998})}\BibitemShut {NoStop}%
\bibitem [{\citenamefont {Rigol}\ \emph {et~al.}(2008{\natexlab{b}})\citenamefont {Rigol}, \citenamefont {Dunjko},\ and\ \citenamefont {Olshanii}}]{rigol2008thermalization}%
  \BibitemOpen
  \bibfield  {author} {\bibinfo {author} {\bibfnamefont {M.}~\bibnamefont {Rigol}}, \bibinfo {author} {\bibfnamefont {V.}~\bibnamefont {Dunjko}},\ and\ \bibinfo {author} {\bibfnamefont {M.}~\bibnamefont {Olshanii}},\ }\bibfield  {title} {\bibinfo {title} {Thermalization and its mechanism for generic isolated quantum systems},\ }\href@noop {} {\bibfield  {journal} {\bibinfo  {journal} {Nature}\ }\textbf {\bibinfo {volume} {452}},\ \bibinfo {pages} {854} (\bibinfo {year} {2008}{\natexlab{b}})}\BibitemShut {NoStop}%
\bibitem [{\citenamefont {Rigol}\ and\ \citenamefont {Santos}(2010)}]{rigol2010quantum}%
  \BibitemOpen
  \bibfield  {author} {\bibinfo {author} {\bibfnamefont {M.}~\bibnamefont {Rigol}}\ and\ \bibinfo {author} {\bibfnamefont {L.~F.}\ \bibnamefont {Santos}},\ }\bibfield  {title} {\bibinfo {title} {Quantum chaos and thermalization in gapped systems},\ }\href@noop {} {\bibfield  {journal} {\bibinfo  {journal} {Physical Review A}\ }\textbf {\bibinfo {volume} {82}},\ \bibinfo {pages} {011604} (\bibinfo {year} {2010})}\BibitemShut {NoStop}%
\bibitem [{\citenamefont {Torres-Herrera}\ and\ \citenamefont {Santos}(2013)}]{torres2013effects}%
  \BibitemOpen
  \bibfield  {author} {\bibinfo {author} {\bibfnamefont {E.}~\bibnamefont {Torres-Herrera}}\ and\ \bibinfo {author} {\bibfnamefont {L.~F.}\ \bibnamefont {Santos}},\ }\bibfield  {title} {\bibinfo {title} {Effects of the interplay between initial state and hamiltonian on the thermalization of isolated quantum many-body systems},\ }\href@noop {} {\bibfield  {journal} {\bibinfo  {journal} {Physical Review E}\ }\textbf {\bibinfo {volume} {88}},\ \bibinfo {pages} {042121} (\bibinfo {year} {2013})}\BibitemShut {NoStop}%
\bibitem [{\citenamefont {Hayden}\ and\ \citenamefont {Preskill}(2007)}]{hayden2007black}%
  \BibitemOpen
  \bibfield  {author} {\bibinfo {author} {\bibfnamefont {P.}~\bibnamefont {Hayden}}\ and\ \bibinfo {author} {\bibfnamefont {J.}~\bibnamefont {Preskill}},\ }\bibfield  {title} {\bibinfo {title} {Black holes as mirrors: quantum information in random subsystems},\ }\href {https://iopscience.iop.org/article/10.1088/1126-6708/2007/09/120} {\bibfield  {journal} {\bibinfo  {journal} {Journal of high energy physics}\ }\textbf {\bibinfo {volume} {2007}},\ \bibinfo {pages} {120} (\bibinfo {year} {2007})}\BibitemShut {NoStop}%
\bibitem [{\citenamefont {Sekino}\ and\ \citenamefont {Susskind}(2008)}]{sekino2008fast}%
  \BibitemOpen
  \bibfield  {author} {\bibinfo {author} {\bibfnamefont {Y.}~\bibnamefont {Sekino}}\ and\ \bibinfo {author} {\bibfnamefont {L.}~\bibnamefont {Susskind}},\ }\bibfield  {title} {\bibinfo {title} {Fast scramblers},\ }\href {https://iopscience.iop.org/article/10.1088/1126-6708/2008/10/065/meta} {\bibfield  {journal} {\bibinfo  {journal} {Journal of High Energy Physics}\ }\textbf {\bibinfo {volume} {2008}},\ \bibinfo {pages} {065} (\bibinfo {year} {2008})}\BibitemShut {NoStop}%
\bibitem [{\citenamefont {Hosur}\ \emph {et~al.}(2016{\natexlab{a}})\citenamefont {Hosur}, \citenamefont {Qi}, \citenamefont {Roberts},\ and\ \citenamefont {Yoshida}}]{hosur2016chaos}%
  \BibitemOpen
  \bibfield  {author} {\bibinfo {author} {\bibfnamefont {P.}~\bibnamefont {Hosur}}, \bibinfo {author} {\bibfnamefont {X.-L.}\ \bibnamefont {Qi}}, \bibinfo {author} {\bibfnamefont {D.~A.}\ \bibnamefont {Roberts}},\ and\ \bibinfo {author} {\bibfnamefont {B.}~\bibnamefont {Yoshida}},\ }\bibfield  {title} {\bibinfo {title} {Chaos in quantum channels},\ }\href@noop {} {\bibfield  {journal} {\bibinfo  {journal} {Journal of High Energy Physics}\ }\textbf {\bibinfo {volume} {2016}},\ \bibinfo {pages} {4} (\bibinfo {year} {2016}{\natexlab{a}})}\BibitemShut {NoStop}%
\bibitem [{\citenamefont {Shenker}\ and\ \citenamefont {Stanford}(2014{\natexlab{a}})}]{shenker2014black}%
  \BibitemOpen
  \bibfield  {author} {\bibinfo {author} {\bibfnamefont {S.~H.}\ \bibnamefont {Shenker}}\ and\ \bibinfo {author} {\bibfnamefont {D.}~\bibnamefont {Stanford}},\ }\bibfield  {title} {\bibinfo {title} {Black holes and the butterfly effect},\ }\href@noop {} {\bibfield  {journal} {\bibinfo  {journal} {Journal of High Energy Physics}\ }\textbf {\bibinfo {volume} {2014}},\ \bibinfo {pages} {67} (\bibinfo {year} {2014}{\natexlab{a}})}\BibitemShut {NoStop}%
\bibitem [{\citenamefont {McGinley}\ \emph {et~al.}(2022)\citenamefont {McGinley}, \citenamefont {Leontica}, \citenamefont {Garratt}, \citenamefont {Jovanovic},\ and\ \citenamefont {Simon}}]{mcginley2022quantifying}%
  \BibitemOpen
  \bibfield  {author} {\bibinfo {author} {\bibfnamefont {M.}~\bibnamefont {McGinley}}, \bibinfo {author} {\bibfnamefont {S.}~\bibnamefont {Leontica}}, \bibinfo {author} {\bibfnamefont {S.~J.}\ \bibnamefont {Garratt}}, \bibinfo {author} {\bibfnamefont {J.}~\bibnamefont {Jovanovic}},\ and\ \bibinfo {author} {\bibfnamefont {S.~H.}\ \bibnamefont {Simon}},\ }\bibfield  {title} {\bibinfo {title} {Quantifying information scrambling via classical shadow tomography on programmable quantum simulators},\ }\href@noop {} {\bibfield  {journal} {\bibinfo  {journal} {Physical Review A}\ }\textbf {\bibinfo {volume} {106}},\ \bibinfo {pages} {012441} (\bibinfo {year} {2022})}\BibitemShut {NoStop}%
\bibitem [{\citenamefont {Bhattacharyya}\ \emph {et~al.}(2022)\citenamefont {Bhattacharyya}, \citenamefont {Joshi},\ and\ \citenamefont {Sundar}}]{bhattacharyya2022quantum}%
  \BibitemOpen
  \bibfield  {author} {\bibinfo {author} {\bibfnamefont {A.}~\bibnamefont {Bhattacharyya}}, \bibinfo {author} {\bibfnamefont {L.~K.}\ \bibnamefont {Joshi}},\ and\ \bibinfo {author} {\bibfnamefont {B.}~\bibnamefont {Sundar}},\ }\bibfield  {title} {\bibinfo {title} {Quantum information scrambling: from holography to quantum simulators},\ }\href@noop {} {\bibfield  {journal} {\bibinfo  {journal} {The European Physical Journal C}\ }\textbf {\bibinfo {volume} {82}},\ \bibinfo {pages} {458} (\bibinfo {year} {2022})}\BibitemShut {NoStop}%
\bibitem [{\citenamefont {Xu}\ \emph {et~al.}(2020)\citenamefont {Xu}, \citenamefont {Scaffidi},\ and\ \citenamefont {Cao}}]{xu2020does}%
  \BibitemOpen
  \bibfield  {author} {\bibinfo {author} {\bibfnamefont {T.}~\bibnamefont {Xu}}, \bibinfo {author} {\bibfnamefont {T.}~\bibnamefont {Scaffidi}},\ and\ \bibinfo {author} {\bibfnamefont {X.}~\bibnamefont {Cao}},\ }\bibfield  {title} {\bibinfo {title} {Does scrambling equal chaos?},\ }\href {https://journals.aps.org/prl/abstract/10.1103/PhysRevLett.124.140602} {\bibfield  {journal} {\bibinfo  {journal} {Physical review letters}\ }\textbf {\bibinfo {volume} {124}},\ \bibinfo {pages} {140602} (\bibinfo {year} {2020})}\BibitemShut {NoStop}%
\bibitem [{\citenamefont {Rozenbaum}\ \emph {et~al.}(2020)\citenamefont {Rozenbaum}, \citenamefont {Bunimovich},\ and\ \citenamefont {Galitski}}]{rozenbaum2020early}%
  \BibitemOpen
  \bibfield  {author} {\bibinfo {author} {\bibfnamefont {E.~B.}\ \bibnamefont {Rozenbaum}}, \bibinfo {author} {\bibfnamefont {L.~A.}\ \bibnamefont {Bunimovich}},\ and\ \bibinfo {author} {\bibfnamefont {V.}~\bibnamefont {Galitski}},\ }\bibfield  {title} {\bibinfo {title} {Early-time exponential instabilities in nonchaotic quantum systems},\ }\href {https://journals.aps.org/prl/abstract/10.1103/PhysRevLett.125.014101} {\bibfield  {journal} {\bibinfo  {journal} {Physical Review Letters}\ }\textbf {\bibinfo {volume} {125}},\ \bibinfo {pages} {014101} (\bibinfo {year} {2020})}\BibitemShut {NoStop}%
\bibitem [{\citenamefont {Pilatowsky-Cameo}\ \emph {et~al.}(2020)\citenamefont {Pilatowsky-Cameo}, \citenamefont {Ch{\'a}vez-Carlos}, \citenamefont {Bastarrachea-Magnani}, \citenamefont {Str{\'a}nsk{\`y}}, \citenamefont {Lerma-Hern{\'a}ndez}, \citenamefont {Santos},\ and\ \citenamefont {Hirsch}}]{pilatowsky2020positive}%
  \BibitemOpen
  \bibfield  {author} {\bibinfo {author} {\bibfnamefont {S.}~\bibnamefont {Pilatowsky-Cameo}}, \bibinfo {author} {\bibfnamefont {J.}~\bibnamefont {Ch{\'a}vez-Carlos}}, \bibinfo {author} {\bibfnamefont {M.~A.}\ \bibnamefont {Bastarrachea-Magnani}}, \bibinfo {author} {\bibfnamefont {P.}~\bibnamefont {Str{\'a}nsk{\`y}}}, \bibinfo {author} {\bibfnamefont {S.}~\bibnamefont {Lerma-Hern{\'a}ndez}}, \bibinfo {author} {\bibfnamefont {L.~F.}\ \bibnamefont {Santos}},\ and\ \bibinfo {author} {\bibfnamefont {J.~G.}\ \bibnamefont {Hirsch}},\ }\bibfield  {title} {\bibinfo {title} {Positive quantum lyapunov exponents in experimental systems with a regular classical limit},\ }\href {https://journals.aps.org/pre/abstract/10.1103/PhysRevE.101.010202} {\bibfield  {journal} {\bibinfo  {journal} {Physical Review E}\ }\textbf {\bibinfo {volume} {101}},\ \bibinfo {pages} {010202} (\bibinfo {year} {2020})}\BibitemShut {NoStop}%
\bibitem [{\citenamefont {Nahum}\ \emph {et~al.}(2018{\natexlab{a}})\citenamefont {Nahum}, \citenamefont {Ruhman},\ and\ \citenamefont {Huse}}]{nahum2018dynamics}%
  \BibitemOpen
  \bibfield  {author} {\bibinfo {author} {\bibfnamefont {A.}~\bibnamefont {Nahum}}, \bibinfo {author} {\bibfnamefont {J.}~\bibnamefont {Ruhman}},\ and\ \bibinfo {author} {\bibfnamefont {D.~A.}\ \bibnamefont {Huse}},\ }\bibfield  {title} {\bibinfo {title} {Dynamics of entanglement and transport in one-dimensional systems with quenched randomness},\ }\href@noop {} {\bibfield  {journal} {\bibinfo  {journal} {Physical Review B}\ }\textbf {\bibinfo {volume} {98}},\ \bibinfo {pages} {035118} (\bibinfo {year} {2018}{\natexlab{a}})}\BibitemShut {NoStop}%
\bibitem [{\citenamefont {Nahum}\ \emph {et~al.}(2018{\natexlab{b}})\citenamefont {Nahum}, \citenamefont {Vijay},\ and\ \citenamefont {Haah}}]{nahum2018operator}%
  \BibitemOpen
  \bibfield  {author} {\bibinfo {author} {\bibfnamefont {A.}~\bibnamefont {Nahum}}, \bibinfo {author} {\bibfnamefont {S.}~\bibnamefont {Vijay}},\ and\ \bibinfo {author} {\bibfnamefont {J.}~\bibnamefont {Haah}},\ }\bibfield  {title} {\bibinfo {title} {Operator spreading in random unitary circuits},\ }\href@noop {} {\bibfield  {journal} {\bibinfo  {journal} {Physical Review X}\ }\textbf {\bibinfo {volume} {8}},\ \bibinfo {pages} {021014} (\bibinfo {year} {2018}{\natexlab{b}})}\BibitemShut {NoStop}%
\bibitem [{\citenamefont {Khemani}\ \emph {et~al.}(2018{\natexlab{a}})\citenamefont {Khemani}, \citenamefont {Vishwanath},\ and\ \citenamefont {Huse}}]{khemani2018operator}%
  \BibitemOpen
  \bibfield  {author} {\bibinfo {author} {\bibfnamefont {V.}~\bibnamefont {Khemani}}, \bibinfo {author} {\bibfnamefont {A.}~\bibnamefont {Vishwanath}},\ and\ \bibinfo {author} {\bibfnamefont {D.~A.}\ \bibnamefont {Huse}},\ }\bibfield  {title} {\bibinfo {title} {Operator spreading and the emergence of dissipative hydrodynamics under unitary evolution with conservation laws},\ }\href@noop {} {\bibfield  {journal} {\bibinfo  {journal} {Physical Review X}\ }\textbf {\bibinfo {volume} {8}},\ \bibinfo {pages} {031057} (\bibinfo {year} {2018}{\natexlab{a}})}\BibitemShut {NoStop}%
\bibitem [{\citenamefont {Rakovszky}\ \emph {et~al.}(2018{\natexlab{a}})\citenamefont {Rakovszky}, \citenamefont {Pollmann},\ and\ \citenamefont {Von~Keyserlingk}}]{rakovszky2018diffusive}%
  \BibitemOpen
  \bibfield  {author} {\bibinfo {author} {\bibfnamefont {T.}~\bibnamefont {Rakovszky}}, \bibinfo {author} {\bibfnamefont {F.}~\bibnamefont {Pollmann}},\ and\ \bibinfo {author} {\bibfnamefont {C.}~\bibnamefont {Von~Keyserlingk}},\ }\bibfield  {title} {\bibinfo {title} {Diffusive hydrodynamics of out-of-time-ordered correlators with charge conservation},\ }\href@noop {} {\bibfield  {journal} {\bibinfo  {journal} {Physical Review X}\ }\textbf {\bibinfo {volume} {8}},\ \bibinfo {pages} {031058} (\bibinfo {year} {2018}{\natexlab{a}})}\BibitemShut {NoStop}%
\bibitem [{\citenamefont {Roberts}\ and\ \citenamefont {Stanford}(2015)}]{roberts2015diagnosing}%
  \BibitemOpen
  \bibfield  {author} {\bibinfo {author} {\bibfnamefont {D.~A.}\ \bibnamefont {Roberts}}\ and\ \bibinfo {author} {\bibfnamefont {D.}~\bibnamefont {Stanford}},\ }\bibfield  {title} {\bibinfo {title} {Diagnosing chaos using four-point functions in two-dimensional conformal field theory},\ }\href@noop {} {\bibfield  {journal} {\bibinfo  {journal} {Physical review letters}\ }\textbf {\bibinfo {volume} {115}},\ \bibinfo {pages} {131603} (\bibinfo {year} {2015})}\BibitemShut {NoStop}%
\bibitem [{\citenamefont {Stanford}(2016)}]{stanford2016many}%
  \BibitemOpen
  \bibfield  {author} {\bibinfo {author} {\bibfnamefont {D.}~\bibnamefont {Stanford}},\ }\bibfield  {title} {\bibinfo {title} {Many-body chaos at weak coupling},\ }\href@noop {} {\bibfield  {journal} {\bibinfo  {journal} {Journal of High Energy Physics}\ }\textbf {\bibinfo {volume} {2016}},\ \bibinfo {pages} {1} (\bibinfo {year} {2016})}\BibitemShut {NoStop}%
\bibitem [{\citenamefont {Chowdhury}\ and\ \citenamefont {Swingle}(2017)}]{chowdhury2017onset}%
  \BibitemOpen
  \bibfield  {author} {\bibinfo {author} {\bibfnamefont {D.}~\bibnamefont {Chowdhury}}\ and\ \bibinfo {author} {\bibfnamefont {B.}~\bibnamefont {Swingle}},\ }\bibfield  {title} {\bibinfo {title} {Onset of many-body chaos in the o (n) model},\ }\href@noop {} {\bibfield  {journal} {\bibinfo  {journal} {Physical Review D}\ }\textbf {\bibinfo {volume} {96}},\ \bibinfo {pages} {065005} (\bibinfo {year} {2017})}\BibitemShut {NoStop}%
\bibitem [{\citenamefont {Patel}\ \emph {et~al.}(2017)\citenamefont {Patel}, \citenamefont {Chowdhury}, \citenamefont {Sachdev},\ and\ \citenamefont {Swingle}}]{patel2017quantum}%
  \BibitemOpen
  \bibfield  {author} {\bibinfo {author} {\bibfnamefont {A.~A.}\ \bibnamefont {Patel}}, \bibinfo {author} {\bibfnamefont {D.}~\bibnamefont {Chowdhury}}, \bibinfo {author} {\bibfnamefont {S.}~\bibnamefont {Sachdev}},\ and\ \bibinfo {author} {\bibfnamefont {B.}~\bibnamefont {Swingle}},\ }\bibfield  {title} {\bibinfo {title} {Quantum butterfly effect in weakly interacting diffusive metals},\ }\href@noop {} {\bibfield  {journal} {\bibinfo  {journal} {Physical Review X}\ }\textbf {\bibinfo {volume} {7}},\ \bibinfo {pages} {031047} (\bibinfo {year} {2017})}\BibitemShut {NoStop}%
\bibitem [{\citenamefont {Luitz}\ and\ \citenamefont {Lev}(2017)}]{luitz2017information}%
  \BibitemOpen
  \bibfield  {author} {\bibinfo {author} {\bibfnamefont {D.~J.}\ \bibnamefont {Luitz}}\ and\ \bibinfo {author} {\bibfnamefont {Y.~B.}\ \bibnamefont {Lev}},\ }\bibfield  {title} {\bibinfo {title} {Information propagation in isolated quantum systems},\ }\href@noop {} {\bibfield  {journal} {\bibinfo  {journal} {Physical Review B}\ }\textbf {\bibinfo {volume} {96}},\ \bibinfo {pages} {020406} (\bibinfo {year} {2017})}\BibitemShut {NoStop}%
\bibitem [{\citenamefont {Heyl}\ \emph {et~al.}(2018)\citenamefont {Heyl}, \citenamefont {Pollmann},\ and\ \citenamefont {D{\'o}ra}}]{heyl2018detecting}%
  \BibitemOpen
  \bibfield  {author} {\bibinfo {author} {\bibfnamefont {M.}~\bibnamefont {Heyl}}, \bibinfo {author} {\bibfnamefont {F.}~\bibnamefont {Pollmann}},\ and\ \bibinfo {author} {\bibfnamefont {B.}~\bibnamefont {D{\'o}ra}},\ }\bibfield  {title} {\bibinfo {title} {Detecting equilibrium and dynamical quantum phase transitions in ising chains via out-of-time-ordered correlators},\ }\href@noop {} {\bibfield  {journal} {\bibinfo  {journal} {Physical review letters}\ }\textbf {\bibinfo {volume} {121}},\ \bibinfo {pages} {016801} (\bibinfo {year} {2018})}\BibitemShut {NoStop}%
\bibitem [{\citenamefont {Lin}\ and\ \citenamefont {Motrunich}(2018)}]{lin2018out}%
  \BibitemOpen
  \bibfield  {author} {\bibinfo {author} {\bibfnamefont {C.-J.}\ \bibnamefont {Lin}}\ and\ \bibinfo {author} {\bibfnamefont {O.~I.}\ \bibnamefont {Motrunich}},\ }\bibfield  {title} {\bibinfo {title} {Out-of-time-ordered correlators in a quantum ising chain},\ }\href@noop {} {\bibfield  {journal} {\bibinfo  {journal} {Physical Review B}\ }\textbf {\bibinfo {volume} {97}},\ \bibinfo {pages} {144304} (\bibinfo {year} {2018})}\BibitemShut {NoStop}%
\bibitem [{\citenamefont {Geller}\ \emph {et~al.}(2022)\citenamefont {Geller}, \citenamefont {Arrasmith}, \citenamefont {Holmes}, \citenamefont {Yan}, \citenamefont {Coles},\ and\ \citenamefont {Sornborger}}]{geller2022quantum}%
  \BibitemOpen
  \bibfield  {author} {\bibinfo {author} {\bibfnamefont {M.~R.}\ \bibnamefont {Geller}}, \bibinfo {author} {\bibfnamefont {A.}~\bibnamefont {Arrasmith}}, \bibinfo {author} {\bibfnamefont {Z.}~\bibnamefont {Holmes}}, \bibinfo {author} {\bibfnamefont {B.}~\bibnamefont {Yan}}, \bibinfo {author} {\bibfnamefont {P.~J.}\ \bibnamefont {Coles}},\ and\ \bibinfo {author} {\bibfnamefont {A.}~\bibnamefont {Sornborger}},\ }\bibfield  {title} {\bibinfo {title} {Quantum simulation of operator spreading in the chaotic ising model},\ }\href@noop {} {\bibfield  {journal} {\bibinfo  {journal} {Physical Review E}\ }\textbf {\bibinfo {volume} {105}},\ \bibinfo {pages} {035302} (\bibinfo {year} {2022})}\BibitemShut {NoStop}%
\bibitem [{\citenamefont {Moudgalya}\ \emph {et~al.}(2019)\citenamefont {Moudgalya}, \citenamefont {Devakul}, \citenamefont {Von~Keyserlingk},\ and\ \citenamefont {Sondhi}}]{moudgalya2019operator}%
  \BibitemOpen
  \bibfield  {author} {\bibinfo {author} {\bibfnamefont {S.}~\bibnamefont {Moudgalya}}, \bibinfo {author} {\bibfnamefont {T.}~\bibnamefont {Devakul}}, \bibinfo {author} {\bibfnamefont {C.}~\bibnamefont {Von~Keyserlingk}},\ and\ \bibinfo {author} {\bibfnamefont {S.}~\bibnamefont {Sondhi}},\ }\bibfield  {title} {\bibinfo {title} {Operator spreading in quantum maps},\ }\href {https://journals.aps.org/prb/abstract/10.1103/PhysRevB.99.094312} {\bibfield  {journal} {\bibinfo  {journal} {Physical Review B}\ }\textbf {\bibinfo {volume} {99}},\ \bibinfo {pages} {094312} (\bibinfo {year} {2019})}\BibitemShut {NoStop}%
\bibitem [{\citenamefont {Omanakuttan}\ \emph {et~al.}(2023)\citenamefont {Omanakuttan}, \citenamefont {Chinni}, \citenamefont {Blocher},\ and\ \citenamefont {Poggi}}]{omanakuttan2023scrambling}%
  \BibitemOpen
  \bibfield  {author} {\bibinfo {author} {\bibfnamefont {S.}~\bibnamefont {Omanakuttan}}, \bibinfo {author} {\bibfnamefont {K.}~\bibnamefont {Chinni}}, \bibinfo {author} {\bibfnamefont {P.~D.}\ \bibnamefont {Blocher}},\ and\ \bibinfo {author} {\bibfnamefont {P.~M.}\ \bibnamefont {Poggi}},\ }\bibfield  {title} {\bibinfo {title} {Scrambling and quantum chaos indicators from long-time properties of operator distributions},\ }\href@noop {} {\bibfield  {journal} {\bibinfo  {journal} {Physical Review A}\ }\textbf {\bibinfo {volume} {107}},\ \bibinfo {pages} {032418} (\bibinfo {year} {2023})}\BibitemShut {NoStop}%
\bibitem [{\citenamefont {Hahn}(1950)}]{hahn1950spin}%
  \BibitemOpen
  \bibfield  {author} {\bibinfo {author} {\bibfnamefont {E.~L.}\ \bibnamefont {Hahn}},\ }\bibfield  {title} {\bibinfo {title} {Spin echoes},\ }\href@noop {} {\bibfield  {journal} {\bibinfo  {journal} {Physical review}\ }\textbf {\bibinfo {volume} {80}},\ \bibinfo {pages} {580} (\bibinfo {year} {1950})}\BibitemShut {NoStop}%
\bibitem [{\citenamefont {Swingle}(2018)}]{Swingle-2018}%
  \BibitemOpen
  \bibfield  {author} {\bibinfo {author} {\bibfnamefont {B.}~\bibnamefont {Swingle}},\ }\bibfield  {title} {\bibinfo {title} {Unscrambling the physics of out-of-time-order correlators},\ }\href {https://doi.org/10.1038/s41567-018-0295-5} {\bibfield  {journal} {\bibinfo  {journal} {Nature Physics}\ }\textbf {\bibinfo {volume} {14}},\ \bibinfo {pages} {988} (\bibinfo {year} {2018})}\BibitemShut {NoStop}%
\bibitem [{\citenamefont {Yan}\ \emph {et~al.}(2020)\citenamefont {Yan}, \citenamefont {Cincio},\ and\ \citenamefont {Zurek}}]{yan2020information}%
  \BibitemOpen
  \bibfield  {author} {\bibinfo {author} {\bibfnamefont {B.}~\bibnamefont {Yan}}, \bibinfo {author} {\bibfnamefont {L.}~\bibnamefont {Cincio}},\ and\ \bibinfo {author} {\bibfnamefont {W.~H.}\ \bibnamefont {Zurek}},\ }\bibfield  {title} {\bibinfo {title} {Information scrambling and loschmidt echo},\ }\href {https://journals.aps.org/prl/abstract/10.1103/PhysRevLett.124.160603} {\bibfield  {journal} {\bibinfo  {journal} {Physical review letters}\ }\textbf {\bibinfo {volume} {124}},\ \bibinfo {pages} {160603} (\bibinfo {year} {2020})}\BibitemShut {NoStop}%
\bibitem [{\citenamefont {Stephen}(1966)}]{stephen1966brush}%
  \BibitemOpen
  \bibfield  {author} {\bibinfo {author} {\bibfnamefont {G.}~\bibnamefont {Stephen}},\ }\href@noop {} {\bibinfo {title} {Brush. kinetic theory, vol. 2, irreversible processes}} (\bibinfo {year} {1966})\BibitemShut {NoStop}%
\bibitem [{\citenamefont {Zurek}(2001)}]{zurek2001sub}%
  \BibitemOpen
  \bibfield  {author} {\bibinfo {author} {\bibfnamefont {W.~H.}\ \bibnamefont {Zurek}},\ }\bibfield  {title} {\bibinfo {title} {Sub-planck structure in phase space and its relevance for quantum decoherence},\ }\href@noop {} {\bibfield  {journal} {\bibinfo  {journal} {Nature}\ }\textbf {\bibinfo {volume} {412}},\ \bibinfo {pages} {712} (\bibinfo {year} {2001})}\BibitemShut {NoStop}%
\bibitem [{\citenamefont {Cucchietti}\ \emph {et~al.}(2003)\citenamefont {Cucchietti}, \citenamefont {Dalvit}, \citenamefont {Paz},\ and\ \citenamefont {Zurek}}]{cucchietti2003decoherence}%
  \BibitemOpen
  \bibfield  {author} {\bibinfo {author} {\bibfnamefont {F.~M.}\ \bibnamefont {Cucchietti}}, \bibinfo {author} {\bibfnamefont {D.~A.}\ \bibnamefont {Dalvit}}, \bibinfo {author} {\bibfnamefont {J.~P.}\ \bibnamefont {Paz}},\ and\ \bibinfo {author} {\bibfnamefont {W.~H.}\ \bibnamefont {Zurek}},\ }\bibfield  {title} {\bibinfo {title} {Decoherence and the loschmidt echo},\ }\href@noop {} {\bibfield  {journal} {\bibinfo  {journal} {Physical review letters}\ }\textbf {\bibinfo {volume} {91}},\ \bibinfo {pages} {210403} (\bibinfo {year} {2003})}\BibitemShut {NoStop}%
\bibitem [{\citenamefont {Cucchietti}\ \emph {et~al.}(2004)\citenamefont {Cucchietti}, \citenamefont {Pastawski},\ and\ \citenamefont {Jalabert}}]{cucchietti2004universality}%
  \BibitemOpen
  \bibfield  {author} {\bibinfo {author} {\bibfnamefont {F.~M.}\ \bibnamefont {Cucchietti}}, \bibinfo {author} {\bibfnamefont {H.~M.}\ \bibnamefont {Pastawski}},\ and\ \bibinfo {author} {\bibfnamefont {R.~A.}\ \bibnamefont {Jalabert}},\ }\bibfield  {title} {\bibinfo {title} {Universality of the lyapunov regime for the loschmidt echo},\ }\href@noop {} {\bibfield  {journal} {\bibinfo  {journal} {Physical Review B}\ }\textbf {\bibinfo {volume} {70}},\ \bibinfo {pages} {035311} (\bibinfo {year} {2004})}\BibitemShut {NoStop}%
\bibitem [{\citenamefont {Larkin}\ and\ \citenamefont {Ovchinnikov}(1969)}]{larkin}%
  \BibitemOpen
  \bibfield  {author} {\bibinfo {author} {\bibfnamefont {A.}~\bibnamefont {Larkin}}\ and\ \bibinfo {author} {\bibfnamefont {Y.~N.}\ \bibnamefont {Ovchinnikov}},\ }\bibfield  {title} {\bibinfo {title} {Quasiclassical method in the theory of superconductivity},\ }\href@noop {} {\bibfield  {journal} {\bibinfo  {journal} {Sov Phys JETP}\ }\textbf {\bibinfo {volume} {28}},\ \bibinfo {pages} {1200} (\bibinfo {year} {1969})}\BibitemShut {NoStop}%
\bibitem [{\citenamefont {Nahum}\ \emph {et~al.}(2018{\natexlab{c}})\citenamefont {Nahum}, \citenamefont {Vijay},\ and\ \citenamefont {Haah}}]{ope2}%
  \BibitemOpen
  \bibfield  {author} {\bibinfo {author} {\bibfnamefont {A.}~\bibnamefont {Nahum}}, \bibinfo {author} {\bibfnamefont {S.}~\bibnamefont {Vijay}},\ and\ \bibinfo {author} {\bibfnamefont {J.}~\bibnamefont {Haah}},\ }\bibfield  {title} {\bibinfo {title} {Operator spreading in random unitary circuits},\ }\href@noop {} {\bibfield  {journal} {\bibinfo  {journal} {Physical Review X}\ }\textbf {\bibinfo {volume} {8}},\ \bibinfo {pages} {021014} (\bibinfo {year} {2018}{\natexlab{c}})}\BibitemShut {NoStop}%
\bibitem [{\citenamefont {Von~Keyserlingk}\ \emph {et~al.}(2018{\natexlab{b}})\citenamefont {Von~Keyserlingk}, \citenamefont {Rakovszky}, \citenamefont {Pollmann},\ and\ \citenamefont {Sondhi}}]{ope1}%
  \BibitemOpen
  \bibfield  {author} {\bibinfo {author} {\bibfnamefont {C.}~\bibnamefont {Von~Keyserlingk}}, \bibinfo {author} {\bibfnamefont {T.}~\bibnamefont {Rakovszky}}, \bibinfo {author} {\bibfnamefont {F.}~\bibnamefont {Pollmann}},\ and\ \bibinfo {author} {\bibfnamefont {S.~L.}\ \bibnamefont {Sondhi}},\ }\bibfield  {title} {\bibinfo {title} {Operator hydrodynamics, otocs, and entanglement growth in systems without conservation laws},\ }\href@noop {} {\bibfield  {journal} {\bibinfo  {journal} {Physical Review X}\ }\textbf {\bibinfo {volume} {8}},\ \bibinfo {pages} {021013} (\bibinfo {year} {2018}{\natexlab{b}})}\BibitemShut {NoStop}%
\bibitem [{\citenamefont {Khemani}\ \emph {et~al.}(2018{\natexlab{b}})\citenamefont {Khemani}, \citenamefont {Vishwanath},\ and\ \citenamefont {Huse}}]{ope4}%
  \BibitemOpen
  \bibfield  {author} {\bibinfo {author} {\bibfnamefont {V.}~\bibnamefont {Khemani}}, \bibinfo {author} {\bibfnamefont {A.}~\bibnamefont {Vishwanath}},\ and\ \bibinfo {author} {\bibfnamefont {D.~A.}\ \bibnamefont {Huse}},\ }\bibfield  {title} {\bibinfo {title} {Operator spreading and the emergence of dissipative hydrodynamics under unitary evolution with conservation laws},\ }\href@noop {} {\bibfield  {journal} {\bibinfo  {journal} {Physical Review X}\ }\textbf {\bibinfo {volume} {8}},\ \bibinfo {pages} {031057} (\bibinfo {year} {2018}{\natexlab{b}})}\BibitemShut {NoStop}%
\bibitem [{\citenamefont {Rakovszky}\ \emph {et~al.}(2018{\natexlab{b}})\citenamefont {Rakovszky}, \citenamefont {Pollmann},\ and\ \citenamefont {von Keyserlingk}}]{ope5}%
  \BibitemOpen
  \bibfield  {author} {\bibinfo {author} {\bibfnamefont {T.}~\bibnamefont {Rakovszky}}, \bibinfo {author} {\bibfnamefont {F.}~\bibnamefont {Pollmann}},\ and\ \bibinfo {author} {\bibfnamefont {C.}~\bibnamefont {von Keyserlingk}},\ }\bibfield  {title} {\bibinfo {title} {Diffusive hydrodynamics of out-of-time-ordered correlators with charge conservation},\ }\href {https://journals.aps.org/prx/abstract/10.1103/PhysRevX.8.031058} {\bibfield  {journal} {\bibinfo  {journal} {Physical Review X}\ }\textbf {\bibinfo {volume} {8}},\ \bibinfo {pages} {031058} (\bibinfo {year} {2018}{\natexlab{b}})}\BibitemShut {NoStop}%
\bibitem [{\citenamefont {Shukla}\ \emph {et~al.}(2022)\citenamefont {Shukla}, \citenamefont {Lakshminarayan},\ and\ \citenamefont {Mishra}}]{shukla2022out}%
  \BibitemOpen
  \bibfield  {author} {\bibinfo {author} {\bibfnamefont {R.~K.}\ \bibnamefont {Shukla}}, \bibinfo {author} {\bibfnamefont {A.}~\bibnamefont {Lakshminarayan}},\ and\ \bibinfo {author} {\bibfnamefont {S.~K.}\ \bibnamefont {Mishra}},\ }\bibfield  {title} {\bibinfo {title} {Out-of-time-order correlators of nonlocal block-spin and random observables in integrable and nonintegrable spin chains},\ }\href@noop {} {\bibfield  {journal} {\bibinfo  {journal} {Physical Review B}\ }\textbf {\bibinfo {volume} {105}},\ \bibinfo {pages} {224307} (\bibinfo {year} {2022})}\BibitemShut {NoStop}%
\bibitem [{\citenamefont {Maldacena}\ \emph {et~al.}(2016)\citenamefont {Maldacena}, \citenamefont {Shenker},\ and\ \citenamefont {Stanford}}]{chaos1}%
  \BibitemOpen
  \bibfield  {author} {\bibinfo {author} {\bibfnamefont {J.}~\bibnamefont {Maldacena}}, \bibinfo {author} {\bibfnamefont {S.~H.}\ \bibnamefont {Shenker}},\ and\ \bibinfo {author} {\bibfnamefont {D.}~\bibnamefont {Stanford}},\ }\bibfield  {title} {\bibinfo {title} {A bound on chaos},\ }\href@noop {} {\bibfield  {journal} {\bibinfo  {journal} {Journal of High Energy Physics}\ }\textbf {\bibinfo {volume} {2016}},\ \bibinfo {pages} {106} (\bibinfo {year} {2016})}\BibitemShut {NoStop}%
\bibitem [{\citenamefont {Hosur}\ \emph {et~al.}(2016{\natexlab{b}})\citenamefont {Hosur}, \citenamefont {Qi}, \citenamefont {Roberts},\ and\ \citenamefont {Yoshida}}]{pawan}%
  \BibitemOpen
  \bibfield  {author} {\bibinfo {author} {\bibfnamefont {P.}~\bibnamefont {Hosur}}, \bibinfo {author} {\bibfnamefont {X.-L.}\ \bibnamefont {Qi}}, \bibinfo {author} {\bibfnamefont {D.~A.}\ \bibnamefont {Roberts}},\ and\ \bibinfo {author} {\bibfnamefont {B.}~\bibnamefont {Yoshida}},\ }\bibfield  {title} {\bibinfo {title} {Chaos in quantum channels},\ }\href@noop {} {\bibfield  {journal} {\bibinfo  {journal} {Journal of High Energy Physics}\ }\textbf {\bibinfo {volume} {2016}},\ \bibinfo {pages} {4} (\bibinfo {year} {2016}{\natexlab{b}})}\BibitemShut {NoStop}%
\bibitem [{\citenamefont {Seshadri}\ \emph {et~al.}(2018)\citenamefont {Seshadri}, \citenamefont {Madhok},\ and\ \citenamefont {Lakshminarayan}}]{seshadri2018tripartite}%
  \BibitemOpen
  \bibfield  {author} {\bibinfo {author} {\bibfnamefont {A.}~\bibnamefont {Seshadri}}, \bibinfo {author} {\bibfnamefont {V.}~\bibnamefont {Madhok}},\ and\ \bibinfo {author} {\bibfnamefont {A.}~\bibnamefont {Lakshminarayan}},\ }\bibfield  {title} {\bibinfo {title} {Tripartite mutual information, entanglement, and scrambling in permutation symmetric systems with an application to quantum chaos},\ }\href {https://journals.aps.org/pre/abstract/10.1103/PhysRevE.98.052205} {\bibfield  {journal} {\bibinfo  {journal} {Physical Review E}\ }\textbf {\bibinfo {volume} {98}},\ \bibinfo {pages} {052205} (\bibinfo {year} {2018})}\BibitemShut {NoStop}%
\bibitem [{\citenamefont {Lakshminarayan}(2019)}]{lakshminarayan2019out}%
  \BibitemOpen
  \bibfield  {author} {\bibinfo {author} {\bibfnamefont {A.}~\bibnamefont {Lakshminarayan}},\ }\bibfield  {title} {\bibinfo {title} {Out-of-time-ordered correlator in the quantum bakers map and truncated unitary matrices},\ }\href {https://journals.aps.org/pre/abstract/10.1103/PhysRevE.99.012201} {\bibfield  {journal} {\bibinfo  {journal} {Physical Review E}\ }\textbf {\bibinfo {volume} {99}},\ \bibinfo {pages} {012201} (\bibinfo {year} {2019})}\BibitemShut {NoStop}%
\bibitem [{\citenamefont {Shenker}\ and\ \citenamefont {Stanford}(2014{\natexlab{b}})}]{shenker2}%
  \BibitemOpen
  \bibfield  {author} {\bibinfo {author} {\bibfnamefont {S.~H.}\ \bibnamefont {Shenker}}\ and\ \bibinfo {author} {\bibfnamefont {D.}~\bibnamefont {Stanford}},\ }\bibfield  {title} {\bibinfo {title} {Black holes and the butterfly effect},\ }\href@noop {} {\bibfield  {journal} {\bibinfo  {journal} {Journal of High Energy Physics}\ }\textbf {\bibinfo {volume} {2014}},\ \bibinfo {pages} {67} (\bibinfo {year} {2014}{\natexlab{b}})}\BibitemShut {NoStop}%
\bibitem [{\citenamefont {Omanakuttan}\ and\ \citenamefont {Lakshminarayan}(2019)}]{omanakuttan2019out}%
  \BibitemOpen
  \bibfield  {author} {\bibinfo {author} {\bibfnamefont {S.}~\bibnamefont {Omanakuttan}}\ and\ \bibinfo {author} {\bibfnamefont {A.}~\bibnamefont {Lakshminarayan}},\ }\bibfield  {title} {\bibinfo {title} {Out-of-time-ordered correlators and quantum walks},\ }\href {https://journals.aps.org/pre/abstract/10.1103/PhysRevE.99.062128} {\bibfield  {journal} {\bibinfo  {journal} {Physical Review E}\ }\textbf {\bibinfo {volume} {99}},\ \bibinfo {pages} {062128} (\bibinfo {year} {2019})}\BibitemShut {NoStop}%
\bibitem [{\citenamefont {Aleiner}\ \emph {et~al.}(2016)\citenamefont {Aleiner}, \citenamefont {Faoro},\ and\ \citenamefont {Ioffe}}]{manybody2}%
  \BibitemOpen
  \bibfield  {author} {\bibinfo {author} {\bibfnamefont {I.~L.}\ \bibnamefont {Aleiner}}, \bibinfo {author} {\bibfnamefont {L.}~\bibnamefont {Faoro}},\ and\ \bibinfo {author} {\bibfnamefont {L.~B.}\ \bibnamefont {Ioffe}},\ }\bibfield  {title} {\bibinfo {title} {Microscopic model of quantum butterfly effect: out-of-time-order correlators and traveling combustion waves},\ }\href@noop {} {\bibfield  {journal} {\bibinfo  {journal} {Annals of Physics}\ }\textbf {\bibinfo {volume} {375}},\ \bibinfo {pages} {378} (\bibinfo {year} {2016})}\BibitemShut {NoStop}%
\bibitem [{\citenamefont {Kukuljan}\ \emph {et~al.}(2017)\citenamefont {Kukuljan}, \citenamefont {Grozdanov},\ and\ \citenamefont {Prosen}}]{chaos2}%
  \BibitemOpen
  \bibfield  {author} {\bibinfo {author} {\bibfnamefont {I.}~\bibnamefont {Kukuljan}}, \bibinfo {author} {\bibfnamefont {S.}~\bibnamefont {Grozdanov}},\ and\ \bibinfo {author} {\bibfnamefont {T.}~\bibnamefont {Prosen}},\ }\bibfield  {title} {\bibinfo {title} {Weak quantum chaos},\ }\href@noop {} {\bibfield  {journal} {\bibinfo  {journal} {Physical Review B}\ }\textbf {\bibinfo {volume} {96}},\ \bibinfo {pages} {060301} (\bibinfo {year} {2017})}\BibitemShut {NoStop}%
\bibitem [{\citenamefont {Prakash}\ and\ \citenamefont {Lakshminarayan}(2020)}]{prakash2020scrambling}%
  \BibitemOpen
  \bibfield  {author} {\bibinfo {author} {\bibfnamefont {R.}~\bibnamefont {Prakash}}\ and\ \bibinfo {author} {\bibfnamefont {A.}~\bibnamefont {Lakshminarayan}},\ }\bibfield  {title} {\bibinfo {title} {Scrambling in strongly chaotic weakly coupled bipartite systems: Universality beyond the ehrenfest timescale},\ }\href {https://journals.aps.org/prb/abstract/10.1103/PhysRevB.101.121108} {\bibfield  {journal} {\bibinfo  {journal} {Physical Review B}\ }\textbf {\bibinfo {volume} {101}},\ \bibinfo {pages} {121108} (\bibinfo {year} {2020})}\BibitemShut {NoStop}%
\bibitem [{\citenamefont {Prakash}\ and\ \citenamefont {Lakshminarayan}(2019)}]{prakash2019out}%
  \BibitemOpen
  \bibfield  {author} {\bibinfo {author} {\bibfnamefont {R.}~\bibnamefont {Prakash}}\ and\ \bibinfo {author} {\bibfnamefont {A.}~\bibnamefont {Lakshminarayan}},\ }\bibfield  {title} {\bibinfo {title} {Out-of-time-order correlators in bipartite nonintegrable systems},\ }\href@noop {} {\bibfield  {journal} {\bibinfo  {journal} {arXiv preprint arXiv:1911.02829}\ } (\bibinfo {year} {2019})}\BibitemShut {NoStop}%
\bibitem [{\citenamefont {Varikuti}\ and\ \citenamefont {Madhok}(2024)}]{varikuti2022out}%
  \BibitemOpen
  \bibfield  {author} {\bibinfo {author} {\bibfnamefont {N.~D.}\ \bibnamefont {Varikuti}}\ and\ \bibinfo {author} {\bibfnamefont {V.}~\bibnamefont {Madhok}},\ }\bibfield  {title} {\bibinfo {title} {Out-of-time ordered correlators in kicked coupled tops: Information scrambling in mixed phase space and the role of conserved quantities},\ }\href@noop {} {\bibfield  {journal} {\bibinfo  {journal} {Chaos: An Interdisciplinary Journal of Nonlinear Science}\ }\textbf {\bibinfo {volume} {34}} (\bibinfo {year} {2024})}\BibitemShut {NoStop}%
\bibitem [{\citenamefont {Markovi{\'c}}\ and\ \citenamefont {{\v{C}}ubrovi{\'c}}(2022)}]{markovic2022detecting}%
  \BibitemOpen
  \bibfield  {author} {\bibinfo {author} {\bibfnamefont {D.}~\bibnamefont {Markovi{\'c}}}\ and\ \bibinfo {author} {\bibfnamefont {M.}~\bibnamefont {{\v{C}}ubrovi{\'c}}},\ }\bibfield  {title} {\bibinfo {title} {Detecting few-body quantum chaos: out-of-time ordered correlators at saturation},\ }\href@noop {} {\bibfield  {journal} {\bibinfo  {journal} {Journal of High Energy Physics}\ }\textbf {\bibinfo {volume} {2022}},\ \bibinfo {pages} {1} (\bibinfo {year} {2022})}\BibitemShut {NoStop}%
\bibitem [{\citenamefont {Varikuti}\ \emph {et~al.}(2024)\citenamefont {Varikuti}, \citenamefont {Sahu}, \citenamefont {Lakshminarayan},\ and\ \citenamefont {Madhok}}]{dileep2024}%
  \BibitemOpen
  \bibfield  {author} {\bibinfo {author} {\bibfnamefont {N.~D.}\ \bibnamefont {Varikuti}}, \bibinfo {author} {\bibfnamefont {A.}~\bibnamefont {Sahu}}, \bibinfo {author} {\bibfnamefont {A.}~\bibnamefont {Lakshminarayan}},\ and\ \bibinfo {author} {\bibfnamefont {V.}~\bibnamefont {Madhok}},\ }\bibfield  {title} {\bibinfo {title} {Probing dynamical sensitivity of a non-kolmogorov-arnold-moser system through out-of-time-order correlators},\ }\href {https://doi.org/10.1103/PhysRevE.109.014209} {\bibfield  {journal} {\bibinfo  {journal} {Phys. Rev. E}\ }\textbf {\bibinfo {volume} {109}},\ \bibinfo {pages} {014209} (\bibinfo {year} {2024})}\BibitemShut {NoStop}%
\bibitem [{\citenamefont {Fan}\ \emph {et~al.}(2017)\citenamefont {Fan}, \citenamefont {Zhang}, \citenamefont {Shen},\ and\ \citenamefont {Zhai}}]{manybody3}%
  \BibitemOpen
  \bibfield  {author} {\bibinfo {author} {\bibfnamefont {R.}~\bibnamefont {Fan}}, \bibinfo {author} {\bibfnamefont {P.}~\bibnamefont {Zhang}}, \bibinfo {author} {\bibfnamefont {H.}~\bibnamefont {Shen}},\ and\ \bibinfo {author} {\bibfnamefont {H.}~\bibnamefont {Zhai}},\ }\bibfield  {title} {\bibinfo {title} {Out-of-time-order correlation for many-body localization},\ }\href@noop {} {\bibfield  {journal} {\bibinfo  {journal} {Science bulletin}\ }\textbf {\bibinfo {volume} {62}},\ \bibinfo {pages} {707} (\bibinfo {year} {2017})}\BibitemShut {NoStop}%
\bibitem [{\citenamefont {Chen}(2016)}]{manybody4}%
  \BibitemOpen
  \bibfield  {author} {\bibinfo {author} {\bibfnamefont {Y.}~\bibnamefont {Chen}},\ }\bibfield  {title} {\bibinfo {title} {Universal logarithmic scrambling in many body localization},\ }\href@noop {} {\bibfield  {journal} {\bibinfo  {journal} {arXiv preprint arXiv:1608.02765}\ } (\bibinfo {year} {2016})}\BibitemShut {NoStop}%
\bibitem [{\citenamefont {Swingle}\ and\ \citenamefont {Chowdhury}(2017)}]{manybody1}%
  \BibitemOpen
  \bibfield  {author} {\bibinfo {author} {\bibfnamefont {B.}~\bibnamefont {Swingle}}\ and\ \bibinfo {author} {\bibfnamefont {D.}~\bibnamefont {Chowdhury}},\ }\bibfield  {title} {\bibinfo {title} {Slow scrambling in disordered quantum systems},\ }\href@noop {} {\bibfield  {journal} {\bibinfo  {journal} {Physical Review B}\ }\textbf {\bibinfo {volume} {95}},\ \bibinfo {pages} {060201} (\bibinfo {year} {2017})}\BibitemShut {NoStop}%
\bibitem [{\citenamefont {Huang}\ \emph {et~al.}(2017)\citenamefont {Huang}, \citenamefont {Zhang},\ and\ \citenamefont {Chen}}]{huang2017out}%
  \BibitemOpen
  \bibfield  {author} {\bibinfo {author} {\bibfnamefont {Y.}~\bibnamefont {Huang}}, \bibinfo {author} {\bibfnamefont {Y.-L.}\ \bibnamefont {Zhang}},\ and\ \bibinfo {author} {\bibfnamefont {X.}~\bibnamefont {Chen}},\ }\bibfield  {title} {\bibinfo {title} {Out-of-time-ordered correlators in many-body localized systems},\ }\href@noop {} {\bibfield  {journal} {\bibinfo  {journal} {Annalen der Physik}\ }\textbf {\bibinfo {volume} {529}},\ \bibinfo {pages} {1600318} (\bibinfo {year} {2017})}\BibitemShut {NoStop}%
\bibitem [{\citenamefont {Roberts}\ \emph {et~al.}(2015)\citenamefont {Roberts}, \citenamefont {Stanford},\ and\ \citenamefont {Susskind}}]{shock1}%
  \BibitemOpen
  \bibfield  {author} {\bibinfo {author} {\bibfnamefont {D.~A.}\ \bibnamefont {Roberts}}, \bibinfo {author} {\bibfnamefont {D.}~\bibnamefont {Stanford}},\ and\ \bibinfo {author} {\bibfnamefont {L.}~\bibnamefont {Susskind}},\ }\bibfield  {title} {\bibinfo {title} {Localized shocks},\ }\href@noop {} {\bibfield  {journal} {\bibinfo  {journal} {Journal of High Energy Physics}\ }\textbf {\bibinfo {volume} {2015}},\ \bibinfo {pages} {51} (\bibinfo {year} {2015})}\BibitemShut {NoStop}%
\bibitem [{\citenamefont {Shenker}\ and\ \citenamefont {Stanford}(2014{\natexlab{c}})}]{shenker3}%
  \BibitemOpen
  \bibfield  {author} {\bibinfo {author} {\bibfnamefont {S.~H.}\ \bibnamefont {Shenker}}\ and\ \bibinfo {author} {\bibfnamefont {D.}~\bibnamefont {Stanford}},\ }\bibfield  {title} {\bibinfo {title} {Multiple shocks},\ }\href@noop {} {\bibfield  {journal} {\bibinfo  {journal} {Journal of High Energy Physics}\ }\textbf {\bibinfo {volume} {2014}},\ \bibinfo {pages} {46} (\bibinfo {year} {2014}{\natexlab{c}})}\BibitemShut {NoStop}%
\bibitem [{\citenamefont {Schubert}\ \emph {et~al.}(2012)\citenamefont {Schubert}, \citenamefont {Vallejos},\ and\ \citenamefont {Toscano}}]{schubert2012wave}%
  \BibitemOpen
  \bibfield  {author} {\bibinfo {author} {\bibfnamefont {R.}~\bibnamefont {Schubert}}, \bibinfo {author} {\bibfnamefont {R.~O.}\ \bibnamefont {Vallejos}},\ and\ \bibinfo {author} {\bibfnamefont {F.}~\bibnamefont {Toscano}},\ }\bibfield  {title} {\bibinfo {title} {How do wave packets spread? time evolution on ehrenfest time scales},\ }\href@noop {} {\bibfield  {journal} {\bibinfo  {journal} {Journal of Physics A: Mathematical and Theoretical}\ }\textbf {\bibinfo {volume} {45}},\ \bibinfo {pages} {215307} (\bibinfo {year} {2012})}\BibitemShut {NoStop}%
\bibitem [{\citenamefont {Jalabert}\ \emph {et~al.}(2018)\citenamefont {Jalabert}, \citenamefont {Garc{\'\i}a-Mata},\ and\ \citenamefont {Wisniacki}}]{jalabert2018semiclassical}%
  \BibitemOpen
  \bibfield  {author} {\bibinfo {author} {\bibfnamefont {R.~A.}\ \bibnamefont {Jalabert}}, \bibinfo {author} {\bibfnamefont {I.}~\bibnamefont {Garc{\'\i}a-Mata}},\ and\ \bibinfo {author} {\bibfnamefont {D.~A.}\ \bibnamefont {Wisniacki}},\ }\bibfield  {title} {\bibinfo {title} {Semiclassical theory of out-of-time-order correlators for low-dimensional classically chaotic systems},\ }\href {https://journals.aps.org/pre/abstract/10.1103/PhysRevE.98.062218} {\bibfield  {journal} {\bibinfo  {journal} {Physical Review E}\ }\textbf {\bibinfo {volume} {98}},\ \bibinfo {pages} {062218} (\bibinfo {year} {2018})}\BibitemShut {NoStop}%
\bibitem [{\citenamefont {Chen}\ and\ \citenamefont {Zhou}(2018)}]{chen2018operator}%
  \BibitemOpen
  \bibfield  {author} {\bibinfo {author} {\bibfnamefont {X.}~\bibnamefont {Chen}}\ and\ \bibinfo {author} {\bibfnamefont {T.}~\bibnamefont {Zhou}},\ }\bibfield  {title} {\bibinfo {title} {Operator scrambling and quantum chaos},\ }\href {https://arxiv.org/abs/1804.08655} {\bibfield  {journal} {\bibinfo  {journal} {arXiv preprint arXiv:1804.08655}\ } (\bibinfo {year} {2018})}\BibitemShut {NoStop}%
\bibitem [{\citenamefont {Pappalardi}\ \emph {et~al.}(2018)\citenamefont {Pappalardi}, \citenamefont {Russomanno}, \citenamefont {{\v{Z}}unkovi{\v{c}}}, \citenamefont {Iemini}, \citenamefont {Silva},\ and\ \citenamefont {Fazio}}]{pappalardi2018scrambling}%
  \BibitemOpen
  \bibfield  {author} {\bibinfo {author} {\bibfnamefont {S.}~\bibnamefont {Pappalardi}}, \bibinfo {author} {\bibfnamefont {A.}~\bibnamefont {Russomanno}}, \bibinfo {author} {\bibfnamefont {B.}~\bibnamefont {{\v{Z}}unkovi{\v{c}}}}, \bibinfo {author} {\bibfnamefont {F.}~\bibnamefont {Iemini}}, \bibinfo {author} {\bibfnamefont {A.}~\bibnamefont {Silva}},\ and\ \bibinfo {author} {\bibfnamefont {R.}~\bibnamefont {Fazio}},\ }\bibfield  {title} {\bibinfo {title} {Scrambling and entanglement spreading in long-range spin chains},\ }\href {https://journals.aps.org/prb/abstract/10.1103/PhysRevB.98.134303} {\bibfield  {journal} {\bibinfo  {journal} {Physical Review B}\ }\textbf {\bibinfo {volume} {98}},\ \bibinfo {pages} {134303} (\bibinfo {year} {2018})}\BibitemShut {NoStop}%
\bibitem [{\citenamefont {Hashimoto}\ \emph {et~al.}(2020)\citenamefont {Hashimoto}, \citenamefont {Huh}, \citenamefont {Kim},\ and\ \citenamefont {Watanabe}}]{hashimoto2020exponential}%
  \BibitemOpen
  \bibfield  {author} {\bibinfo {author} {\bibfnamefont {K.}~\bibnamefont {Hashimoto}}, \bibinfo {author} {\bibfnamefont {K.-B.}\ \bibnamefont {Huh}}, \bibinfo {author} {\bibfnamefont {K.-Y.}\ \bibnamefont {Kim}},\ and\ \bibinfo {author} {\bibfnamefont {R.}~\bibnamefont {Watanabe}},\ }\bibfield  {title} {\bibinfo {title} {Exponential growth of out-of-time-order correlator without chaos: inverted harmonic oscillator},\ }\href {https://link.springer.com/article/10.1007/JHEP11(2020)068} {\bibfield  {journal} {\bibinfo  {journal} {Journal of High Energy Physics}\ }\textbf {\bibinfo {volume} {2020}},\ \bibinfo {pages} {1} (\bibinfo {year} {2020})}\BibitemShut {NoStop}%
\bibitem [{\citenamefont {Hummel}\ \emph {et~al.}(2019)\citenamefont {Hummel}, \citenamefont {Geiger}, \citenamefont {Urbina},\ and\ \citenamefont {Richter}}]{hummel2019reversible}%
  \BibitemOpen
  \bibfield  {author} {\bibinfo {author} {\bibfnamefont {Q.}~\bibnamefont {Hummel}}, \bibinfo {author} {\bibfnamefont {B.}~\bibnamefont {Geiger}}, \bibinfo {author} {\bibfnamefont {J.~D.}\ \bibnamefont {Urbina}},\ and\ \bibinfo {author} {\bibfnamefont {K.}~\bibnamefont {Richter}},\ }\bibfield  {title} {\bibinfo {title} {Reversible quantum information spreading in many-body systems near criticality},\ }\href@noop {} {\bibfield  {journal} {\bibinfo  {journal} {Physical review letters}\ }\textbf {\bibinfo {volume} {123}},\ \bibinfo {pages} {160401} (\bibinfo {year} {2019})}\BibitemShut {NoStop}%
\bibitem [{\citenamefont {Steinhuber}\ \emph {et~al.}(2023)\citenamefont {Steinhuber}, \citenamefont {Schlagheck}, \citenamefont {Urbina},\ and\ \citenamefont {Richter}}]{steinhuber2023dynamical}%
  \BibitemOpen
  \bibfield  {author} {\bibinfo {author} {\bibfnamefont {M.}~\bibnamefont {Steinhuber}}, \bibinfo {author} {\bibfnamefont {P.}~\bibnamefont {Schlagheck}}, \bibinfo {author} {\bibfnamefont {J.~D.}\ \bibnamefont {Urbina}},\ and\ \bibinfo {author} {\bibfnamefont {K.}~\bibnamefont {Richter}},\ }\bibfield  {title} {\bibinfo {title} {Dynamical transition from localized to uniform scrambling in locally hyperbolic systems},\ }\href@noop {} {\bibfield  {journal} {\bibinfo  {journal} {Physical Review E}\ }\textbf {\bibinfo {volume} {108}},\ \bibinfo {pages} {024216} (\bibinfo {year} {2023})}\BibitemShut {NoStop}%
\bibitem [{\citenamefont {Styliaris}\ \emph {et~al.}(2021)\citenamefont {Styliaris}, \citenamefont {Anand},\ and\ \citenamefont {Zanardi}}]{styliaris2021information}%
  \BibitemOpen
  \bibfield  {author} {\bibinfo {author} {\bibfnamefont {G.}~\bibnamefont {Styliaris}}, \bibinfo {author} {\bibfnamefont {N.}~\bibnamefont {Anand}},\ and\ \bibinfo {author} {\bibfnamefont {P.}~\bibnamefont {Zanardi}},\ }\bibfield  {title} {\bibinfo {title} {Information scrambling over bipartitions: Equilibration, entropy production, and typicality},\ }\href@noop {} {\bibfield  {journal} {\bibinfo  {journal} {Physical Review Letters}\ }\textbf {\bibinfo {volume} {126}},\ \bibinfo {pages} {030601} (\bibinfo {year} {2021})}\BibitemShut {NoStop}%
\bibitem [{\citenamefont {Zanardi}\ and\ \citenamefont {Anand}(2021)}]{zanardi2021information}%
  \BibitemOpen
  \bibfield  {author} {\bibinfo {author} {\bibfnamefont {P.}~\bibnamefont {Zanardi}}\ and\ \bibinfo {author} {\bibfnamefont {N.}~\bibnamefont {Anand}},\ }\bibfield  {title} {\bibinfo {title} {Information scrambling and chaos in open quantum systems},\ }\href@noop {} {\bibfield  {journal} {\bibinfo  {journal} {Physical Review A}\ }\textbf {\bibinfo {volume} {103}},\ \bibinfo {pages} {062214} (\bibinfo {year} {2021})}\BibitemShut {NoStop}%
\bibitem [{\citenamefont {Anand}\ \emph {et~al.}(2021)\citenamefont {Anand}, \citenamefont {Styliaris}, \citenamefont {Kumari},\ and\ \citenamefont {Zanardi}}]{anand2021quantum}%
  \BibitemOpen
  \bibfield  {author} {\bibinfo {author} {\bibfnamefont {N.}~\bibnamefont {Anand}}, \bibinfo {author} {\bibfnamefont {G.}~\bibnamefont {Styliaris}}, \bibinfo {author} {\bibfnamefont {M.}~\bibnamefont {Kumari}},\ and\ \bibinfo {author} {\bibfnamefont {P.}~\bibnamefont {Zanardi}},\ }\bibfield  {title} {\bibinfo {title} {Quantum coherence as a signature of chaos},\ }\href@noop {} {\bibfield  {journal} {\bibinfo  {journal} {Physical Review Research}\ }\textbf {\bibinfo {volume} {3}},\ \bibinfo {pages} {023214} (\bibinfo {year} {2021})}\BibitemShut {NoStop}%
\bibitem [{\citenamefont {Sreeram}\ \emph {et~al.}(2021)\citenamefont {Sreeram}, \citenamefont {Madhok},\ and\ \citenamefont {Lakshminarayan}}]{sreeram2021out}%
  \BibitemOpen
  \bibfield  {author} {\bibinfo {author} {\bibfnamefont {P.}~\bibnamefont {Sreeram}}, \bibinfo {author} {\bibfnamefont {V.}~\bibnamefont {Madhok}},\ and\ \bibinfo {author} {\bibfnamefont {A.}~\bibnamefont {Lakshminarayan}},\ }\bibfield  {title} {\bibinfo {title} {Out-of-time-ordered correlators and the loschmidt echo in the quantum kicked top: how low can we go?},\ }\href {https://iopscience.iop.org/article/10.1088/1361-6463/abf8f3/meta} {\bibfield  {journal} {\bibinfo  {journal} {Journal of Physics D: Applied Physics}\ }\textbf {\bibinfo {volume} {54}},\ \bibinfo {pages} {274004} (\bibinfo {year} {2021})}\BibitemShut {NoStop}%
\bibitem [{\citenamefont {PG}\ \emph {et~al.}(2022)\citenamefont {PG}, \citenamefont {Modak},\ and\ \citenamefont {Aravinda}}]{pg2022witnessing}%
  \BibitemOpen
  \bibfield  {author} {\bibinfo {author} {\bibfnamefont {S.}~\bibnamefont {PG}}, \bibinfo {author} {\bibfnamefont {R.}~\bibnamefont {Modak}},\ and\ \bibinfo {author} {\bibfnamefont {S.}~\bibnamefont {Aravinda}},\ }\bibfield  {title} {\bibinfo {title} {Witnessing quantum chaos using observational entropy},\ }\href@noop {} {\bibfield  {journal} {\bibinfo  {journal} {arXiv preprint arXiv:2212.01585}\ } (\bibinfo {year} {2022})}\BibitemShut {NoStop}%
\bibitem [{\citenamefont {{\v{S}}afr{\'a}nek}\ \emph {et~al.}(2019{\natexlab{a}})\citenamefont {{\v{S}}afr{\'a}nek}, \citenamefont {Deutsch},\ and\ \citenamefont {Aguirre}}]{vsafranek2019quantum}%
  \BibitemOpen
  \bibfield  {author} {\bibinfo {author} {\bibfnamefont {D.}~\bibnamefont {{\v{S}}afr{\'a}nek}}, \bibinfo {author} {\bibfnamefont {J.~M.}\ \bibnamefont {Deutsch}},\ and\ \bibinfo {author} {\bibfnamefont {A.}~\bibnamefont {Aguirre}},\ }\bibfield  {title} {\bibinfo {title} {Quantum coarse-grained entropy and thermodynamics},\ }\href@noop {} {\bibfield  {journal} {\bibinfo  {journal} {Physical Review A}\ }\textbf {\bibinfo {volume} {99}},\ \bibinfo {pages} {010101} (\bibinfo {year} {2019}{\natexlab{a}})}\BibitemShut {NoStop}%
\bibitem [{\citenamefont {{\v{S}}afr{\'a}nek}\ \emph {et~al.}(2019{\natexlab{b}})\citenamefont {{\v{S}}afr{\'a}nek}, \citenamefont {Deutsch},\ and\ \citenamefont {Aguirre}}]{vsafranek2019quantum1}%
  \BibitemOpen
  \bibfield  {author} {\bibinfo {author} {\bibfnamefont {D.}~\bibnamefont {{\v{S}}afr{\'a}nek}}, \bibinfo {author} {\bibfnamefont {J.}~\bibnamefont {Deutsch}},\ and\ \bibinfo {author} {\bibfnamefont {A.}~\bibnamefont {Aguirre}},\ }\bibfield  {title} {\bibinfo {title} {Quantum coarse-grained entropy and thermalization in closed systems},\ }\href@noop {} {\bibfield  {journal} {\bibinfo  {journal} {Physical Review A}\ }\textbf {\bibinfo {volume} {99}},\ \bibinfo {pages} {012103} (\bibinfo {year} {2019}{\natexlab{b}})}\BibitemShut {NoStop}%
\bibitem [{\citenamefont {Madhok}\ \emph {et~al.}(2014{\natexlab{a}})\citenamefont {Madhok}, \citenamefont {Riofr\'{\i}o}, \citenamefont {Ghose},\ and\ \citenamefont {Deutsch}}]{PhysRevLett.112.014102}%
  \BibitemOpen
  \bibfield  {author} {\bibinfo {author} {\bibfnamefont {V.}~\bibnamefont {Madhok}}, \bibinfo {author} {\bibfnamefont {C.~A.}\ \bibnamefont {Riofr\'{\i}o}}, \bibinfo {author} {\bibfnamefont {S.}~\bibnamefont {Ghose}},\ and\ \bibinfo {author} {\bibfnamefont {I.~H.}\ \bibnamefont {Deutsch}},\ }\bibfield  {title} {\bibinfo {title} {Information gain in tomography--a quantum signature of chaos},\ }\href {https://doi.org/10.1103/PhysRevLett.112.014102} {\bibfield  {journal} {\bibinfo  {journal} {Phys. Rev. Lett.}\ }\textbf {\bibinfo {volume} {112}},\ \bibinfo {pages} {014102} (\bibinfo {year} {2014}{\natexlab{a}})}\BibitemShut {NoStop}%
\bibitem [{\citenamefont {Silberfarb}\ \emph {et~al.}(2005{\natexlab{a}})\citenamefont {Silberfarb}, \citenamefont {Jessen},\ and\ \citenamefont {Deutsch}}]{PhysRevLett.95.030402}%
  \BibitemOpen
  \bibfield  {author} {\bibinfo {author} {\bibfnamefont {A.}~\bibnamefont {Silberfarb}}, \bibinfo {author} {\bibfnamefont {P.~S.}\ \bibnamefont {Jessen}},\ and\ \bibinfo {author} {\bibfnamefont {I.~H.}\ \bibnamefont {Deutsch}},\ }\bibfield  {title} {\bibinfo {title} {Quantum state reconstruction via continuous measurement},\ }\href {https://doi.org/10.1103/PhysRevLett.95.030402} {\bibfield  {journal} {\bibinfo  {journal} {Phys. Rev. Lett.}\ }\textbf {\bibinfo {volume} {95}},\ \bibinfo {pages} {030402} (\bibinfo {year} {2005}{\natexlab{a}})}\BibitemShut {NoStop}%
\bibitem [{\citenamefont {Massar}\ and\ \citenamefont {Popescu}(2005)}]{massar2005optimal}%
  \BibitemOpen
  \bibfield  {author} {\bibinfo {author} {\bibfnamefont {S.}~\bibnamefont {Massar}}\ and\ \bibinfo {author} {\bibfnamefont {S.}~\bibnamefont {Popescu}},\ }\bibfield  {title} {\bibinfo {title} {Optimal extraction of information from finite quantum ensembles},\ }in\ \href@noop {} {\emph {\bibinfo {booktitle} {Asymptotic Theory Of Quantum Statistical Inference: Selected Papers}}}\ (\bibinfo  {publisher} {World Scientific},\ \bibinfo {year} {2005})\ pp.\ \bibinfo {pages} {356--364}\BibitemShut {NoStop}%
\bibitem [{\citenamefont {Vidal}\ \emph {et~al.}(1999)\citenamefont {Vidal}, \citenamefont {Latorre}, \citenamefont {Pascual},\ and\ \citenamefont {Tarrach}}]{vidal1999optimal}%
  \BibitemOpen
  \bibfield  {author} {\bibinfo {author} {\bibfnamefont {G.}~\bibnamefont {Vidal}}, \bibinfo {author} {\bibfnamefont {J.}~\bibnamefont {Latorre}}, \bibinfo {author} {\bibfnamefont {P.}~\bibnamefont {Pascual}},\ and\ \bibinfo {author} {\bibfnamefont {R.}~\bibnamefont {Tarrach}},\ }\bibfield  {title} {\bibinfo {title} {Optimal minimal measurements of mixed states},\ }\href@noop {} {\bibfield  {journal} {\bibinfo  {journal} {Physical Review A}\ }\textbf {\bibinfo {volume} {60}},\ \bibinfo {pages} {126} (\bibinfo {year} {1999})}\BibitemShut {NoStop}%
\bibitem [{\citenamefont {Gisin}\ and\ \citenamefont {Popescu}(1999)}]{gisin1999spin}%
  \BibitemOpen
  \bibfield  {author} {\bibinfo {author} {\bibfnamefont {N.}~\bibnamefont {Gisin}}\ and\ \bibinfo {author} {\bibfnamefont {S.}~\bibnamefont {Popescu}},\ }\bibfield  {title} {\bibinfo {title} {Spin flips and quantum information for antiparallel spins},\ }\href@noop {} {\bibfield  {journal} {\bibinfo  {journal} {Physical Review Letters}\ }\textbf {\bibinfo {volume} {83}},\ \bibinfo {pages} {432} (\bibinfo {year} {1999})}\BibitemShut {NoStop}%
\bibitem [{\citenamefont {Bagan}\ \emph {et~al.}(2006)\citenamefont {Bagan}, \citenamefont {Ballester}, \citenamefont {Gill}, \citenamefont {Mu{\~n}oz-Tapia},\ and\ \citenamefont {Romero-Isart}}]{bagan2006separable}%
  \BibitemOpen
  \bibfield  {author} {\bibinfo {author} {\bibfnamefont {E.}~\bibnamefont {Bagan}}, \bibinfo {author} {\bibfnamefont {M.}~\bibnamefont {Ballester}}, \bibinfo {author} {\bibfnamefont {R.}~\bibnamefont {Gill}}, \bibinfo {author} {\bibfnamefont {R.}~\bibnamefont {Mu{\~n}oz-Tapia}},\ and\ \bibinfo {author} {\bibfnamefont {O.}~\bibnamefont {Romero-Isart}},\ }\bibfield  {title} {\bibinfo {title} {Separable measurement estimation of density matrices and its fidelity gap with collective protocols},\ }\href@noop {} {\bibfield  {journal} {\bibinfo  {journal} {Physical review letters}\ }\textbf {\bibinfo {volume} {97}},\ \bibinfo {pages} {130501} (\bibinfo {year} {2006})}\BibitemShut {NoStop}%
\bibitem [{\citenamefont {Hou}\ \emph {et~al.}(2018)\citenamefont {Hou}, \citenamefont {Tang}, \citenamefont {Shang}, \citenamefont {Zhu}, \citenamefont {Li}, \citenamefont {Yuan}, \citenamefont {Wu}, \citenamefont {Xiang}, \citenamefont {Li},\ and\ \citenamefont {Guo}}]{hou2018deterministic}%
  \BibitemOpen
  \bibfield  {author} {\bibinfo {author} {\bibfnamefont {Z.}~\bibnamefont {Hou}}, \bibinfo {author} {\bibfnamefont {J.-F.}\ \bibnamefont {Tang}}, \bibinfo {author} {\bibfnamefont {J.}~\bibnamefont {Shang}}, \bibinfo {author} {\bibfnamefont {H.}~\bibnamefont {Zhu}}, \bibinfo {author} {\bibfnamefont {J.}~\bibnamefont {Li}}, \bibinfo {author} {\bibfnamefont {Y.}~\bibnamefont {Yuan}}, \bibinfo {author} {\bibfnamefont {K.-D.}\ \bibnamefont {Wu}}, \bibinfo {author} {\bibfnamefont {G.-Y.}\ \bibnamefont {Xiang}}, \bibinfo {author} {\bibfnamefont {C.-F.}\ \bibnamefont {Li}},\ and\ \bibinfo {author} {\bibfnamefont {G.-C.}\ \bibnamefont {Guo}},\ }\bibfield  {title} {\bibinfo {title} {Deterministic realization of collective measurements via photonic quantum walks},\ }\href@noop {} {\bibfield  {journal} {\bibinfo  {journal} {Nature communications}\ }\textbf {\bibinfo {volume} {9}},\ \bibinfo {pages} {1} (\bibinfo {year} {2018})}\BibitemShut {NoStop}%
\bibitem [{\citenamefont {Bennett}\ \emph {et~al.}(1999)\citenamefont {Bennett}, \citenamefont {DiVincenzo}, \citenamefont {Fuchs}, \citenamefont {Mor}, \citenamefont {Rains}, \citenamefont {Shor}, \citenamefont {Smolin},\ and\ \citenamefont {Wootters}}]{bennett1999quantum}%
  \BibitemOpen
  \bibfield  {author} {\bibinfo {author} {\bibfnamefont {C.~H.}\ \bibnamefont {Bennett}}, \bibinfo {author} {\bibfnamefont {D.~P.}\ \bibnamefont {DiVincenzo}}, \bibinfo {author} {\bibfnamefont {C.~A.}\ \bibnamefont {Fuchs}}, \bibinfo {author} {\bibfnamefont {T.}~\bibnamefont {Mor}}, \bibinfo {author} {\bibfnamefont {E.}~\bibnamefont {Rains}}, \bibinfo {author} {\bibfnamefont {P.~W.}\ \bibnamefont {Shor}}, \bibinfo {author} {\bibfnamefont {J.~A.}\ \bibnamefont {Smolin}},\ and\ \bibinfo {author} {\bibfnamefont {W.~K.}\ \bibnamefont {Wootters}},\ }\bibfield  {title} {\bibinfo {title} {Quantum nonlocality without entanglement},\ }\href@noop {} {\bibfield  {journal} {\bibinfo  {journal} {Physical Review A}\ }\textbf {\bibinfo {volume} {59}},\ \bibinfo {pages} {1070} (\bibinfo {year} {1999})}\BibitemShut {NoStop}%
\bibitem [{\citenamefont {Smith}\ \emph {et~al.}(2006{\natexlab{a}})\citenamefont {Smith}, \citenamefont {Silberfarb}, \citenamefont {Deutsch},\ and\ \citenamefont {Jessen}}]{PhysRevLett.97.180403}%
  \BibitemOpen
  \bibfield  {author} {\bibinfo {author} {\bibfnamefont {G.~A.}\ \bibnamefont {Smith}}, \bibinfo {author} {\bibfnamefont {A.}~\bibnamefont {Silberfarb}}, \bibinfo {author} {\bibfnamefont {I.~H.}\ \bibnamefont {Deutsch}},\ and\ \bibinfo {author} {\bibfnamefont {P.~S.}\ \bibnamefont {Jessen}},\ }\bibfield  {title} {\bibinfo {title} {Efficient quantum-state estimation by continuous weak measurement and dynamical control},\ }\href {https://doi.org/10.1103/PhysRevLett.97.180403} {\bibfield  {journal} {\bibinfo  {journal} {Phys. Rev. Lett.}\ }\textbf {\bibinfo {volume} {97}},\ \bibinfo {pages} {180403} (\bibinfo {year} {2006}{\natexlab{a}})}\BibitemShut {NoStop}%
\bibitem [{\citenamefont {Silberfarb}\ \emph {et~al.}(2005{\natexlab{b}})\citenamefont {Silberfarb}, \citenamefont {Jessen},\ and\ \citenamefont {Deutsch}}]{silberfarb2005quantum}%
  \BibitemOpen
  \bibfield  {author} {\bibinfo {author} {\bibfnamefont {A.}~\bibnamefont {Silberfarb}}, \bibinfo {author} {\bibfnamefont {P.~S.}\ \bibnamefont {Jessen}},\ and\ \bibinfo {author} {\bibfnamefont {I.~H.}\ \bibnamefont {Deutsch}},\ }\bibfield  {title} {\bibinfo {title} {Quantum state reconstruction via continuous measurement},\ }\href@noop {} {\bibfield  {journal} {\bibinfo  {journal} {Physical review letters}\ }\textbf {\bibinfo {volume} {95}},\ \bibinfo {pages} {030402} (\bibinfo {year} {2005}{\natexlab{b}})}\BibitemShut {NoStop}%
\bibitem [{\citenamefont {Madhok}\ \emph {et~al.}(2014{\natexlab{b}})\citenamefont {Madhok}, \citenamefont {Riofr{\'\i}o}, \citenamefont {Ghose},\ and\ \citenamefont {Deutsch}}]{madhok2014information}%
  \BibitemOpen
  \bibfield  {author} {\bibinfo {author} {\bibfnamefont {V.}~\bibnamefont {Madhok}}, \bibinfo {author} {\bibfnamefont {C.~A.}\ \bibnamefont {Riofr{\'\i}o}}, \bibinfo {author} {\bibfnamefont {S.}~\bibnamefont {Ghose}},\ and\ \bibinfo {author} {\bibfnamefont {I.~H.}\ \bibnamefont {Deutsch}},\ }\bibfield  {title} {\bibinfo {title} {Information gain in tomography--a quantum signature of chaos},\ }\href@noop {} {\bibfield  {journal} {\bibinfo  {journal} {Physical review letters}\ }\textbf {\bibinfo {volume} {112}},\ \bibinfo {pages} {014102} (\bibinfo {year} {2014}{\natexlab{b}})}\BibitemShut {NoStop}%
\bibitem [{\citenamefont {Smith}\ \emph {et~al.}(2013)\citenamefont {Smith}, \citenamefont {Riofr{\'\i}o}, \citenamefont {Anderson}, \citenamefont {Sosa-Martinez}, \citenamefont {Deutsch},\ and\ \citenamefont {Jessen}}]{smith2013quantum}%
  \BibitemOpen
  \bibfield  {author} {\bibinfo {author} {\bibfnamefont {A.}~\bibnamefont {Smith}}, \bibinfo {author} {\bibfnamefont {C.}~\bibnamefont {Riofr{\'\i}o}}, \bibinfo {author} {\bibfnamefont {B.}~\bibnamefont {Anderson}}, \bibinfo {author} {\bibfnamefont {H.}~\bibnamefont {Sosa-Martinez}}, \bibinfo {author} {\bibfnamefont {I.}~\bibnamefont {Deutsch}},\ and\ \bibinfo {author} {\bibfnamefont {P.}~\bibnamefont {Jessen}},\ }\bibfield  {title} {\bibinfo {title} {Quantum state tomography by continuous measurement and compressed sensing},\ }\href {https://journals.aps.org/pra/abstract/10.1103/PhysRevA.87.030102} {\bibfield  {journal} {\bibinfo  {journal} {Physical Review A}\ }\textbf {\bibinfo {volume} {87}},\ \bibinfo {pages} {030102} (\bibinfo {year} {2013})}\BibitemShut {NoStop}%
\bibitem [{\citenamefont {Deutsch}\ and\ \citenamefont {Jessen}(2010)}]{deutsch2010quantum}%
  \BibitemOpen
  \bibfield  {author} {\bibinfo {author} {\bibfnamefont {I.~H.}\ \bibnamefont {Deutsch}}\ and\ \bibinfo {author} {\bibfnamefont {P.~S.}\ \bibnamefont {Jessen}},\ }\bibfield  {title} {\bibinfo {title} {Quantum control and measurement of atomic spins in polarization spectroscopy},\ }\href@noop {} {\bibfield  {journal} {\bibinfo  {journal} {Optics Communications}\ }\textbf {\bibinfo {volume} {283}},\ \bibinfo {pages} {681} (\bibinfo {year} {2010})}\BibitemShut {NoStop}%
\bibitem [{\citenamefont {Smith}\ \emph {et~al.}(2006{\natexlab{b}})\citenamefont {Smith}, \citenamefont {Silberfarb}, \citenamefont {Deutsch},\ and\ \citenamefont {Jessen}}]{smith2006efficient}%
  \BibitemOpen
  \bibfield  {author} {\bibinfo {author} {\bibfnamefont {G.~A.}\ \bibnamefont {Smith}}, \bibinfo {author} {\bibfnamefont {A.}~\bibnamefont {Silberfarb}}, \bibinfo {author} {\bibfnamefont {I.~H.}\ \bibnamefont {Deutsch}},\ and\ \bibinfo {author} {\bibfnamefont {P.~S.}\ \bibnamefont {Jessen}},\ }\bibfield  {title} {\bibinfo {title} {Efficient quantum-state estimation by continuous weak measurement and dynamical control},\ }\href@noop {} {\bibfield  {journal} {\bibinfo  {journal} {Physical review letters}\ }\textbf {\bibinfo {volume} {97}},\ \bibinfo {pages} {180403} (\bibinfo {year} {2006}{\natexlab{b}})}\BibitemShut {NoStop}%
\bibitem [{\citenamefont {Ben-Israel}\ and\ \citenamefont {Greville}(2003)}]{ben2003generalized}%
  \BibitemOpen
  \bibfield  {author} {\bibinfo {author} {\bibfnamefont {A.}~\bibnamefont {Ben-Israel}}\ and\ \bibinfo {author} {\bibfnamefont {T.~N.}\ \bibnamefont {Greville}},\ }\href@noop {} {\emph {\bibinfo {title} {Generalized inverses: theory and applications}}},\ Vol.~\bibinfo {volume} {15}\ (\bibinfo  {publisher} {Springer Science \& Business Media},\ \bibinfo {year} {2003})\BibitemShut {NoStop}%
\bibitem [{\citenamefont {Baldwin}\ \emph {et~al.}(2016)\citenamefont {Baldwin}, \citenamefont {Deutsch},\ and\ \citenamefont {Kalev}}]{baldwin2016strictly}%
  \BibitemOpen
  \bibfield  {author} {\bibinfo {author} {\bibfnamefont {C.~H.}\ \bibnamefont {Baldwin}}, \bibinfo {author} {\bibfnamefont {I.~H.}\ \bibnamefont {Deutsch}},\ and\ \bibinfo {author} {\bibfnamefont {A.}~\bibnamefont {Kalev}},\ }\bibfield  {title} {\bibinfo {title} {Strictly-complete measurements for bounded-rank quantum-state tomography},\ }\href@noop {} {\bibfield  {journal} {\bibinfo  {journal} {Physical Review A}\ }\textbf {\bibinfo {volume} {93}},\ \bibinfo {pages} {052105} (\bibinfo {year} {2016})}\BibitemShut {NoStop}%
\bibitem [{\citenamefont {Vandenberghe}\ and\ \citenamefont {Boyd}(1996)}]{vandenberghe1996semidefinite}%
  \BibitemOpen
  \bibfield  {author} {\bibinfo {author} {\bibfnamefont {L.}~\bibnamefont {Vandenberghe}}\ and\ \bibinfo {author} {\bibfnamefont {S.}~\bibnamefont {Boyd}},\ }\bibfield  {title} {\bibinfo {title} {Semidefinite programming},\ }\href@noop {} {\bibfield  {journal} {\bibinfo  {journal} {SIAM review}\ }\textbf {\bibinfo {volume} {38}},\ \bibinfo {pages} {49} (\bibinfo {year} {1996})}\BibitemShut {NoStop}%
\bibitem [{\citenamefont {Kalev}\ \emph {et~al.}(2015)\citenamefont {Kalev}, \citenamefont {Kosut},\ and\ \citenamefont {Deutsch}}]{kalev2015quantum}%
  \BibitemOpen
  \bibfield  {author} {\bibinfo {author} {\bibfnamefont {A.}~\bibnamefont {Kalev}}, \bibinfo {author} {\bibfnamefont {R.~L.}\ \bibnamefont {Kosut}},\ and\ \bibinfo {author} {\bibfnamefont {I.~H.}\ \bibnamefont {Deutsch}},\ }\bibfield  {title} {\bibinfo {title} {Quantum tomography protocols with positivity are compressed sensing protocols},\ }\href@noop {} {\bibfield  {journal} {\bibinfo  {journal} {npj Quantum Information}\ }\textbf {\bibinfo {volume} {1}},\ \bibinfo {pages} {1} (\bibinfo {year} {2015})}\BibitemShut {NoStop}%
\bibitem [{\citenamefont {Merkel}\ \emph {et~al.}(2010)\citenamefont {Merkel}, \citenamefont {Riofrio}, \citenamefont {Flammia},\ and\ \citenamefont {Deutsch}}]{merkel2010random}%
  \BibitemOpen
  \bibfield  {author} {\bibinfo {author} {\bibfnamefont {S.~T.}\ \bibnamefont {Merkel}}, \bibinfo {author} {\bibfnamefont {C.~A.}\ \bibnamefont {Riofrio}}, \bibinfo {author} {\bibfnamefont {S.~T.}\ \bibnamefont {Flammia}},\ and\ \bibinfo {author} {\bibfnamefont {I.~H.}\ \bibnamefont {Deutsch}},\ }\bibfield  {title} {\bibinfo {title} {Random unitary maps for quantum state reconstruction},\ }\href {https://journals.aps.org/pra/abstract/10.1103/PhysRevA.81.032126} {\bibfield  {journal} {\bibinfo  {journal} {Physical Review A}\ }\textbf {\bibinfo {volume} {81}},\ \bibinfo {pages} {032126} (\bibinfo {year} {2010})}\BibitemShut {NoStop}%
\bibitem [{\citenamefont {Sreeram}\ and\ \citenamefont {Madhok}(2021)}]{sreeram2021quantum}%
  \BibitemOpen
  \bibfield  {author} {\bibinfo {author} {\bibfnamefont {P.}~\bibnamefont {Sreeram}}\ and\ \bibinfo {author} {\bibfnamefont {V.}~\bibnamefont {Madhok}},\ }\bibfield  {title} {\bibinfo {title} {Quantum tomography with random diagonal unitary maps and statistical bounds on information generation using random matrix theory},\ }\href {https://journals.aps.org/pra/abstract/10.1103/PhysRevA.104.032404} {\bibfield  {journal} {\bibinfo  {journal} {Physical Review A}\ }\textbf {\bibinfo {volume} {104}},\ \bibinfo {pages} {032404} (\bibinfo {year} {2021})}\BibitemShut {NoStop}%
\bibitem [{\citenamefont {Sahu}\ \emph {et~al.}(2022{\natexlab{a}})\citenamefont {Sahu}, \citenamefont {Sreeram},\ and\ \citenamefont {Madhok}}]{sahu2022effect}%
  \BibitemOpen
  \bibfield  {author} {\bibinfo {author} {\bibfnamefont {A.}~\bibnamefont {Sahu}}, \bibinfo {author} {\bibfnamefont {P.}~\bibnamefont {Sreeram}},\ and\ \bibinfo {author} {\bibfnamefont {V.}~\bibnamefont {Madhok}},\ }\bibfield  {title} {\bibinfo {title} {Effect of chaos on information gain in quantum tomography},\ }\href {https://journals.aps.org/pre/abstract/10.1103/PhysRevE.106.024209} {\bibfield  {journal} {\bibinfo  {journal} {Physical Review E}\ }\textbf {\bibinfo {volume} {106}},\ \bibinfo {pages} {024209} (\bibinfo {year} {2022}{\natexlab{a}})}\BibitemShut {NoStop}%
\bibitem [{\citenamefont {Sahu}\ \emph {et~al.}(2022{\natexlab{b}})\citenamefont {Sahu}, \citenamefont {Varikuti},\ and\ \citenamefont {Madhok}}]{sahu2022quantum}%
  \BibitemOpen
  \bibfield  {author} {\bibinfo {author} {\bibfnamefont {A.}~\bibnamefont {Sahu}}, \bibinfo {author} {\bibfnamefont {N.~D.}\ \bibnamefont {Varikuti}},\ and\ \bibinfo {author} {\bibfnamefont {V.}~\bibnamefont {Madhok}},\ }\bibfield  {title} {\bibinfo {title} {Quantum tomography under perturbed hamiltonian evolution and scrambling of errors--a quantum signature of chaos},\ }\href@noop {} {\bibfield  {journal} {\bibinfo  {journal} {arXiv preprint arXiv:2211.11221}\ } (\bibinfo {year} {2022}{\natexlab{b}})}\BibitemShut {NoStop}%
\bibitem [{\citenamefont {Shannon}(1948)}]{shannon1948mathematical}%
  \BibitemOpen
  \bibfield  {author} {\bibinfo {author} {\bibfnamefont {C.~E.}\ \bibnamefont {Shannon}},\ }\bibfield  {title} {\bibinfo {title} {A mathematical theory of communication, bell systems technol},\ }\href@noop {} {\bibfield  {journal} {\bibinfo  {journal} {J}\ }\textbf {\bibinfo {volume} {27}},\ \bibinfo {pages} {379} (\bibinfo {year} {1948})}\BibitemShut {NoStop}%
\bibitem [{\citenamefont {Thomas M.~Cover}(2006)}]{coverm2006elements}%
  \BibitemOpen
  \bibfield  {author} {\bibinfo {author} {\bibfnamefont {J.~A.~T.}\ \bibnamefont {Thomas M.~Cover}},\ }\href@noop {} {\emph {\bibinfo {title} {Elements of Information Theory}}}\ (\bibinfo  {publisher} {John Wiley \& Sons, Ltd},\ \bibinfo {year} {2006})\BibitemShut {NoStop}%
\bibitem [{\citenamefont {Ng}(2004)}]{ng2004feature}%
  \BibitemOpen
  \bibfield  {author} {\bibinfo {author} {\bibfnamefont {A.~Y.}\ \bibnamefont {Ng}},\ }\bibfield  {title} {\bibinfo {title} {Feature selection, l 1 vs. l 2 regularization, and rotational invariance},\ }in\ \href@noop {} {\emph {\bibinfo {booktitle} {Proceedings of the twenty-first international conference on Machine learning}}}\ (\bibinfo {year} {2004})\ p.~\bibinfo {pages} {78}\BibitemShut {NoStop}%
\bibitem [{\citenamefont {Fisher}(1922)}]{fisher1922mathematical}%
  \BibitemOpen
  \bibfield  {author} {\bibinfo {author} {\bibfnamefont {R.~A.}\ \bibnamefont {Fisher}},\ }\bibfield  {title} {\bibinfo {title} {On the mathematical foundations of theoretical statistics},\ }\href@noop {} {\bibfield  {journal} {\bibinfo  {journal} {Philosophical Transactions of the Royal Society of London. Series A, Containing Papers of a Mathematical or Physical Character}\ }\textbf {\bibinfo {volume} {222}},\ \bibinfo {pages} {309} (\bibinfo {year} {1922})}\BibitemShut {NoStop}%
\bibitem [{\citenamefont {Cover}\ and\ \citenamefont {Thomas}(2012)}]{cover2012elements}%
  \BibitemOpen
  \bibfield  {author} {\bibinfo {author} {\bibfnamefont {T.~M.}\ \bibnamefont {Cover}}\ and\ \bibinfo {author} {\bibfnamefont {J.~A.}\ \bibnamefont {Thomas}},\ }\href@noop {} {\emph {\bibinfo {title} {Elements of information theory}}}\ (\bibinfo  {publisher} {John Wiley \& Sons},\ \bibinfo {year} {2012})\BibitemShut {NoStop}%
\bibitem [{\citenamefont {Cramir}(1946)}]{cramir1946mathematical}%
  \BibitemOpen
  \bibfield  {author} {\bibinfo {author} {\bibfnamefont {H.}~\bibnamefont {Cramir}},\ }\bibfield  {title} {\bibinfo {title} {Mathematical methods of statistics},\ }\href@noop {} {\bibfield  {journal} {\bibinfo  {journal} {Princeton U. Press, Princeton}\ ,\ \bibinfo {pages} {500}} (\bibinfo {year} {1946})}\BibitemShut {NoStop}%
\bibitem [{\citenamefont {Rao}(1992)}]{rao1992information}%
  \BibitemOpen
  \bibfield  {author} {\bibinfo {author} {\bibfnamefont {C.~R.}\ \bibnamefont {Rao}},\ }\bibfield  {title} {\bibinfo {title} {Information and the accuracy attainable in the estimation of statistical parameters},\ }in\ \href@noop {} {\emph {\bibinfo {booktitle} {Breakthroughs in Statistics: Foundations and basic theory}}}\ (\bibinfo  {publisher} {Springer},\ \bibinfo {year} {1992})\ pp.\ \bibinfo {pages} {235--247}\BibitemShut {NoStop}%
\bibitem [{\citenamefont {\ifmmode \check{R}\else \v{R}\fi{}eh\'a\ifmmode~\check{c}\else \v{c}\fi{}ek}\ and\ \citenamefont {Hradil}(2002)}]{hradil02}%
  \BibitemOpen
  \bibfield  {author} {\bibinfo {author} {\bibfnamefont {J.}~\bibnamefont {\ifmmode \check{R}\else \v{R}\fi{}eh\'a\ifmmode~\check{c}\else \v{c}\fi{}ek}}\ and\ \bibinfo {author} {\bibfnamefont {Z.}~\bibnamefont {Hradil}},\ }\bibfield  {title} {\bibinfo {title} {Invariant information and quantum state estimation},\ }\href {https://doi.org/10.1103/PhysRevLett.88.130401} {\bibfield  {journal} {\bibinfo  {journal} {Phys. Rev. Lett.}\ }\textbf {\bibinfo {volume} {88}},\ \bibinfo {pages} {130401} (\bibinfo {year} {2002})}\BibitemShut {NoStop}%
\bibitem [{\citenamefont {Vermersch}\ \emph {et~al.}(2019)\citenamefont {Vermersch}, \citenamefont {Elben}, \citenamefont {Sieberer}, \citenamefont {Yao},\ and\ \citenamefont {Zoller}}]{vermersch2019probing}%
  \BibitemOpen
  \bibfield  {author} {\bibinfo {author} {\bibfnamefont {B.}~\bibnamefont {Vermersch}}, \bibinfo {author} {\bibfnamefont {A.}~\bibnamefont {Elben}}, \bibinfo {author} {\bibfnamefont {L.~M.}\ \bibnamefont {Sieberer}}, \bibinfo {author} {\bibfnamefont {N.~Y.}\ \bibnamefont {Yao}},\ and\ \bibinfo {author} {\bibfnamefont {P.}~\bibnamefont {Zoller}},\ }\bibfield  {title} {\bibinfo {title} {Probing scrambling using statistical correlations between randomized measurements},\ }\href {https://journals.aps.org/prx/abstract/10.1103/PhysRevX.9.021061} {\bibfield  {journal} {\bibinfo  {journal} {Physical Review X}\ }\textbf {\bibinfo {volume} {9}},\ \bibinfo {pages} {021061} (\bibinfo {year} {2019})}\BibitemShut {NoStop}%
\bibitem [{\citenamefont {Blocher}\ \emph {et~al.}(2022)\citenamefont {Blocher}, \citenamefont {Asaad}, \citenamefont {Mourik}, \citenamefont {Johnson}, \citenamefont {Morello},\ and\ \citenamefont {M{\o}lmer}}]{blocher2022measuring}%
  \BibitemOpen
  \bibfield  {author} {\bibinfo {author} {\bibfnamefont {P.~D.}\ \bibnamefont {Blocher}}, \bibinfo {author} {\bibfnamefont {S.}~\bibnamefont {Asaad}}, \bibinfo {author} {\bibfnamefont {V.}~\bibnamefont {Mourik}}, \bibinfo {author} {\bibfnamefont {M.~A.}\ \bibnamefont {Johnson}}, \bibinfo {author} {\bibfnamefont {A.}~\bibnamefont {Morello}},\ and\ \bibinfo {author} {\bibfnamefont {K.}~\bibnamefont {M{\o}lmer}},\ }\bibfield  {title} {\bibinfo {title} {Measuring out-of-time-ordered correlation functions without reversing time evolution},\ }\href@noop {} {\bibfield  {journal} {\bibinfo  {journal} {Physical Review A}\ }\textbf {\bibinfo {volume} {106}},\ \bibinfo {pages} {042429} (\bibinfo {year} {2022})}\BibitemShut {NoStop}%
\bibitem [{\citenamefont {Sundar}\ \emph {et~al.}(2022)\citenamefont {Sundar}, \citenamefont {Elben}, \citenamefont {Joshi},\ and\ \citenamefont {Zache}}]{sundar2022proposal}%
  \BibitemOpen
  \bibfield  {author} {\bibinfo {author} {\bibfnamefont {B.}~\bibnamefont {Sundar}}, \bibinfo {author} {\bibfnamefont {A.}~\bibnamefont {Elben}}, \bibinfo {author} {\bibfnamefont {L.~K.}\ \bibnamefont {Joshi}},\ and\ \bibinfo {author} {\bibfnamefont {T.~V.}\ \bibnamefont {Zache}},\ }\bibfield  {title} {\bibinfo {title} {Proposal for measuring out-of-time-ordered correlators at finite temperature with coupled spin chains},\ }\href@noop {} {\bibfield  {journal} {\bibinfo  {journal} {New Journal of Physics}\ }\textbf {\bibinfo {volume} {24}},\ \bibinfo {pages} {023037} (\bibinfo {year} {2022})}\BibitemShut {NoStop}%
\bibitem [{\citenamefont {Parker}\ \emph {et~al.}(2019)\citenamefont {Parker}, \citenamefont {Cao}, \citenamefont {Avdoshkin}, \citenamefont {Scaffidi},\ and\ \citenamefont {Altman}}]{parker2019universal}%
  \BibitemOpen
  \bibfield  {author} {\bibinfo {author} {\bibfnamefont {D.~E.}\ \bibnamefont {Parker}}, \bibinfo {author} {\bibfnamefont {X.}~\bibnamefont {Cao}}, \bibinfo {author} {\bibfnamefont {A.}~\bibnamefont {Avdoshkin}}, \bibinfo {author} {\bibfnamefont {T.}~\bibnamefont {Scaffidi}},\ and\ \bibinfo {author} {\bibfnamefont {E.}~\bibnamefont {Altman}},\ }\bibfield  {title} {\bibinfo {title} {A universal operator growth hypothesis},\ }\href@noop {} {\bibfield  {journal} {\bibinfo  {journal} {Physical Review X}\ }\textbf {\bibinfo {volume} {9}},\ \bibinfo {pages} {041017} (\bibinfo {year} {2019})}\BibitemShut {NoStop}%
\bibitem [{\citenamefont {Yates}\ and\ \citenamefont {Mitra}(2021)}]{yates2021strong}%
  \BibitemOpen
  \bibfield  {author} {\bibinfo {author} {\bibfnamefont {D.~J.}\ \bibnamefont {Yates}}\ and\ \bibinfo {author} {\bibfnamefont {A.}~\bibnamefont {Mitra}},\ }\bibfield  {title} {\bibinfo {title} {Strong and almost strong modes of floquet spin chains in krylov subspaces},\ }\href@noop {} {\bibfield  {journal} {\bibinfo  {journal} {Physical Review B}\ }\textbf {\bibinfo {volume} {104}},\ \bibinfo {pages} {195121} (\bibinfo {year} {2021})}\BibitemShut {NoStop}%
\bibitem [{\citenamefont {Rabinovici}\ \emph {et~al.}(2021)\citenamefont {Rabinovici}, \citenamefont {S{\'a}nchez-Garrido}, \citenamefont {Shir},\ and\ \citenamefont {Sonner}}]{rabinovici2021operator}%
  \BibitemOpen
  \bibfield  {author} {\bibinfo {author} {\bibfnamefont {E.}~\bibnamefont {Rabinovici}}, \bibinfo {author} {\bibfnamefont {A.}~\bibnamefont {S{\'a}nchez-Garrido}}, \bibinfo {author} {\bibfnamefont {R.}~\bibnamefont {Shir}},\ and\ \bibinfo {author} {\bibfnamefont {J.}~\bibnamefont {Sonner}},\ }\bibfield  {title} {\bibinfo {title} {Operator complexity: a journey to the edge of krylov space},\ }\href@noop {} {\bibfield  {journal} {\bibinfo  {journal} {Journal of High Energy Physics}\ }\textbf {\bibinfo {volume} {2021}},\ \bibinfo {pages} {1} (\bibinfo {year} {2021})}\BibitemShut {NoStop}%
\bibitem [{\citenamefont {Noh}(2021)}]{noh2021operator}%
  \BibitemOpen
  \bibfield  {author} {\bibinfo {author} {\bibfnamefont {J.~D.}\ \bibnamefont {Noh}},\ }\bibfield  {title} {\bibinfo {title} {Operator growth in the transverse-field ising spin chain with integrability-breaking longitudinal field},\ }\href@noop {} {\bibfield  {journal} {\bibinfo  {journal} {Physical Review E}\ }\textbf {\bibinfo {volume} {104}},\ \bibinfo {pages} {034112} (\bibinfo {year} {2021})}\BibitemShut {NoStop}%
\bibitem [{\citenamefont {Dymarsky}\ and\ \citenamefont {Smolkin}(2021)}]{dymarsky2021krylov}%
  \BibitemOpen
  \bibfield  {author} {\bibinfo {author} {\bibfnamefont {A.}~\bibnamefont {Dymarsky}}\ and\ \bibinfo {author} {\bibfnamefont {M.}~\bibnamefont {Smolkin}},\ }\bibfield  {title} {\bibinfo {title} {Krylov complexity in conformal field theory},\ }\href@noop {} {\bibfield  {journal} {\bibinfo  {journal} {Physical Review D}\ }\textbf {\bibinfo {volume} {104}},\ \bibinfo {pages} {L081702} (\bibinfo {year} {2021})}\BibitemShut {NoStop}%
\bibitem [{\citenamefont {Caputa}\ \emph {et~al.}(2022)\citenamefont {Caputa}, \citenamefont {Magan},\ and\ \citenamefont {Patramanis}}]{caputa2022geometry}%
  \BibitemOpen
  \bibfield  {author} {\bibinfo {author} {\bibfnamefont {P.}~\bibnamefont {Caputa}}, \bibinfo {author} {\bibfnamefont {J.~M.}\ \bibnamefont {Magan}},\ and\ \bibinfo {author} {\bibfnamefont {D.}~\bibnamefont {Patramanis}},\ }\bibfield  {title} {\bibinfo {title} {Geometry of krylov complexity},\ }\href@noop {} {\bibfield  {journal} {\bibinfo  {journal} {Physical Review Research}\ }\textbf {\bibinfo {volume} {4}},\ \bibinfo {pages} {013041} (\bibinfo {year} {2022})}\BibitemShut {NoStop}%
\bibitem [{\citenamefont {Rabinovici}\ \emph {et~al.}(2022{\natexlab{a}})\citenamefont {Rabinovici}, \citenamefont {S{\'a}nchez-Garrido}, \citenamefont {Shir},\ and\ \citenamefont {Sonner}}]{rabinovici2022krylov}%
  \BibitemOpen
  \bibfield  {author} {\bibinfo {author} {\bibfnamefont {E.}~\bibnamefont {Rabinovici}}, \bibinfo {author} {\bibfnamefont {A.}~\bibnamefont {S{\'a}nchez-Garrido}}, \bibinfo {author} {\bibfnamefont {R.}~\bibnamefont {Shir}},\ and\ \bibinfo {author} {\bibfnamefont {J.}~\bibnamefont {Sonner}},\ }\bibfield  {title} {\bibinfo {title} {Krylov localization and suppression of complexity},\ }\href@noop {} {\bibfield  {journal} {\bibinfo  {journal} {Journal of High Energy Physics}\ }\textbf {\bibinfo {volume} {2022}},\ \bibinfo {pages} {1} (\bibinfo {year} {2022}{\natexlab{a}})}\BibitemShut {NoStop}%
\bibitem [{\citenamefont {Avdoshkin}\ \emph {et~al.}(2022)\citenamefont {Avdoshkin}, \citenamefont {Dymarsky},\ and\ \citenamefont {Smolkin}}]{avdoshkin2022krylov}%
  \BibitemOpen
  \bibfield  {author} {\bibinfo {author} {\bibfnamefont {A.}~\bibnamefont {Avdoshkin}}, \bibinfo {author} {\bibfnamefont {A.}~\bibnamefont {Dymarsky}},\ and\ \bibinfo {author} {\bibfnamefont {M.}~\bibnamefont {Smolkin}},\ }\bibfield  {title} {\bibinfo {title} {Krylov complexity in quantum field theory, and beyond},\ }\href@noop {} {\bibfield  {journal} {\bibinfo  {journal} {arXiv:2212.14429}\ } (\bibinfo {year} {2022})}\BibitemShut {NoStop}%
\bibitem [{\citenamefont {Rabinovici}\ \emph {et~al.}(2022{\natexlab{b}})\citenamefont {Rabinovici}, \citenamefont {S{\'a}nchez-Garrido}, \citenamefont {Shir},\ and\ \citenamefont {Sonner}}]{rabinovici2022k}%
  \BibitemOpen
  \bibfield  {author} {\bibinfo {author} {\bibfnamefont {E.}~\bibnamefont {Rabinovici}}, \bibinfo {author} {\bibfnamefont {A.}~\bibnamefont {S{\'a}nchez-Garrido}}, \bibinfo {author} {\bibfnamefont {R.}~\bibnamefont {Shir}},\ and\ \bibinfo {author} {\bibfnamefont {J.}~\bibnamefont {Sonner}},\ }\bibfield  {title} {\bibinfo {title} {K-complexity from integrability to chaos},\ }\href@noop {} {\bibfield  {journal} {\bibinfo  {journal} {arXiv:2207.07701}\ } (\bibinfo {year} {2022}{\natexlab{b}})}\BibitemShut {NoStop}%
\bibitem [{\citenamefont {Bhattacharya}\ \emph {et~al.}(2022)\citenamefont {Bhattacharya}, \citenamefont {Nandy}, \citenamefont {Nath},\ and\ \citenamefont {Sahu}}]{bhattacharya2022operator}%
  \BibitemOpen
  \bibfield  {author} {\bibinfo {author} {\bibfnamefont {A.}~\bibnamefont {Bhattacharya}}, \bibinfo {author} {\bibfnamefont {P.}~\bibnamefont {Nandy}}, \bibinfo {author} {\bibfnamefont {P.~P.}\ \bibnamefont {Nath}},\ and\ \bibinfo {author} {\bibfnamefont {H.}~\bibnamefont {Sahu}},\ }\bibfield  {title} {\bibinfo {title} {Operator growth and krylov construction in dissipative open quantum systems},\ }\href@noop {} {\bibfield  {journal} {\bibinfo  {journal} {Journal of High Energy Physics}\ }\textbf {\bibinfo {volume} {2022}},\ \bibinfo {pages} {1} (\bibinfo {year} {2022})}\BibitemShut {NoStop}%
\bibitem [{\citenamefont {Bhattacharya}\ \emph {et~al.}(2023)\citenamefont {Bhattacharya}, \citenamefont {Nandy}, \citenamefont {Nath},\ and\ \citenamefont {Sahu}}]{bhattacharya2023krylov}%
  \BibitemOpen
  \bibfield  {author} {\bibinfo {author} {\bibfnamefont {A.}~\bibnamefont {Bhattacharya}}, \bibinfo {author} {\bibfnamefont {P.}~\bibnamefont {Nandy}}, \bibinfo {author} {\bibfnamefont {P.~P.}\ \bibnamefont {Nath}},\ and\ \bibinfo {author} {\bibfnamefont {H.}~\bibnamefont {Sahu}},\ }\bibfield  {title} {\bibinfo {title} {On krylov complexity in open systems: an approach via bi-lanczos algorithm},\ }\href@noop {} {\bibfield  {journal} {\bibinfo  {journal} {arXiv:2303.04175}\ } (\bibinfo {year} {2023})}\BibitemShut {NoStop}%
\bibitem [{\citenamefont {Suchsland}\ \emph {et~al.}(2023)\citenamefont {Suchsland}, \citenamefont {Moessner},\ and\ \citenamefont {Claeys}}]{suchsland2023krylov}%
  \BibitemOpen
  \bibfield  {author} {\bibinfo {author} {\bibfnamefont {P.}~\bibnamefont {Suchsland}}, \bibinfo {author} {\bibfnamefont {R.}~\bibnamefont {Moessner}},\ and\ \bibinfo {author} {\bibfnamefont {P.~W.}\ \bibnamefont {Claeys}},\ }\bibfield  {title} {\bibinfo {title} {Krylov complexity and trotter transitions in unitary circuit dynamics},\ }\href@noop {} {\bibfield  {journal} {\bibinfo  {journal} {arXiv:2308.03851}\ } (\bibinfo {year} {2023})}\BibitemShut {NoStop}%
\bibitem [{\citenamefont {Nie}\ \emph {et~al.}(2019)\citenamefont {Nie}, \citenamefont {Nozaki}, \citenamefont {Ryu},\ and\ \citenamefont {Tan}}]{nie2019signature}%
  \BibitemOpen
  \bibfield  {author} {\bibinfo {author} {\bibfnamefont {L.}~\bibnamefont {Nie}}, \bibinfo {author} {\bibfnamefont {M.}~\bibnamefont {Nozaki}}, \bibinfo {author} {\bibfnamefont {S.}~\bibnamefont {Ryu}},\ and\ \bibinfo {author} {\bibfnamefont {M.~T.}\ \bibnamefont {Tan}},\ }\bibfield  {title} {\bibinfo {title} {Signature of quantum chaos in operator entanglement in 2d cfts},\ }\href@noop {} {\bibfield  {journal} {\bibinfo  {journal} {Journal of Statistical Mechanics: Theory and Experiment}\ }\textbf {\bibinfo {volume} {2019}},\ \bibinfo {pages} {093107} (\bibinfo {year} {2019})}\BibitemShut {NoStop}%
\bibitem [{\citenamefont {Wang}\ and\ \citenamefont {Zhou}(2019)}]{wang2019barrier}%
  \BibitemOpen
  \bibfield  {author} {\bibinfo {author} {\bibfnamefont {H.}~\bibnamefont {Wang}}\ and\ \bibinfo {author} {\bibfnamefont {T.}~\bibnamefont {Zhou}},\ }\bibfield  {title} {\bibinfo {title} {Barrier from chaos: operator entanglement dynamics of the reduced density matrix},\ }\href@noop {} {\bibfield  {journal} {\bibinfo  {journal} {Journal of High Energy Physics}\ }\textbf {\bibinfo {volume} {2019}},\ \bibinfo {pages} {1} (\bibinfo {year} {2019})}\BibitemShut {NoStop}%
\bibitem [{\citenamefont {Alba}\ \emph {et~al.}(2019)\citenamefont {Alba}, \citenamefont {Dubail},\ and\ \citenamefont {Medenjak}}]{alba2019operator}%
  \BibitemOpen
  \bibfield  {author} {\bibinfo {author} {\bibfnamefont {V.}~\bibnamefont {Alba}}, \bibinfo {author} {\bibfnamefont {J.}~\bibnamefont {Dubail}},\ and\ \bibinfo {author} {\bibfnamefont {M.}~\bibnamefont {Medenjak}},\ }\bibfield  {title} {\bibinfo {title} {Operator entanglement in interacting integrable quantum systems: the case of the rule 54 chain},\ }\href@noop {} {\bibfield  {journal} {\bibinfo  {journal} {Physical review letters}\ }\textbf {\bibinfo {volume} {122}},\ \bibinfo {pages} {250603} (\bibinfo {year} {2019})}\BibitemShut {NoStop}%
\bibitem [{\citenamefont {McCulloch}\ and\ \citenamefont {Von~Keyserlingk}(2022)}]{mcculloch2022operator}%
  \BibitemOpen
  \bibfield  {author} {\bibinfo {author} {\bibfnamefont {E.}~\bibnamefont {McCulloch}}\ and\ \bibinfo {author} {\bibfnamefont {C.}~\bibnamefont {Von~Keyserlingk}},\ }\bibfield  {title} {\bibinfo {title} {Operator spreading in the memory matrix formalism},\ }\href@noop {} {\bibfield  {journal} {\bibinfo  {journal} {Journal of Physics A: Mathematical and Theoretical}\ }\textbf {\bibinfo {volume} {55}},\ \bibinfo {pages} {274007} (\bibinfo {year} {2022})}\BibitemShut {NoStop}%
\bibitem [{\citenamefont {Balasubramanian}\ \emph {et~al.}(2022)\citenamefont {Balasubramanian}, \citenamefont {Caputa}, \citenamefont {Magan},\ and\ \citenamefont {Wu}}]{balasubramanian2022quantum}%
  \BibitemOpen
  \bibfield  {author} {\bibinfo {author} {\bibfnamefont {V.}~\bibnamefont {Balasubramanian}}, \bibinfo {author} {\bibfnamefont {P.}~\bibnamefont {Caputa}}, \bibinfo {author} {\bibfnamefont {J.~M.}\ \bibnamefont {Magan}},\ and\ \bibinfo {author} {\bibfnamefont {Q.}~\bibnamefont {Wu}},\ }\bibfield  {title} {\bibinfo {title} {Quantum chaos and the complexity of spread of states},\ }\href@noop {} {\bibfield  {journal} {\bibinfo  {journal} {Physical Review D}\ }\textbf {\bibinfo {volume} {106}},\ \bibinfo {pages} {046007} (\bibinfo {year} {2022})}\BibitemShut {NoStop}%
\bibitem [{\citenamefont {Sahu}\ \emph {et~al.}(2023)\citenamefont {Sahu}, \citenamefont {Varikuti}, \citenamefont {Das},\ and\ \citenamefont {Madhok}}]{sahu2023quantifying}%
  \BibitemOpen
  \bibfield  {author} {\bibinfo {author} {\bibfnamefont {A.}~\bibnamefont {Sahu}}, \bibinfo {author} {\bibfnamefont {N.~D.}\ \bibnamefont {Varikuti}}, \bibinfo {author} {\bibfnamefont {B.~K.}\ \bibnamefont {Das}},\ and\ \bibinfo {author} {\bibfnamefont {V.}~\bibnamefont {Madhok}},\ }\bibfield  {title} {\bibinfo {title} {Quantifying operator spreading and chaos in krylov subspaces with quantum state reconstruction},\ }\href@noop {} {\bibfield  {journal} {\bibinfo  {journal} {Physical Review B}\ }\textbf {\bibinfo {volume} {108}},\ \bibinfo {pages} {224306} (\bibinfo {year} {2023})}\BibitemShut {NoStop}%
\bibitem [{\citenamefont {Espa{\~n}ol}\ and\ \citenamefont {Wisniacki}(2023)}]{espanol2023assessing}%
  \BibitemOpen
  \bibfield  {author} {\bibinfo {author} {\bibfnamefont {B.~L.}\ \bibnamefont {Espa{\~n}ol}}\ and\ \bibinfo {author} {\bibfnamefont {D.~A.}\ \bibnamefont {Wisniacki}},\ }\bibfield  {title} {\bibinfo {title} {Assessing the saturation of krylov complexity as a measure of chaos},\ }\href@noop {} {\bibfield  {journal} {\bibinfo  {journal} {Physical Review E}\ }\textbf {\bibinfo {volume} {107}},\ \bibinfo {pages} {024217} (\bibinfo {year} {2023})}\BibitemShut {NoStop}%
\bibitem [{\citenamefont {Pesin}(1977)}]{pesin1977characteristic}%
  \BibitemOpen
  \bibfield  {author} {\bibinfo {author} {\bibfnamefont {Y.~B.}\ \bibnamefont {Pesin}},\ }\bibfield  {title} {\bibinfo {title} {Characteristic lyapunov exponents and smooth ergodic theory},\ }\href@noop {} {\bibfield  {journal} {\bibinfo  {journal} {Russian Mathematical Surveys}\ }\textbf {\bibinfo {volume} {32}},\ \bibinfo {pages} {55} (\bibinfo {year} {1977})}\BibitemShut {NoStop}%
\bibitem [{\citenamefont {Hauke}\ \emph {et~al.}(2012)\citenamefont {Hauke}, \citenamefont {Cucchietti}, \citenamefont {Tagliacozzo}, \citenamefont {Deutsch},\ and\ \citenamefont {Lewenstein}}]{hauke2012can}%
  \BibitemOpen
  \bibfield  {author} {\bibinfo {author} {\bibfnamefont {P.}~\bibnamefont {Hauke}}, \bibinfo {author} {\bibfnamefont {F.~M.}\ \bibnamefont {Cucchietti}}, \bibinfo {author} {\bibfnamefont {L.}~\bibnamefont {Tagliacozzo}}, \bibinfo {author} {\bibfnamefont {I.}~\bibnamefont {Deutsch}},\ and\ \bibinfo {author} {\bibfnamefont {M.}~\bibnamefont {Lewenstein}},\ }\bibfield  {title} {\bibinfo {title} {Can one trust quantum simulators?},\ }\href@noop {} {\bibfield  {journal} {\bibinfo  {journal} {Reports on Progress in Physics}\ }\textbf {\bibinfo {volume} {75}},\ \bibinfo {pages} {082401} (\bibinfo {year} {2012})}\BibitemShut {NoStop}%
\bibitem [{\citenamefont {Chaudhury}\ \emph {et~al.}(2009{\natexlab{b}})\citenamefont {Chaudhury}, \citenamefont {Smith}, \citenamefont {Anderson}, \citenamefont {Ghose},\ and\ \citenamefont {Jessen}}]{chaudhury2009quantum}%
  \BibitemOpen
  \bibfield  {author} {\bibinfo {author} {\bibfnamefont {S.}~\bibnamefont {Chaudhury}}, \bibinfo {author} {\bibfnamefont {A.}~\bibnamefont {Smith}}, \bibinfo {author} {\bibfnamefont {B.}~\bibnamefont {Anderson}}, \bibinfo {author} {\bibfnamefont {S.}~\bibnamefont {Ghose}},\ and\ \bibinfo {author} {\bibfnamefont {P.~S.}\ \bibnamefont {Jessen}},\ }\bibfield  {title} {\bibinfo {title} {Quantum signatures of chaos in a kicked top},\ }\href@noop {} {\bibfield  {journal} {\bibinfo  {journal} {Nature}\ }\textbf {\bibinfo {volume} {461}},\ \bibinfo {pages} {768} (\bibinfo {year} {2009}{\natexlab{b}})}\BibitemShut {NoStop}%
\end{thebibliography}%

\end{document}